\documentclass[
11pt, 
oneside,
english, 
doublespacing,
toctotoc, 
headsepline, 
]{MastersDoctoralThesis} 
\usepackage[utf8]{inputenc} 
\usepackage[T1]{fontenc} 
\usepackage[final]{feynmp}
\usepackage{feynmp-auto}
\usepackage{hyperref} 
\usepackage{amsmath}
\usepackage{amssymb}
\usepackage{graphicx}
\usepackage{subcaption}
\usepackage{tikz}
\usetikzlibrary{decorations.pathmorphing,decorations.markings,trees,arrows,patterns}
\usepackage{wrapfig}
\usepackage{comment}
\usepackage{color}
\usepackage{slashed}
\usepackage{braket}
\usepackage{simplewick}
\usepackage{cleveref}
\usepackage[export]{adjustbox}
\usepackage[backend=bibtex,style=ieee,citestyle=numeric-comp,natbib=true]{biblatex} 
\usepackage{arabtex}
\usepackage{utf8}
\setcode{utf8}
\hypersetup{
    pdffitwindow=true,      
    pdfauthor={Abdullah Khalil},     
    pdfnewwindow=true,      
    bookmarksnumbered=true,
    colorlinks=true,       
    citecolor=blue, 
    urlcolor=cyan,           
}
\usepackage{soul}
\renewbibmacro*{bbx:savehash}{} 
\usepackage[autostyle=true]{csquotes} 
\makeatletter
\newcommand*\rel@kern[1]{\kern#1\dimexpr\macc@kerna}
\newcommand*\widebar[1]{%
  \begingroup
  \def\mathaccent##1##2{%
    \rel@kern{0.8}%
    \overline{\rel@kern{-0.8}\macc@nucleus\rel@kern{0.2}}%
    \rel@kern{-0.2}%
  }%
  \macc@depth\@ne
  \let\math@bgroup\@empty \let\math@egroup\macc@set@skewchar
  \mathsurround\z@ \frozen@everymath{\mathgroup\macc@group\relax}%
  \macc@set@skewchar\relax
  \let\mathaccentV\macc@nested@a
  \macc@nested@a\relax111{#1}%
  \endgroup
}
\makeatother

\thesistitle{NLO Rutherford Scattering and the Kinoshita-Lee-Nauenberg Theorem}
\supervisor{Dr. W. A. Horowitz} 
\examiner{} 
\degree{Master of Science} 
\author{Abdullah Khalil Hassan \textsc{Ibrahim}} 
\addresses{} 

\subject{Theoretical Physics} 
\university{University of Cape Town} 
\department{Department of Physics} 
\group{Theoretical Physics} 
\faculty{Faculty of Science} 

\hypersetup{pdftitle=\ttitle} 
\hypersetup{pdfauthor=\authorname} 
\hypersetup{pdfkeywords=\keywordnames} 

\begin{document}

\frontmatter 

\pagestyle{plain} 


\begin{titlepage}
\begin{center}
\includegraphics[width = 0.3\textwidth]{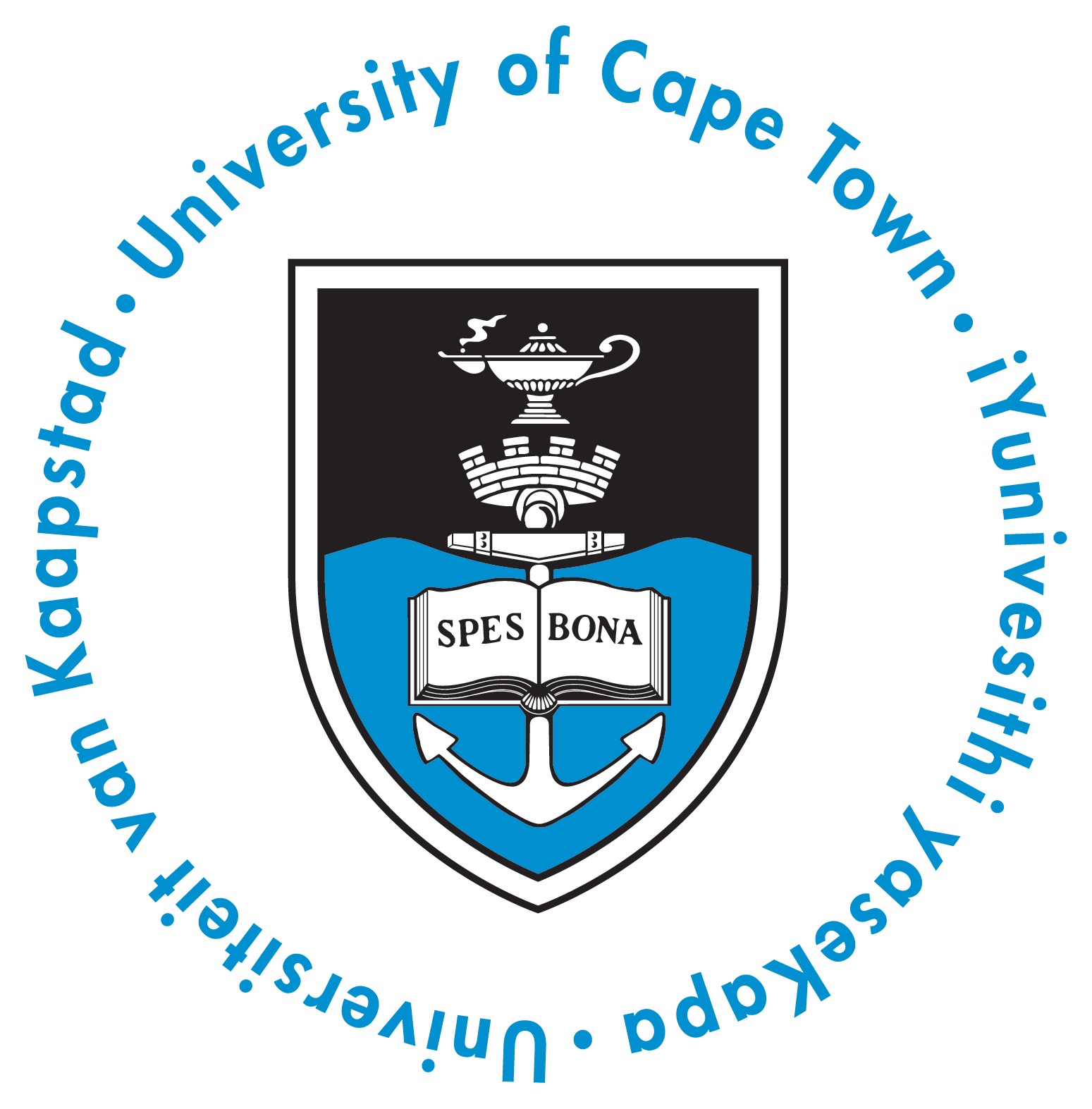}
\end{center}
\begin{center}

{\scshape\LARGE \univname\par}\vspace{1.5cm} 
\textsc{\Large Master's Thesis}\\[0.5cm] 

\HRule \\[0.4cm] 
{\huge \bfseries \ttitle\par}\vspace{0.4cm} 
\HRule \\[1cm] 
 
\begin{minipage}[t]{0.5\textwidth}
\begin{flushleft} \large
\emph{Author:}\\
{\authorname} 
\end{flushleft}
\end{minipage}
\begin{minipage}[t]{0.4\textwidth}
\begin{flushright} \large
\emph{Supervisor:} \\
{\supname} 
\end{flushright}
\end{minipage}\\[2cm]
 
\large \textit{A thesis submitted in fulfillment of the requirements\\ for the degree of {\degreename} in theoretical physics}\\[0.2cm] 
\textit{in the}\\[0.2cm]
\deptname\\[1cm] 
 
{\large March 27, 2017}

\vfill
\end{center}
\end{titlepage}

\cleardoublepage


\begin{abstract}
\addchaptertocentry{\abstractname} 
We calculate to next-to-leading order accuracy the high-energy elastic scattering cross section for an electron off of a classical point source. We use the $\overline{\mathrm{MS}}$ renormalization scheme to tame the ultraviolet divergences while the infrared singularities are dealt with using the well known Kinoshita-Lee-Nauenberg theorem. 

We show for the first time how to correctly apply the Kinoshita-Lee-Nauenberg theorem diagrammatically in a next-to-leading order scattering process. We improve on previous works by including all initial and final state soft radiative processes, including absorption and an infinite sum of partially disconnected amplitudes. Crucially, we exploit the Monotone Convergence Theorem to prove that our delicate rearrangement of this formally divergent series is uniquely correct. This rearrangement yields a factorization of the infinite contribution from the initial state soft photons that then cancels in the physically observable cross section.

 Since we use the $\overline{\mathrm{MS}}$ renormalization scheme, our result is valid up to arbitrarily large momentum transfers between the source and the scattered electron as long as $\alpha \log(1/\delta)\ll 1$ and $\alpha \log(1/\delta)\log(\Delta/E)\ll 1$, where $\Delta$ and $\delta$ are the experimental energy and angular resolutions, respectively, and $E$ is the energy of the scattered electron. Our work aims at computing the NLO corrections to the energy loss of a high energetic parton propagating in a quark-gluon plasma. 
\end{abstract}


\begin{acknowledgements}
\addchaptertocentry{\acknowledgementname} 
I would like to express my great gratitude to my supervisor, Dr.~W.~A.~Horowitz for his guidance and encouragement throughout this work which helped me to advance my research abilities. Many thanks to the African Institute for Mathematical Sciences (AIMS) and the University of Cape Town for their financial support. I would also like to thank the South African National Research Foundation and the SA-CERN consortium for their support to attend the workshops and conferences that helped me completing this research. I am grateful to Raju Venugopalan, Larry McLerran, Robert de Mello Koch, and Stanley Brodsky for useful discussions during my work. Finally, I would especially thank my wife Hager Elboghdady, my parents, brothers, and sisters for their immense support and love. 
\begin{RLtext} 
الحمد لله الذي بنعمته تتم الصالحات.
\end{RLtext}
\end{acknowledgements}


\tableofcontents 



\mainmatter 

\pagestyle{thesis} 


\chapter{Introduction} 

\label{Chapter1} 


\newcommand{\keyword}[1]{\textbf{#1}}
\newcommand{\tabhead}[1]{\textbf{#1}}
\newcommand{\code}[1]{\texttt{#1}}
\newcommand{\file}[1]{\texttt{\bfseries#1}}
\newcommand{\option}[1]{\texttt{\itshape#1}}
\newcommand{\fig}[1]{Fig.~\ref{fig:#1}}
\newcommand{\figtwo}[2]{Fig.\ \ref{#1} (\subref{#2})}


\section{Motivation and Objectives}
Hadrons are made up of quarks and anti-quarks bound together by the strong interaction through exchanging gluons. Quantum chromodynamics (QCD) has been known for a long time to be the accepted theory to describe this interaction \cite{Fritzsch:1972jv,Weinberg:1973un,Marciano:1977su}. QCD is an elegant and self-consistent theory where the coupling strength becomes \emph{weaker} as quarks approach one another. A well-known behavior is known as the asymptotic freedom \cite{Politzer:1973fx,Gross:1973id,Gross:1973ju,Gross:1974cs} opened the road to defining the theory completely, at short distances, in terms of the fundamental microscopic degrees of freedom: quarks and gluons. 

Asymptotic freedom becomes more important at high temperatures, where one of the fundamental results from QCD is the existence of a new state of matter called the quark-gluon plasma (QGP) \cite{Collins:1974ky,Baym:1976yu,Shuryak:1980tp}. The QGP is predicted to be formed at very high energy densities, exceeding the energy density inside the atomic nuclei by an order of magnitude 1-10 GeV$/fm^3$, and at temperatures of order $\sim 170$ MeV \cite{Karsch:2001cy,Karsch:2003jg}. 
\begin{figure}[h]
\centering
\includegraphics[scale=0.5]{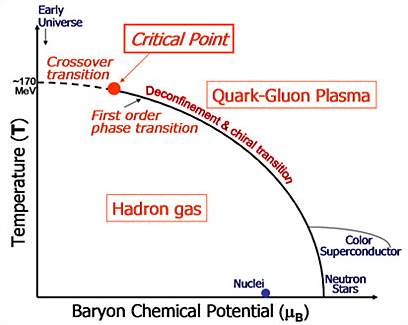}
\caption{A schematic of the QCD phase diagram of nuclear matter \cite{Nayak:2009jp}.}
\label{fig:QCDphd}
\end{figure}
The transition from the deconfined state of quarks and gluons to the QGP state is described by the QCD phase diagram as shown in \Cref{fig:QCDphd}, which shows a phase transition happened above the critical temperature ($T_c$). It is believed that the QGP was the state of the universe a few microseconds after the Big Bang \cite{Shuryak:1980tp,Stock:1989jp}, which made the study of the QCD under these extreme conditions very important.

\begin{figure}[h]
\centering
\includegraphics[scale=0.8]{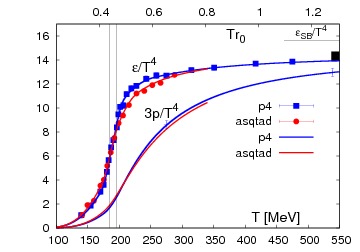}
\caption{Dependence of the energy density as a function of the temperature of the hadronic matter at null baryonic potential given by lattice QCD calculations at finite temperature \cite{Bazavov:2009zn}.}
\label{EDT}
\end{figure}

One of the methods to study the QCD phase transition is on the lattice \cite{Karsch:2001cy,Karsch:2003jg}. The numerical calculations of the lattice QCD predicted the temperature dependence of the energy density $\varepsilon$ at zero baryon density $\mu_B$. Stefan-Boltzmann law predicted that $\varepsilon/T^4$ is proportional to the number of degrees of freedom of a given thermal system. \Cref{EDT} shows that this ratio, as predicted by the lattice QCD, experience a rapid change near the critical temperature which has been interpreted as the change in the number of degrees of freedom in the system. Well below $T_c$, there are three hadronic degrees of freedom due to the three lightest hadrons: $\pi^+$, $\pi^-$ and $\pi^0$. Well above $T_c$, there are $2(N_c^2 - 1)$ gluon degrees of freedom and $2\times 2\times N_c\times N_f$ quark degrees of freedom from the fundamental gluons and quarks of the theory. One can also note from \Cref{EDT} that there are about $\sim 40 \approx 52 \times 80\ \%$ degrees of freedom in the region of $(1-3)T_c$ predicted by lattice QCD, with a $20 \ \%$ reduction compared to the Stephan-Boltzmann limit of zero coupling ideal gas.

There are two potentially analytically accessible limits for the dynamics of the QGP produced at RHIC and LHC. First is the strongly coupled limit, working in the strongly coupled limit gives a good estimate for the dynamics of particles at low $p_{\perp}$, where $p_{\perp}$ is the component of the particle's momentum transverse to the beam direction \cite{Gale:2012rq, Kovtun:2004de}. Second is the weakly coupled limit, which is due to the asymptotic freedom of QCD appears to describe the physics associated with high $p_\perp$ particles. In particular, the study of high $p_\perp$ particles falls under the term `jet quenching' or `jet tomography,' which is one postulated means of investigating the degrees of freedom in a QGP in detail \cite{Majumder:2010qh,Collins:2011zzd,Rak:2013yta}. 

The high $p_\perp$ data from the Relativistic Heavy Ion Collider (RHIC) at the Brookhaven National Laboratory and the Large Hadron Collider (LHC) at CERN have been interpreted as evidence that jet quenching is due to final state energy loss, which is qualitatively well described by leading-order pQCD methods \cite{Baier:2000mf,Gyulassy:2000er,Djordjevic:2003zk,Djordjevic:2007at,Djordjevic:2009cr,Majumder:2010qh,vanLeeuwen:2010ti}. We wish to check the self-consistency of these pQCD results and to make the pQCD calculation more quantitative. To accomplish these two goals, we must compute the next-to-leading order contribution to the energy loss of partons in a QGP. As a first step towards this NLO pQCD calculation, we compute the NLO corrections to the elastic scattering of an electron off of a static source. 

During the rest of this chapter, we give an introduction to the obstacles that faces the NLO calculations such as the usual ultraviolet (UV) and infrared (IR) divergences as well as the different methods to deal with these infinities. In \autoref{Chapter2} we provide the general formalism of the system by calculating a general formula for the differential cross section in terms of the Feynman amplitudes. In \autoref{Chapter3} we renormalize the QED Lagrangian density in order to remove the UV divergences. In \autoref{Chapter4} we give an overview of the possible approaches that have been used to get rid of the IR divergences, while in \autoref{Chapter5} we give for the first time a complete diagrammatic way to tame the IR divergences through the implementation of the Kinoshita-Lee-Nauenberg theorem and collect all the contributions to the differential cross section at NLO. Furthermore, we provide in \autoref{Chapter6} a non-trivial check of the validity of the final formula of the differential cross section through the application of the Callan-Symanzik equation. Finally, we give our concluding remarks in \autoref{Chapter7}.  

\section{Singularities in Perturbative Field Theories}
In perturbative quantum field theories, the tree-level contribution is finite while the next-to-leading (NLO) order contributions diverge in the ultraviolet and infrared limits \cite{Bjorken:1965zz}. These divergences appear either from the momentum loop integrals or the emission or absorption of soft particles at NLO corrections. 
\subsection{Ultraviolet Divergences}
The leading term (tree level) of the perturbation theory consists of diagrams where all momenta of the internal propagators are well defined in terms of the external momenta. However, as we go further in the perturbation series, the Feynman diagrams become topologically more complicated and may contain internal propagators whose momenta are not defined in terms of the external momenta in the form of a loop propagator. In this case, the Feynman amplitudes may lead to a divergence at very high energies (i.e. loop-momentum $\rightarrow \infty$) \cite{Bjorken:1965zz}. This kind of divergence is called an ultraviolet divergence (UV) due to the contribution of the very high energy particles in the process. The degree of the UV divergence depends on the number of internal propagators whose momenta are not determined in terms of the external lines momenta. In QED, the degree of divergence can be determined in terms of the number of external lines (electrons or photons). 

A physical interpretation of the UV divergences is that the fields and parameters defined in the Lagrangian are not the physical ones. The UV divergences elimination require matching between the Lagrangian fields and parameters and the observable ones. 
\subsection{Infra-red Divergences}
Infrared divergences (IR) in gauge theories arise in two forms: soft, due to the massless nature of the radiation (e.g. the massless photon in QED), and collinear, which comes from treating the radiating particle as massless (e.g. the electron in QED) \cite{Bjorken:1965zz}. The soft divergences appear when the radiation energy is less than some experimental energy resolution $\Delta$ in such a way it escapes the detection. While the collinear singularities appear when it is absorbed or emitted collinearly from the radiator so that it can not be distinguished from the radiator.

\begin{figure}[h]
\centering
\includegraphics[scale=0.7]{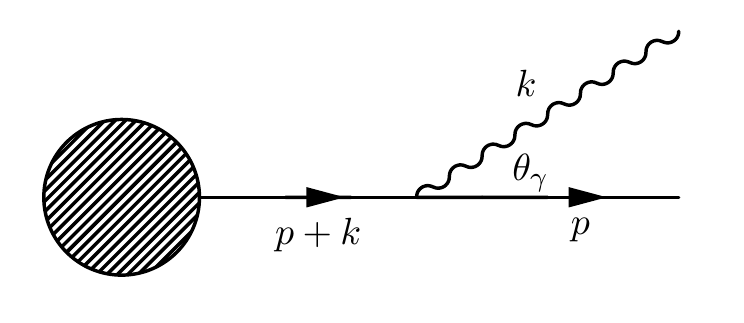}
\caption[Electron bremsstrahlung]{The electron bremsstrahlung process.}
\label{fig:ebrem}
\end{figure}
\Cref{fig:ebrem} shows the emission of a photon from a fast moving electron, for $k\rightarrow 0$  the electron propagator behaves as $\frac{1}{p\cdot k}$ and this causes the soft divergence discussed above. However, if the electron is massless the propagator becomes $\frac{1}{\left|\vec{p}\right|\,\left|\vec{k}\right|\,(1-\cos \theta_{\gamma})}$ where $\theta_{\gamma}$ is the angle between the electron and photon three momenta $\vec{p}$ and $\vec{k}$ respectively. There is now a double singularity, one as $k\rightarrow 0$ and the other as $\theta_{\gamma} \rightarrow 0$. Which means that the mass singularity happened when the photon is emitted or absorbed collinearly with the electron even if the photon is hard (high-energy photon), we sometimes call this singularity as the collinear divergence. Practically, this can be treated by considering only physically observable cross sections. We give a detailed description of the different approaches made to get rid of the IR divergences in \autoref{Chapter4}.

\section{Mathematical Tools in Removing the Infinities}
As discussed in the previous section, infinities in loop corrections are ubiquitous. In order to eliminate these divergences, we follow two main steps. First, we render the divergent integrals finite by introducing a regulator. Second, we apply a renormalization scheme to remove the regulated UV divergences, and thus correct the desired quantities physically observable cross sections. 
\subsection{Regularization Schemes}
Regularization is a mathematical technique which renders divergent Feynman amplitudes finite. The divergent integrals are then said to be regularized. We provide here an overview of the most prevalent regularization schemes focusing on the ones we use in this thesis.
\begin{itemize}
\item[\emph{(a)}]\emph{Pauli-Villars Regularization:}

Pauli and Villars \cite{Pauli:1949zm} proposed one of the first regularization procedures. They introduced an auxiliary mass as a regulator, which allowed them to rewrite Feynman propagators in such a way that the Feynman amplitudes beyond the leading order were finite. This auxiliary mass has no physical meaning which means that the method is only for defining the divergent integrals and the regulator must disappear in the final result of the cross section.

\item[\emph{(b)}]\emph{Analytic Regularization:}

 Analytic regularization is a procedure in which one replaces the Feynman propagator $\frac{1}{p^2-m^2+i\epsilon}$ of a particle of four-momentum $p$ and mass $m$ by $\frac{1}{(p^2-m^2+i\epsilon)^{\alpha}}$ where $\alpha \in \mathbb{C}$ is the regulator; the result then has a pole at $\alpha = 1$. This procedure leads to a convergent result of the Feynman amplitude as a well behaved analytic function of $\alpha$. Analytic regularization was first introduced by Bollini et al.\ in \cite{bollini1964analytic} and investigated in further details by Speer \cite{speer1968analytic}; this method has also been modified in such a way that it gives a gauge invariant result to all orders in perturbation theory \cite{breitenlohner1968analytic}.
\item[\emph{(c)}]\emph{Dimensional Regularization:}
 
G.~'t Hooft and J.~G.~Veltman \cite{tHooft:1972tcz} came up with an elegant regularization procedure based on the fact that the Ward identity holds without regard to the number of space dimensions. The idea is to let the loop momentum variables have $d$-components and then to calculate the loop integrals in $d$-dimensions which lead to well-defined S-matrix elements in the limit $d\rightarrow 4$. The power of using dimensional regularization is that it preserves Lorentz invariance, gauge invariance, and the Ward identity.  
\end{itemize}

In this project, we use two different procedures to regularize the UV and IR divergences. We regularize the soft IR divergences by introducing a fictitious photon mass $m_{\gamma}$. We thus replace the photon propagator $\frac{-g_{\mu\nu}}{k^2+i\epsilon}$ with $\frac{-g_{\mu\nu}}{k^2-m_{\gamma}^2+i\epsilon}$. We also keep the electron mass $m_e$ finite for the moment to regularize the expected collinear divergences. We regularize the UV divergences using dimensional regularization in which we replace the loop momentum integration in $4$-dimensions $\int\frac{d^4p}{(2\pi)^4}$ by an integral in $d$-dimensions $\int\frac{d^dp}{(2\pi)^d}$.

It is important to emphasize that in $d$-dimensions the electron charge $e$ has the dimension of mass to the power $\left(\frac{4-d}{2}\right)$. We must then ensure that $e$ remains dimensionless by choosing an arbitrary mass scale $\mu$; notice that the physical observables should not depend on this mass scale \cite{itzykson2006quantum}. So we set $e \longrightarrow \mu^{(4-d)/2}e$, and as $d\rightarrow4$ we write $e^2\longrightarrow \mu^{\epsilon} e^2$. 
\subsection{Renormalization and Renormalization Schemes}
Renormalization is the study of how a system changes under change of the observation scale. Renormalization rescales the various parameters of the theory (e.g.\ masses, coupling constant, etc.) in order to remove UV divergences. Renormalization theory then ensures that the expressions for the Green functions are finite when expressed in terms of the physical (observed) quantities \cite{Sterman:1994ce}.

In other words, one may set two different objectives for the renormalization process: The first is a mathematical objective where it eliminates the UV divergences from the loop integrals for a given theory in the higher orders in perturbation. The second is the physical objective by matching the observed quantities and the parameters that appear in that given theory. Dyson \cite{dyson1949radiation,dyson1736dyson} and Salam \cite{salam1951overlapping,Salam:1951sj} introduced the first successful renormalization technique by matching the mass and charge in QED to their observed values. In \autoref{Chapter3}, we give a brief comparison between the on-shell (OS) and the modified minimal subtraction $(\mathrm{\overline{MS}})$ renormalization schemes as the most common schemes used in perturbative field theories. We focus on the $\mathrm{\overline{MS}}$ renormalization scheme because we are interested in the very high energy limit (i.e.\ massless limit). 

\chapter{The General Formalism} 

\label{Chapter2}
\newcommand{\eq}[1]{Eq.~(\ref{eq:#1})}
\newcommand{\normord}[1]{:\mathrel{#1}:}


\section{External Field Approximation}
A well-known approximation in QED is the external field approximation in which we expand the photon field $A_{\mu}$ around a non-zero value. We then are able to consider a scattering process of, for example, an electron off of a classical current source $J_{\nu}$. It is shown that this is equivalent to scattering with a heavy charged particle	 \cite{Schwinger:1949ra}. We use this approximation to construct the Lagrangian that describes the formalism of a QED system.

Consider an electron scattering off of a static point charge described by the current $J^{\mu}(x) = V^{\mu} \delta^{(3)}(\vec{x}-\vec{V}x^0)$, where $V^{\mu}= \left(1,\vec{0}\right)^{\mu}$ is the unit time-like velocity vector. The Lagrangian density then becomes the Lagrangian of the normal QED process with a modified interaction term, given by
\begin{equation}
\mathcal{L} = -\frac{1}{4}F^{\mu \nu} F_{\mu \nu}+\widebar{\psi}\left(i \slashed \partial-m\right)\psi-e\bar{\psi}\gamma^{\mu} \psi A_{\mu}+e J^{\mu}A_{\mu}.
\label{qedlagrangian}
\end{equation}    
\section{Leading Term Calculations}
\subsection{Tree Level Amplitude}
Let $\ket{\vec{p}\,',s'}$ and $\ket{\vec{p},s}$ be the final and the initial state with spins $s'$ and $s$ respectively. Then the elements of the scattering matrix are given in terms of the transition matrix elements \cite{Peskin:1995ev}
\begin{equation}
\bra{\vec{p}\,',s'}S\ket{\vec{p},s} = \left<\vec{p}\,',s'|\vec{p},s\right>+\bra{\vec{p}\,',s'}i\mathcal{T}\ket{\vec{p},s}.
\label{stmatrix}
\end{equation}
However the elements of the scattering matrix are given also in terms of the interaction Hamiltonian $\mathcal{H}_I(x)$ as follows \cite{Peskin:1995ev}

\setlength{\abovedisplayskip}{-2pt}
\begin{align}
\bra{\vec{p}\,',s'}S\ket{\vec{p},s}
&=\bra{\vec{p}\,',s'}T\left[\exp\left(-i \int d^4x \mathcal{H}_I\right)\right]\ket{\vec{p},s}\nonumber\\
& \approx \braket{\vec{p}\,',s'|\vec{p},s}-\frac{1}{2}\bra{\vec{p}\,',s'}T\left[\int d^4x_1\int d^4x_2 \mathcal{H}_I(x_1)\mathcal{H}_I(x_2)\right]\ket{\vec{p},s},
\label{smatrix}
\end{align}
\setlength{\abovedisplayskip}{10pt}
where $T$ is the time-ordered product while the odd terms of the expansion in \Cref{smatrix} vanish because they contain an odd number of the field $A_{\mu}$ which contracts with each other leaving one non-contracted field. We have also used the perturbation theory to neglect the higher order in the series given in \Cref{smatrix} where these terms become higher order in the electromagnetic coupling constant $\alpha_e$. From \Cref{stmatrix,smatrix} we find
\begin{equation}
\bra{\vec{p}\,',s'}i\mathcal{T}\ket{\vec{p},s} \approx -\frac{1}{2}\bra{\vec{p}\,',s'}T\left[\int d^4x_1\int d^4x_2 \mathcal{H}_I(x_1)\mathcal{H}_I(x_2)\right]\ket{\vec{p},s}.
\label{tmatrix}
\end{equation}
From the given Lagrangian in \Cref{qedlagrangian}, the interaction Hamiltonian is expressed by
\begin{align}
\mathcal{H}_I(x_1) &= e\left[\widebar{\psi}(x_1)\gamma^{\mu}\psi(x_1)A_{\mu}(x_1)-J^{\mu}(x_1)A_{\mu}(x_1)\right].
\label{hamiltonian}
\end{align}

\begin{figure}
\centering
\includegraphics[scale=1]{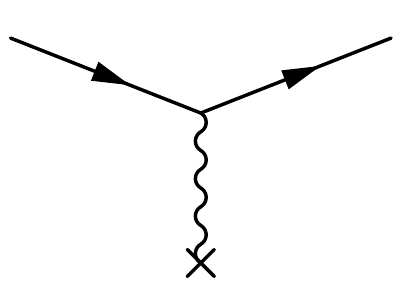}
\caption{The tree level Feynman diagram of an electron scattered off of an external source.}
\label{fig:tlevel}
\end{figure}

In this section we are interested in the tree level amplitude which describes the scattering process shown in \Cref{fig:tlevel}. Thus one may use \Cref{hamiltonian} to rewrite \Cref{tmatrix}, where the first term of \Cref{hamiltonian} squared describes a two-to-two scattering process in which we are not interested in, while the interference between the first and the second terms will leave one non-contracted field which becomes zero. Finally, one can find a contribution only from the second term squared, then \Cref{tmatrix} becomes  
\begin{equation}
 \bra{\vec{p}\,',s'}i\mathcal{T}\ket{\vec{p},s} \approx e^2 \bra{\vec{p}\,',s'}T\left[\int d^4x_1\int d^4x_2\,\widebar{\psi}(x_1)\gamma^{\mu} \psi(x_1)A_{\mu}(x_1)J^{\nu}(x_2) A_{\nu}(x_2)\right]\ket{\vec{p},s}.
 \label{tmatrixm}
 \end{equation}
Then we use Wick's theorem to express the time ordering in terms of the contracted fields \cite{Peskin:1995ev}
\begin{multline}
\label{wickcontrac}
T\left[\widebar{\psi}(x_1)\gamma^{\mu} \psi(x_1)A_{\mu}(x_1)J^{\nu}(x_2) A_{\nu}(x_2)\right] \\=\normord{\widebar{\psi}(x_1)\gamma^{\mu} \psi(x_1)A_{\mu}(x_1)J^{\nu}(x_2) A_{\nu}(x_2)}+\normord{\widebar{\psi}(x_1)\gamma^{\mu} \psi(x_1)\contraction{}{A_{\mu}(x_1)}{J^{\nu}(x_2)}{A_{\nu}(x_2)}
A_{\mu}(x_1)J^{\nu}(x_2)A_{\nu}(x_2)}\\
+\normord{\contraction{}{\widebar{\psi}(x_1)}{\gamma^{\mu}}{\psi(x_1)}
\widebar{\psi}(x_1)\gamma^{\mu}\psi(x_1)A_{\mu}(x_1)J^{\nu}(x_2) A_{\nu}(x_2)}+\dots \text{all other possible contractions},
\end{multline}
where the symbol $\normord{}$ describes the normal ordering of the contracted fields while the contraction between two fields $A$ and $B$ is given by $\contraction{}{A}{\ }{B)}
A\ B$ \cite{Peskin:1995ev}. The second term of \Cref{wickcontrac} gives the only contribution to the tree level of the interested process where the electron is scattered with the source by exchanging a photon, then we have
\begin{align}
\bra{\vec{p}\,',s'}i\mathcal{T}\ket{\vec{p},s} 
&\approx e^2\int d^4x_1 \int d^4x_2\, J^{\nu}(x_2)\left[iD_{\mu \nu}(x_1-x_2)\right]\contraction{}{\widebar G}{{}_1(\mathbf{q}_1)\,{}}{\widebar G}
\bra{\vec{p}\,',s'}\,{}\widebar \psi(x_1)\gamma^{\mu}\contraction{}{\widebar G}{{}_1(\mathbf{q}_1)\,{}}{\widebar G}
\psi(x_1)\,{}\ket{\vec{p},s},
\label{tmatrixtree}
\end{align}
where $D_{\mu\nu}(x_1-x_2)$ is the photon propagator in the position space. The remaining contractions in \Cref{tmatrixtree} are given by
\begin{equation}
\begin{aligned}
       \contraction{}{\widebar G}{{}_1(\mathbf{q}_1)\,{}}{\widebar G}
\bra{\vec{p}\,',s'}\,{}\widebar \psi(x_1)
&=\bar{u}^{s'}(p')\, e^{ip'\cdot x_1},\\
\contraction{}{\widebar G}{{}_1(\mathbf{q}_1)\,{}}{\widebar G}
\psi(x_1)\,{}\ket{\vec{p},s}
&=u^{s}(p)\, e^{-ip\cdot x_1},
       \end{aligned}
\label{econtractions}
\end{equation}
while $u^{s}(p)$ and $\bar{u}^{s'}(p')$ are the fields for the incoming and the outgoing electrons. We may also use the Fourier transform of the photon propagator $D_{\mu\nu}(q)$ with momentum $q$ \cite{Peskin:1995ev}, which when we substitute into \Cref{tmatrixtree}, gives 
\begin{align}
\bra{\vec{p}\,',s'}i\mathcal{T}\ket{\vec{p},s} &\approx ie^2\int \frac{d^4q}{(2\pi)^4} D_{\mu\nu}(q)\bar{u}^{s'}(p')\gamma^{\mu} u^{s}(p)\int d^4x_1 \int d^4x_2\, e^{ix_1\cdot(p'-p)}
 e^{-iq\cdot (x_1-x_2)}J^{\nu}(x_2)\nonumber\\
&\approx ie^2\,\bar{u}^{s'}(p')\gamma^{\mu} u^{s}(p)\int d^4q\, D_{\mu\nu}(q)\tilde{J}^{\nu}(q) \,\delta^{(4)}(p'-p-q)\nonumber\\
&\approx ie^2 \bar{u}^{s'}(p')\gamma^{\mu} u^{s}(p)\, D_{\mu\nu}(p'-p)\,\tilde{J}^{\nu}(p'-p)\nonumber\\
&\approx
\frac{-ie^2}{(p'-p)^2}\,\bar{u}^{s'}(p')\gamma_{\nu} u^{s}(p)\tilde{J}^{\nu}(p'-p),
\label{tmatrixtreem}
\end{align}
given $\tilde{J}_{\nu}(q)$ to be the Fourier transform of the current $J^{\nu}(x)$. 
However the current $J$ has only a temporal component which is given by $J^0(x) = \delta^{(3)}(\vec{x})$. Then the Fourier transform of the current produces $2\pi \, \delta (E_{p'}-E_p)$. The delta functions in appeared in the previous derivation ensures that the momentum transfer $q=p'-p$ and the energy of the scattered electron is conserved (i.e $E_{p'} = E_p = E$). The transition matrix elements becomes
\begin{align}
\bra{\vec{p}\,',s'}i\mathcal{T}\ket{\vec{p},s} &\approx \frac{-ie^2}{(p'-p)^2}\,\bar{u}^{s'}(p')\gamma^{0} u^{s}(p)\ 2\pi \delta(E_{p'}-E_p)\nonumber\\&\approx 2\pi \delta(E_{p'}-E_p)\times i\mathcal{M}_{\texttt{0}},
\label{tmatrixtreef}
\end{align}
where $\mathcal{M}_{\texttt{0}}$ is the Feynman scattering amplitude of the tree level of an electron scattered off of a static point charge. The Feynman rules for the given process may now be extracted, which become the same rules as for the normal QED process in addition to a new rule for each source, where we write 
\begin{equation} \text{For each external source:} \hspace{0.3cm}
\parbox{25mm}{\begin{fmffile}{frsource}
\begin{fmfgraph*}(60,40) 
\fmfleft{t1}
\fmfright{b1}
\fmf{photon}{t1,b1}
\fmfv{d.sh=cross,d.f=empty,d.si=.1w}{b1}
\end{fmfgraph*}
\end{fmffile}}
 = -ie\, V^{\mu}.
\label{srule}
\end{equation} 
The complete Feynman rules for the process are given in \autoref{AppendixB}.
\subsection{General Form of The Differential Cross Section}
The cross section is the most significant physical quantity for describing a scattering process where it describes the effective area for collision giving an intuition for the probability of an initial state $\ket{\vec{p},s}$ to scatter and become a final state $\ket{\vec{p}',s'}$. The differential cross section is principally defined as the ratio of the number of particles scattered into a specific direction per unit time per unit solid angle divided by the incident flux. In terms of the impact parameter $\vec{b}$ \cite{Peskin:1995ev}, it is given by
\begin{equation}
\sigma = \int d^2b\,\, \mathcal{W}(\vec{b}),
\label{xsection}
\end{equation}
where $\mathcal{W}$ is the probability of finding the system in the final state $\ket{\vec{p}\,',s'}$ given by 
\begin{equation}
d\mathcal{W}(\vec{b}) = \frac{d^3p'}{(2\pi)^3}\frac{1}{2E_{p'}}\left|\braket{\vec{p}\,',s'|i\mathcal{T}|\psi_{in}}\right|^2.
\label{prob}
\end{equation}

Here we define the incoming state in the wave packet approach instead of the normal plane wave description to avoid the singularities in the normalization of the incoming state. Let the incoming electron wavepacket $\phi(\vec{p})$ to be uniformly distributed in the impact parameter $\vec{b}$
\begin{equation}
\ket{\psi_{in}} = \int \frac{d^3q}{(2\pi)^3}\frac{1}{\sqrt{2E_q}}\phi(\vec{q}) e^{-i\vec{b}\cdot \vec{q}}\ket{\vec{q},s}.
\label{instate}
\end{equation}
Now we can relate the probability of scattering to the Feynman amplitudes in which we can get all the interesting physics from the scattering process
\begin{multline}
d\mathcal{W}(\vec{b}) 
= \frac{d^3p'}{(2\pi)^3}\frac{1}{2E_{p'}}\int \frac{d^3q}{(2\pi)^3}\frac{1}{\sqrt{2E_q}}\phi(\vec{q}) e^{-i\vec{b}\cdot \vec{q}}\braket{\vec{p}\,',s'|i\mathcal{T}|\vec{q},s}\\\times\int \frac{d^3r}{(2\pi)^3}\frac{1}{\sqrt{2E_r}}\phi^*(\vec{r}) e^{i\vec{b}\cdot \vec{r}}\braket{\vec{p}\,',s'|i\mathcal{T}|\vec{r},s}^*,
\label{prob2}
\end{multline}
where the superscript $*$ denotes the complex conjugate. Substituting \Cref{tmatrixtreef} into \Cref{prob2}, the probability will be given by
\begin{multline}
d\mathcal{W}(\vec{b}) = \frac{d^3p'}{(2\pi)^3}\int \frac{d^3q\,d^3r}{(2\pi)^4\sqrt{4E_qE_r}}\,\,\phi(\vec{q})\phi^*(\vec{r})e^{-i\vec{b}\cdot (\vec{q}-\vec{r})}\,\delta(E_{p'}-E_q)\,\delta(E_{p'}-E_r)\\
\times\mathcal{M}(q\rightarrow p') \mathcal{M}^*(r\rightarrow p').
\label{prob3}
\end{multline}
We then substitute \Cref{prob3} into \Cref{xsection} to calculate the differential cross section
\begin{multline}
d\sigma = \frac{d^3p'}{(2\pi)^3}\int \frac{d^3q\,d^3r}{(2\pi)^4\sqrt{4E_qE_r}}\,\,\phi(\vec{q})\,\phi^*(\vec{r})\,\delta(E_{p'}-E_q)\\
\times \delta(E_{p'}-E_r)\mathcal{M}(q\rightarrow p') \mathcal{M}^*(r\rightarrow p')\int d^2b\,\,e^{-i\vec{b}\cdot (\vec{q}-\vec{r})}.
\label{xsectionm}
\end{multline}

The integral over the impact parameter yields $(2\pi)^2 \, \delta^{(2)}(q_{\bot}-r_{\bot})$ \cite{Peskin:1995ev}. We use also from the conservation of energy $
\delta(E_{p'}-E_r) = \delta(E_q-E_r) = \frac{E_r}{r_z}\delta(q_z-r_z) \approx \frac{1}{v_i}\delta(q_z-r_z)$, where $v_i$ is the incoming velocity. The differential cross section is then given by
\begin{align}
d\sigma &= \frac{p^{\prime\,2}dp'\, d\Omega}{(2\pi)^3}\frac{1}{2E_{p'}v_i}\int \frac{d^3q\,d^3r}{(2\pi)^2\sqrt{4E_qE_r}}\,\phi(\vec{q})\phi^*(\vec{r})\,\delta(E_{p'}-E_q)\nonumber\\
&\hspace{3cm}\times\delta(p_z-r_z)\delta^{(2)}(q_{\bot}-r_{\bot})\mathcal{M}(q\rightarrow p') \mathcal{M}^*(r\rightarrow p')\nonumber\\
& =\frac{p^{\prime\,2}dp'\, d\Omega}{(2\pi)^3}\frac{1}{2E_{p'}v_i}\int \frac{d^3q}{(2\pi)^2\,2E_q}\,\phi(\vec{q})\phi^*(\vec{q})\,\delta(E_{p'}-E_q)\, \mathcal{M}(q\rightarrow p') \mathcal{M}^*(q\rightarrow p') ,
\label{xsectionm2}
\end{align}
where we used the recombination of the delta functions $\delta(q_z-r_z)\,\delta^{(2)}(q_{\bot}-r_{\bot}) = \delta^{(3)}(\vec{q}-\vec{r})$. Since the wavepacket $\phi$ is localized and peaked at $\vec{p}$, we can approximate $\mathcal{M}(q\rightarrow p') \mathcal{M}^*(q\rightarrow p')$ and $E_q$ with their values at the central external momentum $p$ and pull them out of the integral. We also use the normalization of the wave packet $\int \frac{d^3q}{(2\pi)^3}\left|\phi(\vec{q})\right|^2=1$. Then we have 
\begin{align}
d\sigma
& = \frac{p^{\prime\,2}dp'\, d\Omega}{(2\pi)^2}\frac{1}{2E_{p'}v_i2E_p}\,\, \delta(E_{p'}-E_p)\left|\mathcal{M}(p\rightarrow p')\right|^2.
\label{xsectionm21}
\end{align} 
 
Integrating \Cref{xsectionm21} over $p'$, we find
\begin{align}
\frac{d\sigma}{d\Omega} &= \int \frac{p^{\prime\,2}\,dp'}{(2\pi)^2}\frac{1}{2E_{p'}v_i2E_p}\,\, \delta(E_{p'}-E_p)\left|\mathcal{M}\right|^2.
\label{xsectionm3}
\end{align}
However $ 
\delta(E_{p'}-E_p) = \frac{E_{p'}}{p'}\delta(p'-p)$, we also sum over all spins $s'$ and $s$. We finally get the general formula for the differential cross section of a process where an electron is scattered by an external point charge
\begin{align}
\frac{d\sigma}{d\Omega} 
& =\frac{1}{16\pi^2}\frac{1}{2}\sum_{s,s'}\left|\mathcal{M}\right|^2.
\label{xsectionf}
\end{align}
\subsection{Leading Term of The Differential Cross Section}
To derive the differential cross section of the leading term in perturbation series from the amplitude of the tree level given in \Cref{tmatrixtreef}, we use the general formula for the differential cross section given in \Cref{xsectionf} beside the so-called Feynman trace technology \cite{Peskin:1995ev} which uses the algebraic properties of $\gamma$-matrices. The differential cross section for the leading term can be first written as following
\begin{align}
\left(\frac{d\sigma}{d\Omega}\right)_{\texttt{0}} & = \frac{e^4}{32 \pi^2(p'-p)^4}\sum_{s,s'} \sum_{a,b}\bar{u}^s_a(p)\left[\gamma^0u^{s'}(p')\bar{u}^{s'}(p')\gamma^0\right]_{ab}u^s_b(p).
\label{loxsec}
\end{align}

The trace technology allows us to replace the sum over the parameters $a$ and $b$ in \Cref{loxsec} by the trace of a number of matrices. We also recall the identity $\sum_s u^s(p)\,\bar{u}^s(p) = \slashed{p}+m$, and using the properties of the $\gamma$-matrices given in \autoref{AppendixA}, \Cref{loxsec} becomes
\begin{align}
\left(\frac{d\sigma}{d\Omega}\right)_{\texttt{0}} & =\frac{e^4}{32 \pi^2(p'-p)^4} \sum_{a,b}\left[\slashed{p}+m\right]_{ba}\left[\gamma^0\left(\slashed{p}'+m\right)\gamma^0\right]_{ab}\nonumber\\
& = \frac{e^4}{32 \pi^2(p'-p)^4} \text{Tr}\big[\gamma^0(\slashed{p}'+m)
\gamma^0(\slashed{p}+m)\big]\nonumber\\
&= \frac{e^4}{8\pi^2(p'-p)^4}\left(2E^2-p'\cdot p+m^2\right).
\label{loxsecm}
\end{align}
Defining the electromagnetic coupling constant $\alpha_e= \frac{e^2}{4\pi}$ and recall that $q = p'-p$. The differential cross section for the leading term in perturbation of an electron scattered off of a point charge is then given by
\begin{equation}
\left(\frac{d\sigma}{d\Omega}\right)_{\texttt{0}} = \frac{\alpha_e^2}{q^4}\,(4E^2+q^2).
\label{loxsecf}
\end{equation}
\subsection{Non-Relativistic Approach}
Now let us rewrite the differential cross section in terms of the scattering angle $\theta$ given by $\vec{p}\cdot \vec{p}\,'= |\vec{p}\,'|\,|\vec{p}| \cos \theta$. We choose the laboratory reference frame in which we define
\begin{equation}
\begin{aligned}
\vec{p} &= |\vec{p}|\, \hat{z},\\
 \vec{p}\,' &= |\vec{p}\,'|\,(\cos \theta\, \hat{z}+\sin \theta\, \hat{x}).
\end{aligned}
\label{labframe}
\end{equation}
Energy conservation implies $E_{p'}=E_p=E$ from which it follows that $|\vec{p}\,'|= |\vec{p}|$. \Cref{labframe} then allows us to write
\begin{align}
q^2 & = 2\,|\vec{p}|^4\,(1-\cos\theta)^2 = 4\,|\vec{p}|^2 \, \sin^2\frac{\theta}{2}.
\label{simp}
\end{align}
Let $\frac{|p|^2}{E^2} = \beta^2$, then \Cref{loxsecf} becomes the well know Mott scattering formula \cite{Mandl:1985bg}
\begin{equation}
\left(\frac{d\sigma}{d\Omega}\right)_{\texttt{0}} = \frac{\alpha_e^2\left(1-\beta^2 \sin^2 \frac{\theta}{2}\right)}{4\,|\vec{p}|^2\, \beta^2 \sin^4 \frac{\theta}{2}}.
\label{mott}
\end{equation}
In the high energy limit $-q^2\gg m^2$, we can set $\beta^2 \approx 1$ and \Cref{mott} becomes
\begin{equation}
\left(\frac{d\sigma}{d\Omega}\right)_{\texttt{0}} = \frac{\alpha_e^2\left(1- \sin^2 \frac{\theta}{2}\right)}{4E^2 \sin^4 \frac{\theta}{2}},
\label{rmott}
\end{equation}
while in the non-relativistic limit $\beta^2 << 1$ (equivalently low energies), \Cref{mott} reduces to the Rutherford formula \cite{Mandl:1985bg}
\begin{equation}
\left. \left(\frac{d\sigma}{d\Omega}\right)_{\texttt{0}}\right|_{E\approx m} = \frac{\alpha_e^2}{4\,|\vec{p}|^2\, \beta^2 \sin^4 \frac{\theta}{2}}.
\label{lorutherford}
\end{equation}
 
\chapter{NLO Rutherford Scattering} 

\label{Chapter3}


In \autoref{Chapter2} we calculated the first term of the perturbation series which is trivially in $\mathcal{O}(\alpha_e^2)$. Corrections to the differential cross section at Next-to-leading order require including diagrams such as the vertex, vacuum polarization, box, etc., which contain either fermion or photon loops. In this chapter we face the UV divergences discussed in \autoref{Chapter1} due to the high momentum scale which appears in the 4-dimensional loop integrals in the NLO diagrams. We first use the dimensional regularization to render the UV divergences finite. Then we imitate the systematic renormalization procedure to renormalize the Lagrangian of the system \cite{Peskin:1995ev}. Finally, we apply the appropriate renormalization scheme to omit these UV divergences.

\section{Renormalizing The Lagrangian}
Let us define the Lagrangian from \Cref{qedlagrangian} in terms of the bare parameters and fields, where we give them a subscript $0$ to distinguish them from the physical ones, as follows
\begin{equation}
\mathcal{L}_{\texttt{0}} = -\frac{1}{4}F_0^{\mu\nu}F_{0\mu\nu} + \bar{\psi}_0(i \slashed{\partial}-m_0)\psi_0 -e_0\bar{\psi}_0\gamma^{\mu} \psi_{\texttt{0}} A_{0\mu} +e_0J_{0\mu}A_0^{\mu}.
\label{barelag}
\end{equation}
We first relate these bare fields $A_0$ and $\psi_0$ to the renormalized ones $A$ and $\psi$ by defining the renormalization scales $Z_A$ and $Z_{\psi}$ respectively
\begin{equation}
\begin{aligned}
\psi_0 &= Z_{\psi}^{\frac{1}{2}} \,\psi,\\ A_0^{\mu} &= Z_A^{\frac{1}{2}}\,A^{\mu}.
\end{aligned}
\label{fieldrenorm}
\end{equation}
We also need to match the bare parameters $e_0,\,m_0\,\text{and}\, J_{0}$ to the renormalized ones $e,\ m\ \text{and}\ J$, so we define
\begin{equation}
\begin{aligned}
Z_{\psi}\,m_0 &= Z_m \,m,\\ \mu^{-\frac{\epsilon}{2}} e_0\,Z_{\psi} Z_A^{\frac{1}{2}} &= Z_e\,e,\\  \frac{Z_e}{Z_{\psi}}\,J_{0}^{\mu} &= Z_J\,J^{\mu}.
\label{parrenorm}
\end{aligned}
\end{equation}
Where $Z_m$, $Z_e$, and $Z_J$ are the renormalization scales for the mass, electron charge, and the current source respectively. The Lagrangian density after is rescaling is then now
\begin{equation}
\mathcal{L} = -\frac{1}{4}Z_A\,F^{\mu\nu}F_{\mu\nu} +Z_{\psi}\, \bar{\psi}\ i \slashed{\partial}\ \psi-Z_m\,m\bar{\psi}\psi - e \mu^{\frac{4-d}{2}}\,Z_{e}\,\bar{\psi}\gamma^{\mu} \psi A_{\mu} +e \mu^{\frac{4-d}{2}}\,Z_J\,J_{\mu}A^{\mu}.
\label{renormlag}
\end{equation}
Next we expand each renormalization scale $Z$ in terms of a corresponding counter term $\delta$ 
\begin{equation}
 \begin{aligned}
        Z_{\psi} & = 1+\delta_{\psi},\\
        Z_{A}  &= 1+\delta_{A},\\
        Z_{e}  &= 1+\delta_{e},\\
        Z_{m}  &= 1+\delta_{m},\\
        Z_{J}  &= 1+\delta_{J}.
       \end{aligned}
       \label{counter}
\end{equation}
Each of the previous counter terms must be fixed by the renormalization scheme to define the renormalized fields and parameters. In terms of the renormalized parameters and the counter terms the Lagrangian density becomes
\begin{align}
\mathcal{L}& = -\frac{1}{4}F_{\mu\nu}F^{\mu\nu}+\bar{\psi}(i\slashed{\partial}-m)\psi-e \mu^{\frac{4-d}{2}}\bar{\psi}\gamma^{\mu}\psi A_{\mu}+e \mu^{\frac{4-d}{2}}J_{\mu}A^{\mu}\nonumber\\
&\quad -\frac{1}{4}\delta_A F_{\mu\nu}F^{\mu\nu}+\bar{\psi}(i\delta_{\psi}\slashed{\partial}-m\delta_m)\psi-e \mu^{\frac{4-d}{2}}\delta_e\bar{\psi}\gamma^{\mu}\psi A_{\mu}+e \mu^{\frac{4-d}{2}}\delta_j J_{\mu}A^{\mu}.
\label{counterlag}
\end{align}

The next step is to determine the Feynman diagrams whose amplitudes contain UV divergences by defining the superficial degree of divergence $D$ in terms of the number of external electrons $N_e$ and external photons $N_{\gamma}$ which characterize each diagram. Loop momentum integrals in Feynman amplitudes diverge in the high momentum scale when there are more powers of momentum in the denominator than in the numerator. Hence diagrams with $D\geq 0$ are said to be divergent in the UV limit. One can show from the previous definition that $D$ is given by \cite{Peskin:1995ev}
\begin{equation}
D = 4-\frac{3}{2}N_e-N_{\gamma}.
\label{divdegree}
\end{equation}

\begin{figure}[ht!]
      \centering
      \begin{subfigure}[b]{0.3\textwidth}              
\includegraphics[scale=1]{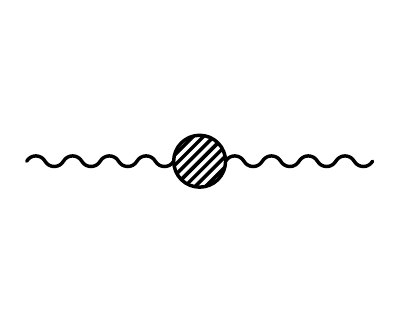}      
      \caption{D = 0}
                \label{diva}
        \end{subfigure}
              \begin{subfigure}[b]{0.3\textwidth}              
\includegraphics[scale=1]{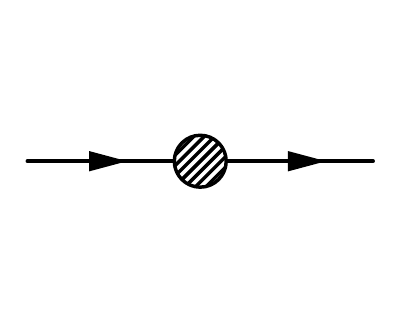}      
      \caption{D = 1}
                \label{divb}
        \end{subfigure}
              \begin{subfigure}[b]{0.3\textwidth}              
\includegraphics[scale=1]{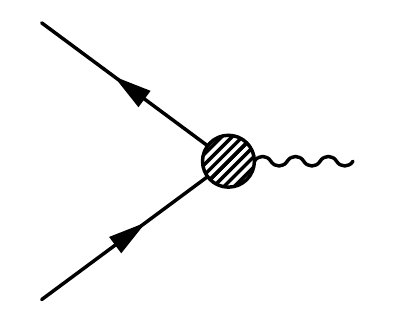}      
      \caption{D = 2}
                \label{divc}
        \end{subfigure}
        \caption{Superficially Divergent 1PI diagrams in QED.}
\label{fig:divdegree}
\end{figure}

The infinite diagrams for the given Lagrangian are shown in \Cref{fig:divdegree}, which do not differ from that of the complete QED process. Here we excluded all the other divergent diagrams because they either do not describe a scattering process or their contributions is zero due to symmetries. The Feynman rules of the renormalized Lagrangian are described in \autoref{AppendixB}. Each of these rules relates the renormalization of the fields and parameters to the counter terms defined in \Cref{counter}. We note that there is no need to renormalize the external source $J^{\mu}$ because it does not contribute any divergences, this means $Z_J = 1$.

The dashed blob indicates that the graphs are one-particle irreducible (1PI). The 1PI is any graph that can not be cut into two different propagators (i.e whose all internal lines have loop momentum integrals). It is shown that all the UV divergences can be eliminated by the counter terms defined in \Cref{counter} corresponding to each 1PI amplitude shown in \Cref{fig:divdegree}, this is known as the BHPZ theorem where the complete proof can be found in \cite{Hepp:1966eg}. 
\section{Vacuum Polarization Correction}\label{polcorrec}
The superficially 1PI diagram in \Cref{diva} includes the amplitude and the counter term that describe the renormalization of the electromagnetic field $A$. The Feynman amplitude of the vacuum polarization diagram and its corresponding counter diagram is given by
\begin{align}
i\mathcal{M}_{\texttt{VP}}&= \begin{gathered}
\includegraphics[scale=1]{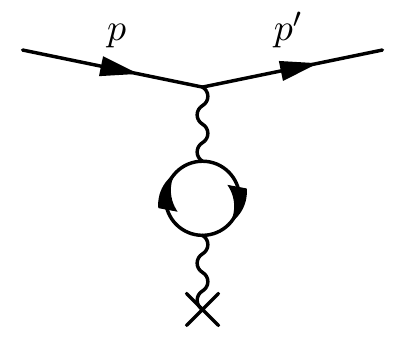}
\end{gathered}\quad+\quad\begin{gathered}
\includegraphics[scale=1]{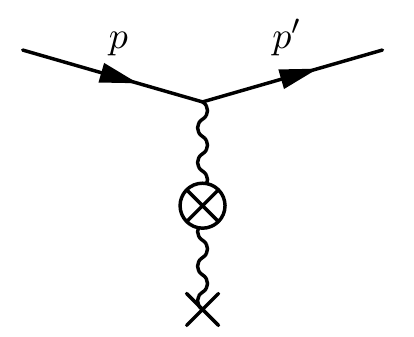}
\end{gathered}
\nonumber\\ &=\bar{u}^{s'}(p') \left[-ie\, \mu^{\frac{4-d}{2}}\gamma^{\mu}\right] u^s(p) \,D_{\mu\alpha}(q)\,\left[i\Pi^{\alpha\beta}(q)\right]\,D_{\beta\nu}(q)\,\left[-ie \, \mu^{\frac{4-d}{2}} V^{\nu}(q)\right],
\label{vpamplitude}
\end{align}
where $\Pi^{\alpha\beta}$ can be written using the Feynman rules defined in \Cref{renormrules} in $d$-dimension as
\begin{align}
i\Pi^{\alpha\beta}(q) &= \begin{gathered}
\includegraphics[scale=1]{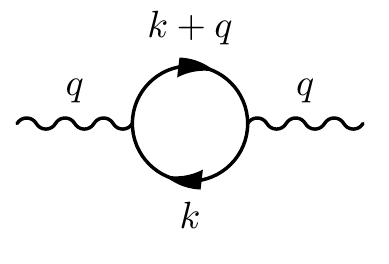}
\end{gathered}\quad+\quad\begin{gathered}
\includegraphics[scale=1]{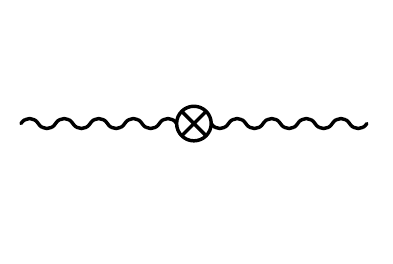}
\end{gathered}
\nonumber\\ &= -e^2 \mu^{4-d} \int \frac{d^dk}{(2\pi)^d}\text{ tr}\left[\gamma^{\alpha}
\frac{(\slashed{k}+m)}{k^2-m^2+i\epsilon}\gamma^{\beta}
\frac{(\slashed{k}+\slashed{q}+m)}{(k+q)^2-m^2+i\epsilon}\right]\nonumber\\
&\hspace{8cm}-i \left[g^{\alpha\beta}q^2-q^{\alpha}q^{\beta}\right] \delta_A. 
\label{piamplitude}
\end{align}

The integral over $k$ in $d$-dimensions can be calculated in several steps. First, we use Feynman parameters trick defined in \Cref{feynpar2} to combine the denominators of \Cref{piamplitude} and then complete the square, we write
\begin{multline}
\frac{1}{\left[k^2-m^2+i\epsilon\right]\left[(k+q)^2-m^2+i\epsilon\right]} \\
 \begin{aligned}&= \int_{\texttt{0}}^1 dx \frac{1}{\left\{x\left[(k+q)^2-m^2+i\epsilon\right]+(1-x)\left[k^2-m^2+i\epsilon\right]\right\}^2}\\ 
  & = \int_{\texttt{0}}^1 dx \frac{1}{\left[(k+xq)^2+x(1-x)q^2 -m^2+i\epsilon\right]^2} \\
 & = \int_{\texttt{0}}^1 dx \frac{1}{\left[\ell^2 -M^2+i\epsilon\right]^2},
 \end{aligned}
\label{vpfeynpar}
 \end{multline}
 where we shifted the momentum to be $\ell = k+xq$ and defined $M^2=m^2-x(1-x)q^2$. We can also simplify the numerator $\mathcal{N}_{\texttt{VP}}$ of equation \Cref{piamplitude} in terms of the new momentum $\ell$ by taking the trace and using the properties of the gamma matrices given in \Cref{contrgamma}, where we have
 \begin{align}
\mathcal{N}_{\texttt{VP}} &= 4\left[k^{\alpha}(k+q)^{\beta}+k^{\beta}(k+q)^{\alpha}-g^{\alpha\beta}\left(k\cdot(k+q)+m^2\right)\right]\nonumber\\
& = 4\left[[2\ell^{\alpha}\ell^{\beta}-2x(1-x)q^{\alpha}q^{\beta}-g^{\alpha\beta}\left(\ell^2-x(1-x)q^2+m^2\right)+\text{linear terms in }\ell\right].
\label{vpnumerator}
\end{align}
The symmetry of the integral over $\ell$ shows that the integrals with linear terms in $\ell$ vanish and allows us to replace $\ell^{\alpha}\ell^{\beta}\rightarrow \frac{1}{d}\ \ell^2 g^{\alpha\beta}$ as in \Cref{tensorellint1,tensorellint2}. This simplifies the numerator $\mathcal{N}_{\texttt{VP}}$ to be written as 
 \begin{align}
\mathcal{N}_{\texttt{VP}} &= 4\left[-g^{\alpha\beta}(1-\frac{2}{d}) \ell^2 -2x(1-x)q^{\alpha}q^{\beta}+g^{\alpha\beta}\left(x(1-x)q^2-m^2\right)\right].
\label{vpnumerators}
\end{align}
The full expression for \Cref{piamplitude} becomes
\begin{multline}
i\Pi^{\alpha\beta}(q)= -4e^2 \mu^{4-d}\int_{\texttt{0}}^1dx\int \frac{d^d\ell}{(2\pi)^d}\,\\
\left\{\frac{-g^{\alpha\beta}(1-\frac{2}{d}) \ell^2 -2x(1-x)q^{\alpha}q^{\beta}+g^{\alpha\beta}\left[x(1-x)q^2-m^2\right]}{(\ell^2-M^2+i\epsilon)^2}\right\}
 -i (g^{\alpha\beta}q^2-q^{\alpha}q^{\beta})\  \delta_A .
\label{piamps}
\end{multline}

Now our main task is to perform the momentum integral in \Cref{piamps}. A trick introduced by Wick can make this integral much easier to calculate, the trick called the Wick rotation in which we rotate the contour counter-clockwise by $\frac{\pi}{2}$ by defining a new 4-momentum variable $\ell_E$ such that $l^0 = i\ell_E^0,\text{ and } \vec{\ell}= \vec{\ell}_E$ \cite{Peskin:1995ev}. \Cref{piamps} will be
\begin{multline}
i\Pi^{\alpha\beta}(q)= -4ie^2 \mu^{4-d}\int_{\texttt{0}}^1 dx\, \left\{\left[-2x(1-x)q^{\alpha}q^{\beta}+g^{\alpha\beta}\left(x(1-x)q^2-m^2\right)\right]\right.\\\times\int \frac{d^d\ell_E}{(2\pi)^d}\,\frac{1}{(\ell_E^2+M^2+i\epsilon)^2}
\left.+(1-\frac{2}{d})g^{\alpha\beta}\int \frac{d^d\ell_E}{(2\pi)^d}\,\frac{\ell_E^2}{(\ell_E^2+M^2+i\epsilon)^2}\right\}\\
-i (g^{\alpha\beta}q^2-q^{\alpha}q^{\beta})\ \delta_A .
\label{piamps2}
\end{multline}
Using the momentum integrals defined in \Cref{scalarellint} and taking the limit $d\rightarrow 4$
\begin{align}
i\Pi^{\alpha\beta}(q)&\ = -8ie^2\mu^{4-d}\,(g^{\alpha\beta}q^2-q^{\alpha}q^{\beta})\int_0^1 dx\ \frac{x(1-x)}{(4\pi)^{d/2}}\frac{\Gamma(2-d/2)}{(M^2)^{(2-d/2)}}-i \,(g^{\alpha\beta}q^2-q^{\alpha}q^{\beta})\, \delta_A\nonumber\\&
\underset{d\rightarrow 4}{=}(g^{\alpha\beta}q^2-q^{\alpha}q^{\beta}) \times i\Pi(q^2),
\label{piamps3}
\end{align}
where 
\begin{equation}
\Pi(q^2) = \frac{-e^2}{2\pi^2}\int_0^1dx\,\,x(1-x)\left(\frac{2}{\epsilon} -\log M^2 -\gamma_E + \log 4\pi +\log \mu^2+O(\epsilon)\right) -\delta_A ,
\label{xi}
\end{equation}
and $d= 4-\epsilon$ is the number of space-time dimensions as defined in \Cref{dimensions}. 

Now it is time to choose the renormalization scheme in order to eliminate the divergence in the form of $\frac{1}{\epsilon}$. Since we used the dimensional regularization, every loop correction takes almost the same form as in \Cref{xi} which makes it easier to apply the renormalization scheme. In order to choose an appropriate scheme for our calculations here, we first make a comparison between the most two common schemes in QFT.
\section*{On-Shell Vs. $\overline{\mathrm{MS}}$ renormalization schemes}
The counterterms defined in \Cref{counter} have divergent and finite pieces. One is free to choose how to fix the finite piece \cite{Sterman:1994ce}. Two common choices are momentum subtraction (i.e.\ on-shell) and the generalized minimal subtraction schemes.
\begin{itemize}
\item[\emph{a)}]\emph{On-Shell Scheme:}

The on-shell (OS) renormalization scheme is a most common scheme used in the QED calculations, in which one fixes the counter terms such that they define the renormalized parameters to be the physical ones. The OS scheme allows us to write a set of renormalization conditions by which we can eliminate the UV divergences in each diagram; these conditions are: 
\begin{enumerate}
\item The Fourier transform of the electron propagator has a pole at the physical mass, equivalently the renormalized mass, which ensures that electron self energy correction at the renormalized mass vanishes (i.e $\Sigma_2(m) = 0)$, where $\Sigma_2$ is the coefficient of the $\frac{i}{\slashed{p}-m}$ in the 1PI contribution from \Cref{divb}.
\item The pole of the electron propagator has a residue 1, which means that the first derivative of the electron self-energy correction at the renormalized mass must vanish (i.e $\Sigma_2'(m)=0$).
\item The Fourier transform of the photon propagator has a pole at  $q^2 = 0$, this pole has a residue 1, which means that the vacuum polarization correction must vanish at $q^2=0$ (i.e $\Pi(q^2 = 0 ) = 0$), where $\Pi(q^2)$ is the coefficient of $\frac{i}{q^2}(-g_{\mu\nu}+q^{\mu}q^{\nu})$ in the 1PI contribution from \Cref{diva}.
\item The electron charge is fixed to be the renormalized charge $e$, which ensure that the amputated vertex correction gives back the normal vertex (i.e $-ie\Gamma^{\mu} (q=0) = -ie \gamma^{\mu}$), where $\Gamma^{\mu}$ is the sum of all 1PI contribution to the 3-point function in \Cref{divc}.
\end{enumerate}
An important remark on the OS renormalization scheme is that the full formula of the differential cross section is expected not to be finite as we send the mass of the electron to zero, equivalently in the high energy limit $-q^2\gg m^2$, which appear as an extra $\log(m)$ from  the vacuum polarization correction \cite{Srednicki:2007qs}. However, the OS renormalization conditions defined above are not the only way to define the counter terms.   
\item[\emph{b)}]\emph{Generalized Minimal Subtraction Scheme:}

Dimensional regularization allows us to write the pole of Feynman diagrams beyond the leading order, at the UV limit, in the form of the number of space-time dimensions $d$. The generalized minimal subtraction scheme defines the counterterms to cancel the $1/\epsilon$ pole at the original dimensionality ($d=4$) \cite{t1973dimensional}.

One of the advantages of the generalized minimal subtraction scheme is that we simply set the finite piece to a convenient value. Two common choices for the finite pieces are: minimal subtraction (MS), in which we choose the finite piece to be zero, and the modified minimal subtraction scheme, in which we choose the finite piece to cancel the common term $\log(4\pi)-\gamma_E$ that arise from using the dimensional regularization \cite{Bardeen:1978yd}.

We note that in the $\overline{\mathrm{MS}}$ scheme the position of the pole of the electron propagator is no longer at the physical mass which means that the physical quantities are not necessarily the renormalized ones and the residue of the pole is no longer $1$ \cite{Srednicki:2007qs}. Aside from the fact that the $\overline{\mathrm{MS}}$  renormalization scheme does not have a physical meaning, it can be considered as a very powerful scheme where it automatically cancels the UV divergences through the counter terms with very convenient calculations. In addition to the avoidance of the subdivergences that may appear from the vacuum polarization diagram which ensures in return a finite formula for the differential cross section at NLO correction when the mass of the electron goes to zero \cite{Srednicki:2007qs}.    
\end{itemize}

Now we apply the $\overline{\mathrm{MS}}$ renormalization scheme on \Cref{xi} which allows us to choose $\delta_A$ such that it removes the infinity and the term $-\gamma_{E}+\log 4\pi$. From now and on we identify the mass scale $\mu_{\overline{\mathrm{MS}}}$ to specify that the scheme we used is the $\overline{\mathrm{MS}}$ scheme, then \Cref{xi} becomes 
\begin{align}
\Pi(q^2) &= \frac{e^2}{2\pi^2}\int_0^1dx\,\,x(1-x)\,\log\left(\frac{m^2- x(1-x)q^2}{\mu_{\overline{\mathrm{MS}}}^2}\right)\nonumber\\
&=\frac{e^2}{2\pi^2}\int_0^1dx\,\,x(1-x)\,\left[\log\left(\frac{m^2- x(1-x)q^2}{m^2}\right)+\log\left(\frac{m^2}{\mu_{\overline{\mathrm{MS}}}^2}\right)\right]\nonumber\\
&= \frac{\alpha}{3\pi}\left[\log\left(\frac{-q^2}{\mu_{\overline{\mathrm{MS}}}^2}\right)-\frac{5}{3}+\mathcal{O}(m^2)\right],\text{ as } -q^2\gg m^2.
\label{xis}
\end{align}
The counter term for the photon field renormalization will be
\begin{align}
\delta_A & = \frac{-e^2}{2\pi^2}\,\left(\frac{2}{\epsilon}-\gamma_E+\log 4\pi \right)\,\int_{\texttt{0}}^1 dx\,\,x(1-x)  = \frac{-\alpha}{3\pi}\,\left(\frac{2}{\epsilon}-\gamma_E+\log 4\pi\right).
\label{deltaa}
\end{align}
Since $V^{\nu}$ contributes only with the temporal part, \Cref{piamps3} will also contribute with $\Pi^{00}$ term. We also recall that $q^0 = p^{'0}-p^0  = 0$, the $q^{\alpha}q^{\beta}$ term vanishes and the vacuum polarization amplitude becomes
\begin{align}
i\mathcal{M}_{\texttt{VP}} &= \frac{e^2\,\mu^{\epsilon}}{q^4} \bar{u}^{s'}(p') \gamma^0 u^s(p) \ i\Pi^{00}(q)\nonumber\\
&= i\mathcal{M}_{\texttt{0}}\ \Pi(q^2)+\mathcal{O}(\epsilon)\nonumber\\
& =i \mathcal{M}_{\texttt{0}}\, \frac{\alpha}{\pi}\left[ \frac{1}{3}\log \left(\frac{-q^2}{\mu_{\overline{\mathrm{MS}}}^2}\right)-\frac{5}{9}+\mathcal{O}(m^2)\right].
\label{vpampf}
\end{align} 
The contribution to the differential cross section will be
\begin{align}
\left(\frac{d\sigma}{d\Omega}\right)_{\texttt{VP}} &= \frac{1}{32 \pi^2} \sum_{s,s'}\left[\mathcal{M}_{\texttt{0}}\mathcal{M}_{\texttt{VP}}^*
+\mathcal{M}_{\texttt{VP}}\mathcal{M}_{\texttt{0}}^*\right]\nonumber\\
& =\left(\frac{d\sigma}{d\Omega}\right)_{\texttt{0}} \frac{\alpha}{\pi}\left[ \frac{2}{3}\log \left(\frac{-q^2}{\mu_{\overline{\mathrm{MS}}}^2}\right)-\frac{10}{9}+\mathcal{O}(m^2)\right].
\label{vpxsection}
\end{align}
\section{Vertex Correction}
The vertex diagram is one of the diagrams that contribute to the differential cross section at NLO corresponding to the superficially 1PI divergent diagram in \Cref{divc}. This diagram gives the correction to the electron charge $e$ where the amplitude corresponding to such diagram is given by
\begin{align}
i\mathcal{M}_{\texttt{V}} &= 
\begin{gathered}
\includegraphics[scale=1]{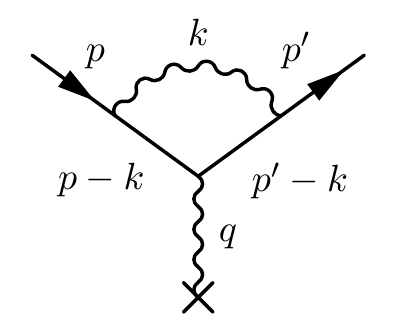}
\end{gathered}\quad+\quad
\begin{gathered}
\includegraphics[scale=1]{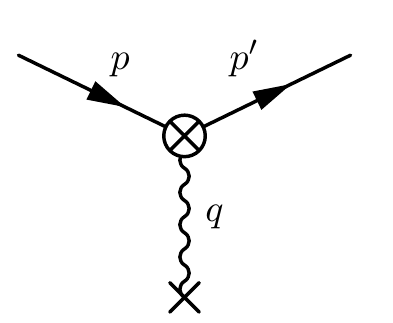}
\end{gathered}
\nonumber\\
 &= \bar{u}^{s'}(p')\left[-ie\ \mu^{\frac{4-d}{2}}\delta\Gamma^{\mu}\right] u^s(p)\,D_{\mu\nu}(q)\,\left[-ie\  \mu^{\frac{4-d}{2}} V^{\nu}(q)\right],
\label{vertexamp}
\end{align}
where
\begin{multline}
-ie\ \delta\Gamma^{\mu}
= (-ie)^3 \mu^{4-d}\int \frac{d^dk}{(2\pi)^d} \left[\gamma^{\alpha}\frac{i(\slashed{p}'
-\slashed{k}+m)}{(p'-k)^2-m^2+i\epsilon}\gamma^{\mu}\frac{i(\slashed{p}-\slashed{k}+m)
}{(p-k)^2-m^2+i\epsilon}\gamma^{\beta}\right.\\
\left.\frac{-ig_{\alpha\beta}}{k^2-m_{\gamma}^2+i\epsilon}\right]
- i e \ \gamma^{\mu} \delta_e.
\label{gammaamp}
\end{multline}

We emphasize here the use of the photon mass  $m_{\gamma}$ in the photon propagator to regularize the expected soft IR divergence due to the emission and absorption of the virtual photon which might be soft. Using Feynman parameters, we write
\begin{multline}
\frac{1}{x\left[(p-k)^2-m^2+i\epsilon\right]+y\left[(p'-k)^2
-m^2+i\epsilon\right]+z(k^2-m_{\gamma}^2+i\epsilon)} \\= \int_0^1 \frac{2\, \delta(x+y+z-1)\,dx\,dy\,dz}{\left\{x\left[(p-k)^2-m^2+i\epsilon\right]+y\left[(p'-k)^2
-m^2+i\epsilon\right]+z(k^2-m_{\gamma}^2+i\epsilon)\right\}^3}\\
=\int dF_3\  \frac{2}{\left(\ell^2-M^2+i\epsilon\right)^3},
\label{vertxdenom}
\end{multline}
where we used the condition for the on shell momenta $p^2 = p'^2 = m^2$ to combine the denominators and rewrite the whole denominator in terms of $\ell = k-(xp+yp')$ and $M^2=m^2(1-z)^2 -xy q^2+zm_{\gamma}^2$. We also defined $\int dF_3 = \int \delta(x+y+z-1) \ dx\ dy\ dz$ for simple writing. Let us now simplify the numerator $\mathcal{N}_{\texttt{V}}$ of equation \Cref{gammaamp} by using the properties of the gamma matrices in $d$-dimensions
\begin{align}
\mathcal{N}_{\texttt{V}}& = \gamma^{\alpha}[(\slashed{p}'-\slashed{k})+m]\gamma^{\mu}[(\slashed{p}-\slashed{k})+m]\gamma_{\alpha} \nonumber\\
& = -2 (\slashed{p} - \slashed{k}) \gamma^{\mu}(\slashed{p}' - \slashed{k}) + 4m ( p' + p  - 2 k )^{\mu}  - 2 m^2 \gamma^{\mu}  \nonumber\\
&\hspace{0.5cm}+ (4-d)\big[ (\slashed{p}' - \slashed{k}+m)\gamma^{\mu}(\slashed{p}-\slashed{k} + m )\big]\nonumber\\
& = -2 \slashed{\ell} \gamma^{\mu}\slashed{\ell} - 2 (\slashed{p} - x\slashed{p} - y \slashed{p}')\gamma^{\mu}(\slashed{p}' - x\slashed{p} - y \slashed{p}')-2 m^2 \gamma^{\mu}+ 4m (p^{' \mu}+p^{\mu} -2x p^{\mu} -2y p^{'\mu} ) \nonumber\\
& \quad + (4-d) \slashed{\ell}\gamma^{\mu} \slashed{\ell} + (4-d)(\slashed{p}'-x\slashed{p}-y\slashed{p}' -m ) \gamma^{\mu} (\slashed{p}-x\slashed{p}-y\slashed{p}' -m )+ \text{linear terms in }\ell.
\label{vertxnum}
\end{align}

Before simplifying the numerator even more we can put an expectation for what the function $\delta\Gamma^{\mu}$ is going to look like which will help us to put a goal for every step in the manipulation process. We recall the fact that $\delta\Gamma^{\mu} = \gamma^{\mu}$ at leading order, this means that $\delta\Gamma^{\mu}$ should include $\gamma^{\mu}$ and some other functions of $q^2$. We use the trick $2xp^{\mu} + 2y p^{'\mu} = (x+y)(p^{\mu}+p^{' \mu })+(x-y)p^{\mu}-p^{' \mu}$ such that we write the fourth term of \Cref{vertxnum} as 
\begin{align}
4m\, (p^{' \mu}+p^{\mu} -2x p^{\mu} -2y p^{'\mu} ) & = 4m\,\big[(p^{'\mu}+p^{\mu}) - (x+y)(p^{'\mu}+p^{\mu}+(x-y)(p^{'\mu}-p^{\mu}))\big]\nonumber\\
& = 4mz(p^{'\mu}+p^{\mu})+4x(x-y)q^{\mu}\nonumber.
\label{num4}
\end{align}
We can also rewrite the second and last terms of \Cref{vertxnum} in different forms, using the identities $q = p'-p$ and $x+y+z=1$, to get
\begin{equation}
\begin{aligned}
\slashed{p} - x\slashed{p} - y \slashed{p}^{' } & = (1-x)(\slashed{p}'-\slashed{q})-y\slashed{p}'=z \slashed{p}' - (1-x) \slashed{q},\\
\slashed{p}' - x\slashed{p} - y \slashed{p}'
&=(1-y)(\slashed{p}+\slashed{q})-x\slashed{p}=z\slashed{p} +(1-y) \slashed{q},\\
\slashed{p}'-x\slashed{p}-y\slashed{p}' & =\slashed{p}'-(x\slashed{p}'-\slashed{q})-y\slashed{p}'=z\slashed{p}'+x\slashed{q},\\
\slashed{p}-x\slashed{p}-y\slashed{p}'
&=\slashed{p}-x\slashed{p}-y(\slashed{p}+\slashed{q})=  z\slashed{p} -y \slashed{q}.
\label{numall}
\end{aligned}
\end{equation}
The numerator now becomes
\begin{multline}
\mathcal{N} _{\texttt{V}}= (2-d) \slashed{\ell} \gamma^{\mu} \slashed{\ell} - 2 \left[z\slashed{p}' - (1-x) \slashed{q}\right] \gamma^{\mu}\left[z\slashed{p} +(1-y)\slashed{q} \right]-2m^2 \gamma^{\mu} + 4mz (p'+p)^{\mu}\\
+4m (x-y)q^{\mu} +(4-d) (z\slashed{p}' +x\slashed{q}-m)\gamma^{\mu} (z\slashed{p} -y \slashed{q}-m ) .
\label{vertxnum2}
\end{multline}

We note that the numerator $\mathcal{N}_{\texttt{V}}$ is sandwiched between $\bar{u}^{s'}(p')$ and $u^s(p)$, so we can use the on shell momenta conditions $\slashed{p}\, u^s(p) =m\, u^s(p)$ and $\bar{u}^{s'}(p')\slashed{p}'=m\ \bar{u}^{s'}(p')$ \cite{Peskin:1995ev}. Using this we can make more simplifications for the numerator, by noting
\begin{align}
\left[z\slashed{p}' - (1-x) \slashed{q}\right] \gamma^{\mu}\left[\slashed{p} +(1-y)\slashed{q} \right]& = \left[zm - (1-x) \slashed{q}\right] \gamma^{\mu}\left[zm +(1-y)\slashed{q} \right]\nonumber\\
& = z^2 m^2 \gamma^{\mu} - (1-x)(1-y) \slashed{q} \gamma^{\mu} \slashed{q} \nonumber\\
&\hspace{0.5cm}+ mz\left([\gamma^{\mu}, \slashed{q}]+x\slashed{q} \gamma^{\mu} -y \gamma^{\mu}\slashed{q}\right)\nonumber\\
&=z^2 m^2 \gamma^{\mu} - (1-x)(1-y) \slashed{q} \gamma^{\mu} \slashed{q}\nonumber\\
&\quad  +\frac{1}{2} mz (2-x-y) [\gamma^{\mu},\slashed{q}]+\frac{1}{2}(x-y)\{\gamma^{\mu},\slashed{q}\}.\label{vertxnum22}
\end{align}
Above we used the following trick to write the last line of \Cref{vertxnum22}
\begin{align}
x\slashed{q} \gamma^{\mu} -y \gamma^{\mu}\slashed{q}& = \frac{1}{2}(x-y) (\slashed{q}\gamma^{\mu}+\gamma^{\mu}\slashed{q})+\frac{1}{2}(x+y) (\slashed{q}\gamma^{\mu}-\gamma^{\mu}\slashed{q})\nonumber\\
& = \frac{1}{2}(x-y)\{\gamma^{\mu},\slashed{q}\}-\frac{1}{2}(x+y)[\gamma^{\mu},\slashed{q}].
\label{numtrick2}
\end{align}

We recall $\frac{1}{2}[\gamma^{\mu},\slashed{q}] = -i \sigma ^{\mu\nu} q_{\nu}$ and $\frac{1}{2}\{\gamma^{\mu},\slashed{q}\} = q^{\mu}$, where $\sigma^{\mu\nu}$ is the generator of the Lorentz group \cite{Peskin:1995ev}, the latter allows us to write $\slashed{q}\gamma^{\mu}\slashed{q}  = \slashed{q}(2q^{\mu}-\slashed{q}\gamma^{\mu}) = - q^2 \gamma^{\mu}$, where we used Dirac equation to write $\bar{u}(p^{\prime})\ \slashed{q} \ u(p) =\bar{u}(p^{\prime})\  (\slashed{p}^{\prime}-\slashed{p})\ u(p) = 0$. Then equation \Cref{vertxnum22} becomes
\begin{multline}
\left[z\slashed{p}' - (1-x) \slashed{q}\right] \gamma^{\mu}\left[\slashed{p} +(1-y)\slashed{q} \right] = z^2 m^2 \gamma^{\mu} + (1-x)(1-y)\ q^2 \gamma^{\mu}  + mz(x-y)q^{\mu}\\ - imz (1+z)\sigma^{\mu\nu}q_{\nu}.
\label{vertxnum23}
\end{multline}
Similarly for the last term of \Cref{vertxnum}, we write
\begin{multline}
(z\slashed{p}' +x\slashed{q}-m)\gamma^{\mu} (z\slashed{p} -y \slashed{q}-m)
 = m^2(z-1)^2 \gamma^{\mu} +xy \ q^2 \gamma^{\mu} - m (1-z)(x-y)q^{\mu}\\ -im (1-z)(x+y)\sigma^{\mu\nu} q_{\nu}.
\label{vertxnum4}
\end{multline}
The Gordon identity \cite{Peskin:1995ev}, which is given by 
\begin{equation}
\bar{u}(p') \gamma^{\mu}u(p) = \bar{u}(p')\bigg(\frac{p^{' \mu}+p^{\mu}}{2m}+\frac{i\sigma^{\mu\nu} q_{\nu}}{2m}\bigg)u(p),
\label{gordon}
\end{equation}
allows us to write $(p' + p )^{\mu} \sim 2m\gamma^{\mu}-i \sigma^{\mu\nu}q_{\nu} $. Using this with \Cref{vertxnum23,vertxnum4} and recall that $\slashed{\ell} \gamma^{\mu} \slashed{\ell}\rightarrow \frac{(2-d)}{d} \ell^2 \gamma^{\mu}$, the numerator becomes
\begin{multline}
\mathcal{N}_{\texttt{V}}
 = \frac{(2-d)^2}{d} \ell^2 \gamma^{\mu} + m^2 \gamma^{\mu} \cdot \left[ 8z-2(1+z)^2 + (4-d)(1-z)^2 \right]\\
 -q^2 \gamma^{\mu}\cdot \left[2(1-x)(1-y)-(4-d) xy\right]+m\,q^{\mu} (x-y) \left[ 4-2z - (4-d) (1-z)^2 \right]\\
 + im\,\sigma^{\mu\nu}q_{\nu}(1-z)\left[2z+(4-d)(1-z)\right].
\label{vertxnumm}
\end{multline}

The Ward identity: $q_{\mu}\delta\Gamma^{\mu}=0$, ensures that the term with coefficient $q^{\mu}$ vanishes \cite{Peskin:1995ev}. Finally after a long journey of simplifications, the numerator can be written as 
\begin{align}
\mathcal{N}_{\texttt{V}}& = \mathcal{N}_{\texttt{V}}^{(1)} \gamma^{\mu} -  \frac{i \sigma^{\mu\nu}q_{\nu}}{2m} \mathcal{N}_{\texttt{V}}^{(2)},
\label{vertxnumf}
\end{align}
where 
\begin{equation}
\mathcal{N}_{\texttt{V}}^{(1)} = \frac{(2-d)^2}{d} \ell^2 + m^2 \big( 8z-2(1+z^2) + (4-d) (1-z)^2 \big) -q^2 \big( 2(1-x) - (4-d) xy \big),
\label{frstnum}
\end{equation}
while
\begin{equation}
\mathcal{N}_{\texttt{V}}^{(2)} = 2m^2 \,(1-z) \left[2z + (4-d) (1-z)\right].
\label{scndnum}
\end{equation}
From \Cref{vertxdenom,vertxnumf} into \Cref{gammaamp}, the function $\Gamma^{\mu}$ can be written as
\begin{align}
 \delta \Gamma^{\mu} &= -2i e^2  \mu^{4-d}   \int dF_3\, \left(\gamma^{\mu}\cdot \int \frac{d^d\ell}{(2\pi)^d} \frac{\mathcal{N}_{\texttt{V}}^{(1)}}{(\ell^2 -M^2)^3} - \frac{i \sigma^{\mu\nu} q_{\nu} }{2m} \int \frac{d^d\ell}{(2\pi)^d} \frac{\mathcal{N}_{\texttt{V}}^{(2)}}{(\ell^2 -M^2)^3}\right) +  \gamma^{\mu} \delta_e\nonumber\\
 &= \gamma^{\mu} F_1(q^2) +\frac{i\sigma^{\mu\nu}q_{\nu}}{2m}\ F_2(q^2).
 \label{gamaampm}
\end{align}

$\Gamma^{\mu}$ has the form exactly as expected earlier in this section, where $F_1(q^2)$ and $F_2(q^2)$ are called the form factors and can be evaluated by using first the Wick rotation trick, the first integral in \Cref{gamaampm} will be given by
\begin{multline}
\int \frac{d^d\ell}{(2\pi)^d} \frac{\mathcal{N}_{\texttt{V}}^{(1)}}{(\ell^2 -M^2)^3}   =  i\frac{(2-d)^2}{d} \int \frac{d^d\ell_E}{(2\pi)^d}\frac{\ell_E^2}{(\ell_E^2 + M^2)^3}-i \left[m^2 \big( 8z - 2 (1+z^2)\right. \\ \left. + (4-d) (1-z)^2 \big)- q^2 \big(2(1-x)(1-y)-(4-d)xy\big)\right] \int \frac{d^d\ell_E}{(2\pi)^d}\frac{1}{(\ell_E^2 +M^2)^3}.
\label{n1int} 
\end{multline}
With the help of \Cref{scalarellint}, we can evaluate the loop momentum integrals and take the limit that $d\rightarrow 4$. The first integral of \Cref{gamaampm} becomes
\begin{multline}
\int \frac{d^d\ell}{(2\pi)^d} \frac{\mathcal{N}_{\texttt{V}}^{(1)}}{(\ell^2 -\Delta)^3}   = i\frac{(2-d)^2}{d}\,\frac{1}{(4\pi)^{d/2}}\frac{d\, \Gamma(2-d/2)}{4(M^2)^{2-d/2}} 
-i\left[m^2\big( 8z - 2 (1+z^2)\right.\\
\left. + (4-d) (1-z)^2 \big)- q^2 \big(2(1-x)(1-y)-(4-d)xy\big)\right] \frac{(4-d)}{4M^2} \frac{1}{(4\pi)^{d/2}} \frac{\Gamma(2-d/2)}{(M^2)^{2-d/2}}\\
\underset{d\rightarrow 4}{=} \frac{i}{(4\pi)^2} \left( \frac{2}{\epsilon}- \log M^2 -\gamma_E + \log 4\pi + \frac{q^2 (1-x)(1-y)+ (1-4z+z^2)m^2}{M^2}-2\right).
\label{n1intf}
\end{multline}
Similarly for the second integral of \Cref{gamaampm}
\begin{align}
\int \frac{d^d\ell}{(2\pi)^d} \frac{\mathcal{N}_2}{(\ell^2 -M^2)^3}& \underset{d\rightarrow 4}{=} \frac{1}{(4\pi)^2}\frac{-2im^2 z(1-z)}{M^2}.
\label{n2int}
\end{align}
The form factors will be given by 
\begin{multline}
F_1(q^2) = \frac{2e^2}{(4\pi)^2}\mu^{\epsilon}\int dF_3 \bigg( \frac{2}{\epsilon}- \log M^2 -\gamma_E + \log 4\pi \\+ \frac{q^2 (1-x)(1-y)+ (1-4z+z^2)m^2}{M^2}
-2 +\mathcal{O}(\epsilon)\bigg)+\delta_e,
\label{f1}
\end{multline}
\begin{align}
F_2(q^2) &=\frac{\alpha}{2\pi} \int dF_3\ \frac{2m^2 z(1-z)}{M^2}.
\label{f2}
\end{align} 
This is the point where we apply the $\overline{\mathrm{MS}}$ renormalization scheme to remove the divergent part of \Cref{f1}, we choose the associated counter term to be
\begin{align}
\delta_e & = \frac{-2e^2 }{(4\pi)^2}\left(\frac{2}{\epsilon}- \gamma_E + \log 4\pi \right) \int dF_3= -\frac{\alpha}{4\pi}\left(\frac{2}{\epsilon}- \gamma_E + \log 4\pi  \right).
\label{deltae}
\end{align} 
Finally, the first form factor will be
\begin{multline}
F_1(q^2)     =\frac{\alpha}{2\pi} \int dF_3\,\left[  \log\left(\frac{\mu^2}{m^2(1-z)^2 - xy q^2+z m_{\gamma}^2 }\right)\right.\\
\left. + \frac{q^2 (1-x)(1-y)+ (1-4z+z^2)m^2}{m^2(1-z)^2 - xy q^2 +z m_{\gamma}^2}-2\right].
\label{f1m}
\end{multline}

Now we evaluate the integrals of $F_1(q^2)$ and $F_2(q^2)$ where first integral of \Cref{f1m} is given by 
\begin{align}
\mathcal{I}_1 =\int dF_3\,\log\left(\frac{\mu^2}{m^2(1-z)^2 - xy q^2+z m_{\gamma}^2 }\right).
\label{f1i1}
\end{align}
We notice that $\mathcal{I}_1$ is finite as we set $m_{\gamma} \rightarrow 0$, so we can safely take that limit in this step. We change the variables from $x,\,y,\,z$ to $w = 1-z$ and $\xi = \frac{x}{x+y}$, then we have $x = w\,\xi$, $y = w(1-\xi)$ and $dF_3 = w\,dw\,d\xi$. The first integral becomes 
\begin{align}
\mathcal{I}_1 &= \int_0^1 d\xi\int_0^1 dw\,w \log \left(\frac{\mu^2 }{m^2w^2-w^2 \xi(1-\xi)q^2}\right)\nonumber\\
& =\int_0^1 d\xi\int_0^1 dw\,w \left[\log \left(\frac{m^2 }{m^2w^2-w^2 \xi(1-\xi)q^2}\right)+ \log\left(\frac{\mu^2}{m^2}\right)\right] \nonumber\\
&=  \frac{3}{2}-\frac{1}{2}\log\left(\frac{-q^2}{\mu^2}\right)+\mathcal{O}(m^2) \qquad,\text{ as $-q^2\gg m^2$}.
\label{f1i1f}
\end{align}

The second integral of \Cref{f1m} is given by
\begin{equation}
\mathcal{I}_2 = \int dF_3\,\frac{q^2 (1-x)(1-y)+ (1-4z+z^2)m^2}{m^2(1-z)^2 - xy q^2 +z m_{\gamma}^2}.
\label{f1i2}
\end{equation}
This integral diverges when $z\rightarrow 1$, We will use a trick to solve this tough integral where we add and subtract the argument of the integral in the region where we set $z=1$ and $x=y=0$ in the numerator and $z=1$ in the $m_{\gamma}^2$ term in the denominator. Then we have two integrals 
\begin{align}
\mathcal{J}_1 &= \int dF_3\left(\frac{q^2(1-x)(1-y)+(1-4z+z^2)m^2}{m^2(1-z)^2 - xy q^2 +z m_{\gamma}^2}-\,\frac{q^2 -2m^2}{m^2(1-z)^2 - xy q^2 + m_{\gamma}^2}\right),\label{f1i2a}\\
\mathcal{J}_2 & = \int dF_3\,\frac{q^2 -2m^2}{m^2(1-z)^2 - xy q^2 + m_{\gamma}^2}.\label{f1i2b}
\end{align}
We see that $\mathcal{J}_1$ is finite as we set $m_{\gamma}$ to be zero and the integral will be
\begin{align}
\mathcal{J}_1 &= \int dF_3\,\frac{q^2\big((1-x)(1-y)-1\big)+(3-4z+z^2)m^2}{m^2(1-z)^2 - xy q^2}\nonumber\\
& = 2 \log\left(\frac{-q^2}{m^2}\right)-\frac{1}{2}+\mathcal{O}(m^2)\qquad, \text{ as $-q^2 \gg m^2$}. \label{f1i2af}
\end{align}
To evaluate $\mathcal{J}_2$ we Change the variables in the same way as in $\mathcal{I}_1$, then $\mathcal{J}_2$ becomes
\begin{align}
\mathcal{J}_2 & = \int_0^1 d\xi\int_0^1 d\omega\, \omega \frac{q^2-2m^2}{m^2\omega^2-\omega^2 \xi(1-\xi)q^2+m_{\gamma}^2}\nonumber\\
& = \int_0^1d\xi\,\frac{q^2-2m^2}{2m^2-2q^2\xi(1-\xi)}\log\left(\frac{m^2+m_{\gamma}^2-q^2\xi(1-\xi)}{m_{\gamma}^2}\right),
\label{f1i2bm}
\end{align}
we can safely neglect the $m_{\gamma}^2$ in the numerator inside the logarithm, then we write equation \eqref{f1i2bm} as follows
\begin{align}
\mathcal{J}_2 & = \int_0^1d\xi\,\frac{q^2-2m^2}{2m^2-2q^2\xi(1-\xi)}\Bigg(\log\left(\frac{m^2-q^2\xi(1-\xi)}{-q^2}\right)+\log\left(\frac{-q^2}{m_{\gamma}^2}\right)\Bigg)\nonumber\\
& = \frac{1}{2}\log^2\left(\frac{-q^2}{m^2}\right)+\frac{\pi^2}{6}-\log\left(\frac{-q^2}{m^2}\right)\log\left(\frac{-q^2}{m_{\gamma}^2}\right)+\mathcal{O}(m^2,m_{\gamma}^2).
\label{f1i2bf}
\end{align}
The integral $\mathcal{I}_2$ becomes
\begin{align}
\mathcal{I}_2 = -\log\left(\frac{-q^2}{m^2}\right)\log\left(\frac{-q^2}{m_{\gamma}^2}\right)+\frac{1}{2}\log^2\left(\frac{-q^2}{m^2}\right)+2 \log\left(\frac{-q^2}{m^2}\right)-\frac{1}{2}+\frac{\pi^2}{6}+\mathcal{O}(m^2,m_{\gamma}^2).
\label{f1i2f}
\end{align}
While the third integral of equation \eqref{f1m} is given by $\mathcal{I}_3 =-2\, \int dF_3 = -1$.
Substituting the values of the three integrals $\mathcal{I}_1$, $\mathcal{I}_2$, and ${\mathcal{I}_3}$ into equation \eqref{f1m}, $F_1(q^2)$ simplifies to

\begin{multline}
F_1(q^2) = \frac{\alpha}{2\pi}\bigg[-\log\left(\frac{-q^2}{m^2}\right)\log\left(\frac{-q^2}{m_{\gamma}^2}\right)+\frac{1}{2}\log^2\left(\frac{-q^2}{m^2}\right)+2 \log\left(\frac{-q^2}{m^2}\right)\\
-\frac{1}{2}\log\left(\frac{-q^2}{\mu^2}\right)+\frac{\pi^2}{6}+\mathcal{O}(m^2,m_{\gamma}^2)\bigg].
\label{f1f}
\end{multline}
$F_2(q^2)$ can be evaluated by doing the same change of variables as in $\mathcal{I}_1$ to find
\begin{align}
F_2(q^2) &=\frac{\alpha}{2\pi} \int_0^1 d\xi\int_0^1 dw\, \frac{2(1-w) m^2}{ m^2 - \xi(1-\xi)q^2}=\frac{\alpha}{\pi}\left[ \frac{m^2}{-q^2}\log\left(\frac{-q^2}{m^2}\right)+\mathcal{O}(m^4)\right].
\label{f2f}
\end{align}
We see that $F_2(q^2)$ is negligible in the limit $m\rightarrow 0$. Finally, the amplitude of the vertex correction is given by
\begin{align}
i \mathcal{M}_{\texttt{V}} & = \frac{i e^2}{q^2} \,\bar{u}^{s'}(p')\,\gamma^0\,u^s(p)\,F_1(q^2) = i\,\mathcal{M}_0 \,F_1(q^2).
\label{vertxampf}
\end{align}
Consequently, the contribution of the vertex correction to the differential cross section at NLO will be
\begin{align}
\left(\frac{d\sigma}{d\Omega}\right)_{\texttt{V}} &= \frac{1}{32 \pi^2} \sum_{s,s'}\left[\mathcal{M}_{\texttt{0}}\mathcal{M}_{\texttt{V}}^*
+\mathcal{M}_{\texttt{V}}\mathcal{M}_{\texttt{0}}^*\right]\nonumber\\
& =
\left(\frac{d\sigma}{d\Omega}\right)_{\texttt{0}}\frac{\alpha}{\pi}\bigg[-\log\left(\frac{-q^2}{m^2}\right)\log\left(\frac{-q^2}{m_{\gamma}^2}\right)+\frac{1}{2}\log^2\left(\frac{-q^2}{m^2}\right)+2 \log\left(\frac{-q^2}{m^2}\right)\nonumber\\
&\hspace{5.5cm}-\frac{1}{2}\log\left(\frac{-q^2}{\mu_{\overline{\mathrm{MS}}}^2}\right)+\frac{\pi^2}{6}+\mathcal{O}(m^2,m_{\gamma}^2)\bigg].
\label{vertxxsection}
\end{align}
\section{Electron Self-Energy Correction}
Power counting implies that the electron self energy contains a UV divergent term corresponding to the 1PI diagram in \Cref{divb}.
The amplitude of the electron self energy is given by
\begin{align}
-i\Sigma_2(p) &=\begin{gathered}
\includegraphics[scale=1]{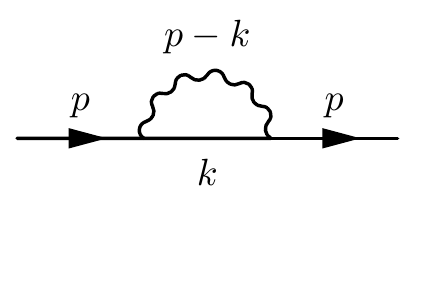}
\end{gathered}\quad+\quad\begin{gathered}
\includegraphics[scale=1]{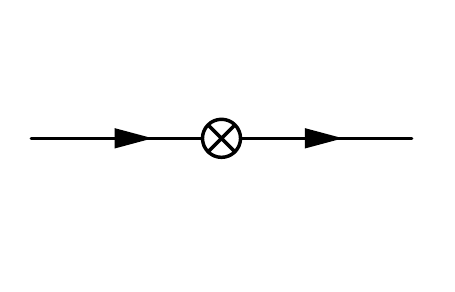}
\end{gathered}
\nonumber\\
& = (-ie\ \mu^{\frac{4-d}{2}})^2 \int \frac{d^dk}{(2\pi)^d}\,\gamma^{\alpha}\frac{i(\slashed{k}+m)}{[k^2-m^2+i\epsilon]}\gamma_{\alpha}\frac{-ig_{\alpha\beta}}{[(p-k)^2-m_{\gamma}^2+i\epsilon]}+i(\slashed{p}\delta_{\psi}-m\delta_m).
\label{eselfamp}
\end{align}
As usual for the calculations of the loop integrals, we use Feynman parameters so that we rewrite 
\begin{multline}
\frac{1}{[(p-k)^2-m_{\gamma}^2+i\epsilon][k^2-m^2+i\epsilon]}\\ = \int_0^1 dx\, \frac{1}{\big(x[(p-k)^2-m_{\gamma}^2+i\epsilon]+(1-x)[k^2-m^2+i\epsilon]\big)^2}\\
 = \int_0^1 dx\,\frac{1}{(\ell^2-M^2)^2+i\epsilon}\label{eselfdenom},
\end{multline}
where $\ell = k - xp$ and $M^2 = (1-x)^2m^2- x(1-x)p^2 + xm_{\gamma}^2$. Let us now simplify the numerator of \Cref{eselfamp} using the properties of gamma matrices in $d$-dimensions to find $ \mathcal{N}_{\texttt{e}} =\gamma^{\alpha}(\slashed{k}+m)\gamma_{\alpha} = (2-d)x\slashed{p}- m\,d + \text{linear terms in $\ell$}$. We also use the Wick rotation to evaluate the momentum integral. The amplitude of the electron self energy becomes
\begin{multline}
-i\Sigma_2(p)= -e^2\ \mu^{4-d} \int_0^1 dx \, [(2-d)x\slashed{p}+m\,d]\cdot \int \frac{d^d\ell}{(2\pi)^d}\,\frac{1}{(\ell^2-M^2+i\epsilon)^2}+i(\slashed{p}\delta_{\psi}-m\delta_m)\\
\hspace{1.6cm}= -ie^2\ \mu^{4-d} \int_0^1 dx \, [(2-d)x\slashed{p}+m\,d]\cdot \bigg(\frac{1}{(4\pi)^{d/2}}\frac{\Gamma(2-d/2)}{(M^2)^{2-d/2}}\bigg)+i(\slashed{p}\delta_{\psi}-m\delta_m)\\
  \underset{d\rightarrow 4}{=}  \frac{-ie^2\ \mu^{\epsilon}}{(4\pi)^2}\int_0^1 dx \,\bigg[-2x\slashed{p} \cdot \left(\frac{2}{\epsilon} -\log M^2 -\gamma_E + \log 4\pi-1+O(\epsilon)\right) \\
 +4m\left(\frac{2}{\epsilon} -\log M^2 -\gamma_E + \log 4\pi-1/2+O(\epsilon)\right)\bigg]+i(\slashed{p}\delta_{\psi}-m\delta_m).
 \label{boxampm}
\end{multline}

Now we apply the $\overline{\mathrm{MS}}$ renormalization scheme by choosing the counter terms $\delta_{\psi}$ and $\delta_m$ to absorb the UV divergent term as well as the constant term $(\log (4\pi) -\gamma_E)$:
\begin{align}
\delta_{\psi}&= \frac{-\alpha}{4\pi}\left(\frac{2}{\epsilon}-\gamma_E+\log 4\pi\right)\label{deltapsi},\\
\delta_m &= \frac{-\alpha}{\pi}\left(\frac{2}{\epsilon}-\gamma_E+\log 4\pi \right).
\label{deltam}
\end{align}
While the amplitude for the electron-self energy becomes
\begin{equation}
\Sigma_2(\slashed{p}) = \frac{\alpha}{4\pi}\left[(\slashed{p}-2m)+ \int_0^1 dx \,(4m-2x\slashed{p})\,\log\left(\frac{\mu^2}{(1-x)m^2-x(1-x)p^2+xm_{\gamma}^2}\right)\right].
\label{eselfampf}
\end{equation}

Before evaluating the integral in \Cref{eselfampf}, one may find an easy way to obtain the contribution from the electron self energy to the differential cross section. Based on a previous discussion we saw that in the $\overline{\mathrm{MS}}$ renormalziation scheme the renormalized parameters are not necessarily the physical ones. The Fourier transform of the two point correlation function of the electron self energy is given by \cite{Peskin:1995ev} 
\begin{align}
\int d^4x \bra{\Omega}T(\psi(x)\widebar{\psi}(0))\ket{\Omega}e^{ip \cdot x} &= \frac{i}{\slashed{p}-m}+\frac{i}{\slashed{p}-m}\left(\frac{\Sigma(\slashed{p})}{\slashed{p}-m}\right)+\frac{i}{\slashed{p}-m}\left(\frac{\Sigma(\slashed{p})}{\slashed{p}-m}\right)^2 + \dots \nonumber\\
& = \frac{i}{\slashed{p}-m-\Sigma(\slashed{p})}.
\label{efullprop}
\end{align} 

\Cref{efullprop} means that the pole is shifted by $\Sigma(\slashed{p})$, so the renormalized mass is not the physical mass and the residue of this pole is no longer one \cite{Srednicki:2007qs}. Our goal now is to find the correction to the residue and the relation between the renormalized mass $m$ and the physical mass $m_{e}$, where the pole should occur exactly at the physical mass. Then we have
\begin{align}
\big(\slashed{p}- m - \Sigma(\slashed{p})\big)\big|_{\slashed{p} = m_{e}} = 0,
\label{maspole}
\end{align}
which implies $ m_{e} = m + \Sigma(m_{e})$. We note that $\Sigma_2(p^2)$ is in $O(\alpha)$, so the difference between $m_{e}$ and $m$ is $O(\alpha)$ and we can replace $m_{e}$ by $m$ and set the error to be $O(\alpha^2)$ \cite{Srednicki:2007qs}. Then we have  
\begin{equation}
m_{e} = m + \Sigma(m)+O(\alpha^2).
\label{physmass}
\end{equation}
We also note that $\Sigma_2(m)$ is finite as we set $m_{\gamma}^2\rightarrow 0$, which means that there is no soft IR divergences in $\Sigma_2$ to worry about, which becomes 
\begin{align}
\Sigma_2(m) & = \frac{\alpha}{4\pi}\bigg(-m+ \int_0^1 dx\, (4-2x)m\, \log\left(\frac{\mu^2}{(1-x)^2 m^2}\right)\bigg)\nonumber\\
& = m\,\frac{\alpha}{4\pi}\left(4+3\log \frac{\mu^2}{m^2}\right).
\label{sigma2}
\end{align} 
Then the relation between the physical mass and the renormalized mass is given by
\begin{align}
m_{e} = m\left[1+\frac{\alpha}{4\pi}\left(4+3\log  \left(\frac{\mu^2}{m^2}\right)\right) + O(\alpha^2)\right].
\label{physmassm}
\end{align}
Again the difference between $m$ and $m_{e}$ is $O(\alpha)$, so we can replace $m^2$ by $m_{e}^2$ in the logarithm
\begin{equation}
m \approx m_{e} \left[1- \frac{\alpha}{4\pi}\left(4+3\log \left(\frac{\mu^2_{\overline{\mathrm{MS}}}}{m_{e}^2}\right)\right)+ O(\alpha^2)\right].
\label{physmassf}
\end{equation}

Now it is time to find the correction to the residue $R$. One could add the contribution from the electron self energy directly by correcting the LSZ reduction formula \cite{Schwartz:2013pla} in which the shifted pole with a non-unity residue exist. The inverse of the residue is given by
\begin{align}
R^{-1} & = \frac{d}{d\slashed{p}}\big(\slashed{p}- m - \Sigma(\slashed{p})\big)\big|_{\slashed{p} = m_{e}}\nonumber\\
& = 1- \Sigma^{\prime}(m_{e}) \nonumber \\
& = 1 - \Sigma^{\prime}(m) + O(\alpha^2)\nonumber\\
& = 1-\frac{\alpha}{4\pi}\left(1+ \int_0^1dx\, \left[ \frac{4x(1-x)(2-x)m^2}{(1-x)^2m^2+xm_{\gamma}^2}-2x \, \log\left(\frac{\mu^2}{(1-x)^2 m^2 + xm_{\gamma}^2}\right)\right]\right).
\label{resid} 
\end{align}
We note that the first integral contains an infrared divergence as $m_{\gamma}\rightarrow 0 $, while the second integral is finite. Then we have as $m_{\gamma}\rightarrow 0$
\begin{align}
\int_0^1dx\, \frac{4x(1-x)(2-x)m^2}{(1-x)^2m^2+xm_{\gamma}^2} &  = 2 \log\left(\frac{m^2}{m_{\gamma}^2}\right)-2+\mathcal{O}(m^2,m_{\gamma}^2)\label{residi1},
\\
\int_0^1dx\,2x \, \log\left(\frac{\mu^2}{(1-x)^2 m^2 }\right) & = \log\left(\frac{\mu^2}{m^2}\right)+3.
\label{residi2}
\end{align}
Finally the inverse of the residue is given by
\begin{equation}
R^{-1} = 1 - \frac{\alpha}{4\pi}\left[2\log\left(\frac{m^2}{m_{\gamma}^2}\right) - \log\left(\frac{\mu^2}{m^2}\right) -4+\mathcal{O}(m^2,m_{\gamma}^2)\right].
\label{residf}
\end{equation}

As we discussed above, the contribution from the electron self-energy can be encapsulated as a correction to the LSZ formula in which we multiply the amplitude by the value of $R^{\frac{1}{2}}$ for each external leg, which means that we directly multiply the differential cross section by $R^2$ \cite{Srednicki:2007qs}. However the residue correction is in $O(\alpha)$, so all the NLO terms will not be affected by this correction to stay in the same order of the perturbation and the only affected term will be the leading order. Then the leading term will be corrected to
\begin{align}
\left(\frac{d\sigma}{d\Omega}\right)_{\texttt{L}} =  R^{2}\left(\frac{d\sigma}{d\Omega}\right)_{\texttt{0}} = \left(\frac{d\sigma}{d\Omega}\right)_0\,\left\{1 + \frac{\alpha}{\pi}\bigg[\log\left(\frac{m^2}{m_{\gamma}^2}\right) - \frac{1}{2}\log\left(\frac{\mu_{\overline{\mathrm{MS}}}^2}{m^2}\right) -2\bigg]\right\}.
\label{xsectionl}
\end{align}
Now we can write the contribution from the tree level amplitude plus the vertex and the self energy corrections in one single equation to give
\begin{multline}
\left(\frac{d\sigma}{d\Omega}\right)_{\texttt{VL}} = \left(\frac{d\sigma}{d\Omega}\right)_{\texttt{0}} \left\{1+\frac{\alpha}{\pi}\left[\log\left(\frac{m^2}{m_{\gamma}^2}\right)\left(1-\log\left(\frac{-q^2}{m^2}\right)\right) -\frac{1}{2}\log^2\left(\frac{-q^2}{m^2}\right)\right.\right.\\
\left.\left.+\frac{3}{2}\log\left(\frac{-q^2}{m^2}\right)+\frac{\pi^2}{6}-2\right] \right\}.
\label{vlxsection}
\end{multline}
\section{Box Correction}
In the last three sections, we calculated, at NLO, the contribution from the diagrams corresponding to the three superficially divergent 1PI diagrams shown in \Cref{divdegree}. Hence we do not expect more divergent diagrams in the UV limit. One of the non-divergent diagrams that contribute to the differential cross section at NLO is the box correction which occurs when the incoming electron interacts twice with the external source. The amplitude of the box diagram is given by
\begin{align}
i\mathcal{M}_{\texttt{BO}} &=\hspace{0.5cm}\begin{gathered}
\includegraphics[scale=1]{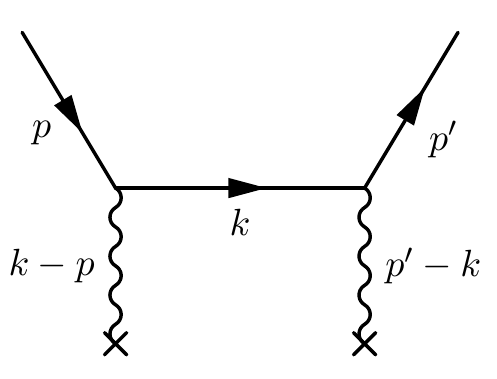}
\end{gathered}
\nonumber\\&= -e^2 \bar{u}^{s'}(p')\,  i\eta_{\mu\nu} (p,p')\, V^{\mu}(p'-k) \,V^{\nu}(k-p)\, u^s(p),
\label{boxamp}
\end{align}
where
\begin{align}
i\eta_{00} (p,p') = ie^2 \int\, \frac{d^4k}{(2\pi)^4} \frac{\gamma^0 (\slashed{k}+m)\gamma^0}{\big[(p'-k)^2 - m_{\gamma}^2\big]\big[(k-p)^2-m_{\gamma}^2\big]\big[k^2 - m^2 +i\epsilon\big]}.
\label{mamp}
\end{align}

It is clear that the box diagram does not contain any ultraviolet divergences. We also note that $\tilde{J}_{0}(p'-k)$ and $\tilde{J}_{0}(k-p)$ yield two delta functions $\delta(p^{' 0}-k^0)$ and $\delta(p^0-k^0)$, which allow us to perform the integral over $k^0$ in \Cref{mamp}, where 
\begin{equation}
\begin{aligned}
(p'-k)^2 &= -(\vec{p'}-\vec{k})^2,\\
(k-p)^2 &= -(\vec{k}-\vec{p})^2, \\
k^2-m^2 &= k^2 - p^2 = \vec{p}^{\ 2} -\vec{k}^{2}.
\end{aligned}
\label{boxmomenta}
\end{equation}
The denominator of \Cref{mamp} becomes
\begin{equation}
\big[(p'-k)^2 - m_{\gamma}^2\big]\big[(k-p)^2-m_{\gamma}^2\big]\big[k^2 - m^2 +i\epsilon\big] = \big[(\vec{p}^{\,'}-\vec{k})^2 - m_{\gamma}^2\big]\big[(\vec{k}-\vec{p})^2-m_{\gamma}^2\big]\big[\vec{p}^{\,2}-\vec{k}^2 +i\epsilon\big].
\label{mdenom}
\end{equation}
We can also simplify the numerator of \Cref{mamp} to be
$ \gamma^0(\slashed{k}+m)\gamma^0 = \gamma^0 E + \vec{\gamma}\cdot \vec{k} + m$. The amplitude of the box diagram can be now rewritten as
\begin{align}
i\mathcal{M}_{\texttt{BO}} & = \frac{-ie^4}{(2\pi)^3} \bar{u}^{s'}(p')\, \bigg( \int d^3k\,\frac{(E\gamma^0+m)+\vec{\gamma}\cdot \vec{k}}{\big[(\vec{p}^{\,'}-\vec{k})^2 + m_{\gamma}^2\big]\big[(\vec{k}-\vec{p})^2+m_{\gamma}^2\big]\big[\vec{p}^{\,2}-\vec{k}^2 +i\epsilon\big]}\bigg) u^s(p)\nonumber\\
& = \frac{-2i\alpha^2}{\pi}\, \bar{u}^{s'}(p')\,\big[(E\gamma^0 +m )I_1+\vec{\gamma}\cdot\vec{I}\big]u^s(p),
\label{boxampm}
\end{align}
where we define
\begin{align}
I_1 &=\int \frac{d^3k}{\big[(\vec{p}^{\,'}-\vec{k})^2 + m_{\gamma}^2\big]\big[(\vec{k}-\vec{p})^2+
m_{\gamma}^2\big]\big[\vec{p}^{\,2}-\vec{k}^2 +i\epsilon\big]}\label{boxint1},\\
\vec{I}& =  \int \frac{\vec{k}\,d^3k}{\big[(\vec{p}^{\,'}-\vec{k})^2 + m_{\gamma}^2\big]\big[(\vec{k}-\vec{p})^2+
m_{\gamma}^2\big]\big[\vec{p}^{\,2}-\vec{k}^2 +i\epsilon\big]}\label{boxint2}.
\end{align}
Now we calculate the integrals $I_1$ and $I_2$, to do this we will use a trick introduced by R.~H.~Dalitz in \cite{Dalitz:1951ah}. This trick makes use of the identity 
\begin{equation}
\frac{1}{A\,B} = \int_{-1}^1 dx \, \frac{2}{\big[A(1+x)+B(1-x)\big]^2},
\label{boxtrick1}
\end{equation}
such that we can write
\begin{multline}
\frac{1}{\big[(\vec{p}^{\,'}-\vec{k})^2 + m_{\gamma}^2\big]\big[(\vec{k}-\vec{p})^2+
m_{\gamma}^2\big]} \\= \int_{-1}^1 dx\, \frac{1}{\bigg(\big[(\vec{p}^{\,'}-\vec{k})^2 + m_{\gamma}^2\big](1+x)+\big[(\vec{k}-\vec{p})^2+
m_{\gamma}^2\big](1-x)\bigg)^2}.
\label{intdenom}
\end{multline}

Now we define a new vector $\vec{\ell} = \frac{1}{2}[(1+x)\vec{p}'+ (1-x)\vec{p}]$ and doing some manipulations to the denominator of \Cref{intdenom}, it becomes $\mathcal{D} = 2\,[(\vec{k}-\vec{\ell})^2 + M^2]$ where $M^2  = \frac{1}{2}(1-x^2)(\vec{p}^{\,2} - \vec{p}^{\,'}\cdot\vec{p}) + m_{\gamma}^2$. Then the integral $I_1$ and the components of the integral $\vec{I}$ become
\begin{align}
I_1 & = \frac{1}{2} \int_{-1}^1dx\,\int \frac{d^3k}{[(\vec{k}-\vec{\ell})^2 + M^2]^2 [\vec{p}^{\,2}-\vec{k}^{\,2}+i\epsilon]},\label{boxint1m}\\
I_r & = \frac{1}{2} \int_{-1}^1dx\,\int \frac{k_r\,d^3k}{[(\vec{k}-\vec{\ell})^2 + M^2]^2 [\vec{p}^{\,2}-\vec{k}^{\,2}+i\epsilon]}.
\label{boxintr}
\end{align}

The integrals in the form of \Cref{boxint1m,boxintr} can be solved by solving an integral in the following form \cite{Dalitz:1951ah}
\begin{align}
J &= \int \frac{d^3k}{[(\vec{k}-\vec{\ell})^2 + M^2] \cdot [\vec{p}^{\,2}-\vec{k}^{\,2}+i\epsilon]}\nonumber\\
&= \pi \int_{-1}^1 d(\cos \theta) \int_{-\infty}^{\infty} \frac{k^2 \,dk}{[k^2 + \ell^2 -2 k\ell\, \cos\theta + M^2] \cdot [p^2-k^2+i\epsilon]}.
\label{boxi1j}
\end{align}
Completing the contour in the upper half-plane and carrying out this integral, gives
\begin{equation}
J = \int \frac{d^3k}{[(\vec{k}-\vec{\ell})^2 + M^2] \cdot [\vec{p}^{\,2}-\vec{k}^{\,2}+i\epsilon]} = \frac{i\pi^2}{\ell} \, \log\left(\frac{\left|\vec{p}\right|-\ell+iM}{\left|\vec{p}\right|+\ell+iM}\right).
\label{boxi1jf}
\end{equation}
By differentiating \Cref{boxi1jf} with respect to $M$ we find
\begin{align}
\int \frac{d^3k}{[(\vec{k}-\vec{\ell})^2 + M^2]^2 [\vec{p}^{\,2}-\vec{k}^{\,2}+i\epsilon]} & = \frac{\pi^2}{M[\vec{p}^{\,2} - \ell^2 +2i\left|\vec{p}\right|M-M^2]},\label{jint}
\end{align}
where $\ell^2 =\frac{1}{2}[(1+x^2)\vec{p}^{\,2}+(1-x^2) \vec{p}\cdot \vec{p}^{\,'}]$.
The denominator of the first integral can be simplified to be $M(2i\left|\vec{p}\right|M-m_{\gamma}^2)$. Let us also define $Q^2=-q^2 = -(p'-p)^2\approx 2(E^2-\vec{p}\cdot \vec{p}\ ')$, in the massless limit, which implies $\vec{p}\cdot \vec{p}\,' = E^2-Q^2/2$. Then we have
\begin{align}
I_1 &= \frac{\pi^2}{2} \int_{-1}^1 dx\, \frac{1}{M\,(2i\left|\vec{p}\right|M - m_{\gamma}^2)}\nonumber\\
& =\frac{-\pi^2}{Q\sqrt{E^2( Q^2+4m_{\gamma}^2)+m_{\gamma}^4}} \left[2i\tan^{-1}\left(\frac{m_{\gamma}^2Q}{2\sqrt{m_{\gamma}^6+E^2 m_{\gamma}^2(Q^2+4m_{\gamma}^2)}}\right)\right.\nonumber\\
&\left.\hspace{5cm}+i\log
\left(\frac{\sqrt{E^2( Q^2+4m_{\gamma}^2)+m_{\gamma}^4}+EQ}{\sqrt{E^2( Q^2+4m_{\gamma}^2)+m_{\gamma}^4}-EQ}\right)\right].
\label{boxint1f}
\end{align}

We note that $I_1$ diverges as $m_{\gamma}\rightarrow 0$. However we are not concerned with this divergent part because we expect to use only the real part of $I_1$ which is exactly zero in the limit $m_{\gamma}\rightarrow 0$. Let us now calculate $I_r$ by differentiating \Cref{boxi1jf} with respect to $\ell_{\mu}$ 
\begin{multline}
\int \frac{k_r\,d^3k}{[(\vec{k}-\vec{\ell})^2 + M^2]^2 [\vec{p}^{\,2}-\vec{k}^{\,2}+i\epsilon]} \\= \pi^2 \ell_r \bigg[\frac{1}{M(\left|\vec{p}\right|-\ell+iM)(\left|\vec{p}\right|+\ell+iM)}+ \frac{i}{2\ell^3}\log\left(\frac{(\left|\vec{p}\right|-\ell+iM)}{(\left|\vec{p}\right|+\ell+iM)}\right)\\
 +\frac{i}{2\ell^2}\left(\frac{1}{(\left|\vec{p}\right|-\ell+iM)}+\frac{1}{(\left|\vec{p}\right|+\ell+iM)}\right)\bigg].
\label{krint}
\end{multline}
Substituting \Cref{krint} into \Cref{boxintr}, $I_r$ becomes
\begin{multline}
I_r = \frac{\pi^2}{2}\int_{-1}^1dx\, 
\ell_r \bigg[\frac{1}{M(\left|\vec{p}\right|-\ell+iM)(\left|\vec{p}\right|+\ell+iM)}+ \frac{i}{2\ell^3}\log\left(\frac{(\left|\vec{p}\right|-\ell+iM)}{(\left|\vec{p}\right|+\ell+iM)}\right)\\
+\frac{i}{2\ell^2}\left(\frac{1}{(\left|\vec{p}\right|-\ell+iM)}+\frac{1}{(\left|\vec{p}\right|+\ell+iM)}\right)\bigg].
\label{boxintr}
\end{multline}
When we plug $\ell_r = \frac{1}{2}[(p'+p)_r+x(p'-p)_r]$ into equation \Cref{boxintr}, the second term vanishes, giving us $I_r = \frac{1}{2}(p+p')_r I_2$, where
\begin{multline}
I_2 =I_1+ \frac{\pi^2}{2}\int_{-1}^1dx\,  \bigg[ \frac{i}{2\ell^3}\log\left(\frac{\left|\vec{p}\right|-\ell+iM}{\left|\vec{p}\right|+\ell+iM}\right)
+\frac{i}{2\ell^2}\left(\frac{1}{(\left|\vec{p}\right|-\ell+iM)}+\frac{1}{(\left|\vec{p}\right|+\ell+iM)}\right)\bigg].
\label{iri2}
\end{multline}
We recall that $\ell^2= \frac{1}{2}\left[p^2(1+x^2)+(1-x^2)\vec{p}\cdot \vec{p}\ '\right] = \left[E^2-(1-x^2)\frac{Q^2}{4}\right]$ and $M^2 = (1-x^2)\frac{Q^2}{4}+m_{\gamma}^2$. The Second integral in \Cref{iri2} is finite in the limit $m_{\gamma}\rightarrow 0$, so we find
\begin{multline}
\int_{-1}^1dx\,  \frac{i}{2\ell^3}\log\left(\frac{\left|\vec{p}\right|-\ell+iM}{\left|\vec{p}\right|+\ell+iM}\right)\\ =\frac{\pi^2}{EQ^2(4E^2-Q^2)^{3/2}}\Bigg\{\pi Q\left[2E\sqrt{4E^2-Q^2}-4E^2+Q^2\right]\\+iQ\left[(2Q^2-8E^2)\tan^{-1}\left(\frac{Q}{\sqrt{4E^2-Q^2}}\right)+Q\sqrt{4E^2-Q^2}\left(\log\frac{Q^2}{4E^2}+i\pi\right)\right]. 
\label{i2part1}
\end{multline}

Similarly, the third integral of \Cref{iri2} is given by
\begin{align}
\frac{\pi^2}{2}\int_{-1}^1dx\, \frac{i}{2\ell^2}\left(\frac{1}{(\left|\vec{p}\right|-\ell+iM)}+\frac{1}{(\left|\vec{p}\right|+\ell+iM)}\right)& = \frac{\pi^2}{2}\int_{-1}^1dx\, \frac{i}{2\ell^2}\left(\frac{1}{i M}+\frac{1}{\left|\vec{p}\right|}\right)\nonumber\\
& = \pi^2\left[\frac{\pi+2 i \tan^{-1}\left(\frac{Q}{\sqrt{4E^2-Q^2}}\right)}{EQ\sqrt{4E^2-Q^2}}\right].
\end{align}
The real part of $I_2$ will be
\begin{equation}
\mathrm{Re}(I_2)= \frac{\pi^3}{Q^2E\left(\frac{2E}{Q}+1\right)}.
\label{ReI2}
\end{equation}
One can now write the box amplitude in terms of $I_1$ and $I_2$ to be
\begin{align}
i\mathcal{M}_{\texttt{BO}}&  = \frac{-2i\alpha^2}{\pi} \ \bar{u}^{s'}(p')\left[(E\gamma^0+m)I_1+\frac{1}{2} (\vec{p}+\vec{p}\ ')\cdot \vec{\gamma}\ I_2\right] u^s(p).
\label{boxampm2}
\end{align}
Using the on-shell conditions to write
\begin{equation}
\begin{aligned}
\vec{\gamma}\cdot \vec{p}\ u^s(p) &= (E\gamma^0-m)u^s(p),\ u^{s'}(p')\ \vec{\gamma}\cdot \vec{p}\ ' &=u^{s'}(p')\  (E\gamma^0-m).
\end{aligned}
\label{boxampsim}
\end{equation}
The box amplitude is now
\begin{align}
i\mathcal{M}_{\texttt{BO}}&  = \frac{-2i\alpha^2}{\pi} \ \bar{u}^{s'}(p')\left[(E\gamma^0(I_1+I_2)+m (I_1-I_2)\right] u^s(p).
\label{boxampm3}
\end{align}
We can now omit the second term of \Cref{boxampm3} in the limit $m\rightarrow 0$ to finally find
\begin{align}
i\mathcal{M}_{\texttt{BO}}&  = -i \mathcal{M}_{\texttt{0}} \frac{\alpha}{\pi}\cdot \frac{Q^2 E(I_1+I_2)}{2\pi}.
\label{boxampf}
\end{align}

After evaluating all the required integrals for the box correction, we get back to \Cref{boxampm} where we calculate the contribution from the box correction to the differential cross section from the interference between the leading term and the box amplitudes. This interference is giving by
\begin{equation}
\begin{aligned}
\mathcal{M}_{\texttt{0}}^* \mathcal{M}_{\texttt{BO}}
&= \left|\mathcal{M}_{\texttt{0}}\right|^2\frac{-\alpha}{\pi}\cdot\frac{Q^2E(I_1+I_2)}{2\pi},\\
\mathcal{M}_{\texttt{BO}}^* \mathcal{M}_{\texttt{0}} & = \left|\mathcal{M}_{\texttt{0}}\right|^2\frac{-\alpha}{\pi}\cdot\frac{Q^2E(I_1^*+I_2^*)}{2\pi}.
\end{aligned}
\label{boxinterf}
\end{equation}

We recall the real value of $I_2$ from \Cref{ReI2}. Finally, the contribution from the box diagram to the differential cross section at NLO is given by
\begin{align}
\left(\frac{d\sigma}{d\Omega}\right)_{\texttt{BO}}
& = \frac{1}{32\pi^2}\sum_{s,s'}\left[\mathcal{M}_{\texttt{0}}^* \mathcal{M}_{\texttt{BO}}+\mathcal{M}_{\texttt{BO}}^* \mathcal{M}_{\texttt{0}}\right]\nonumber\\
&=\left(\frac{d\sigma}{d\Omega}\right)_{\texttt{0}}\frac{-\alpha}{\pi}\cdot \frac{Q^2E}{\pi}\, \operatorname{Re}(I_1+I_2)\nonumber\\
& =\left(\frac{d\sigma}{d\Omega}\right)_{\texttt{0}}\frac{\alpha}{\pi}\left[\frac{-\pi^2}{\left(\frac{2E}{Q}+1\right)}\right].\label{boxxsection} 
\end{align}
\chapter{IR Cancellation}
\label{Chapter4}
The infrared problem in purely massless gauge theories has been understood and dealt with in two different approaches, the first is the coherent state approach introduced by Chung \cite{Chung:1965zza} and used later in QED by Curci and Greco \cite{Curci:1978kj} and QCD by Kibble \cite{Kibble:1969kd} and Nelson \cite{Nelson:1980qs}; the main idea is to define a representation for the photon states other than the usual Fock representation in which the $S$-matrix has no IR divergences. The second is by applying quantum mechanical based theories in which we sum over the physically indistinguishable degenerate states where this sum becomes free of any IR divergences. We will follow the latter approach.
\section{Bloch-Nordsieck Theorem}
The Bloch-Nordsieck (BN) theorem was the first theoretical attempt to solve the IR problem. The BN theorem has been first introduced by Bloch and Nordsieck \cite{Bloch:1937pw} and generalized later by D.~R.~Yennie, S.~C.~Frautschi, and H.~Suura \cite{Yennie:1961ad} in which they showed that it is impossible to specify exactly a final state with a charged particle or a charged particle plus a photon soft enough to be below the detector resolution. In a modern language, BN proved that summing over indistinguishable final states gives in return a formula which is free of all IR divergences.

It has been known for long time that applying the BN theorem solves the soft IR problem by adding the bremsstrahlung correction in which a soft photon is emitted from either the incoming or the outgoing electron. Let us review the BN cancellation.
\subsection{Soft Bremsstrahlung Corrections}
The scattering of any charged particle leads to the emission of radiation. This process is known as bremsstrahlung radiation where such a process is important for the cancellation of the soft IR divergences from the vertex correction according to the BN theorem. The amplitude for the emission of a photon from both the incoming and the outgoing electrons, as a final state $f$, is given by
\begin{align}
i\mathcal{M}_{\texttt{B}}^f & =\begin{gathered}
\includegraphics[scale=1]{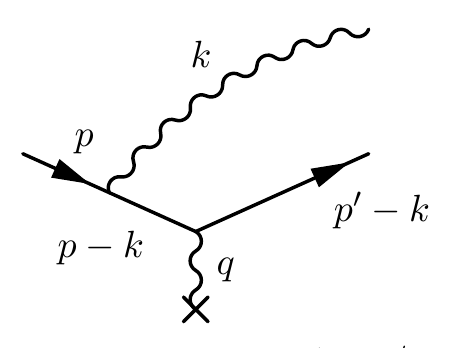}
\end{gathered}\quad+\quad
\begin{gathered}
\includegraphics[scale=1]{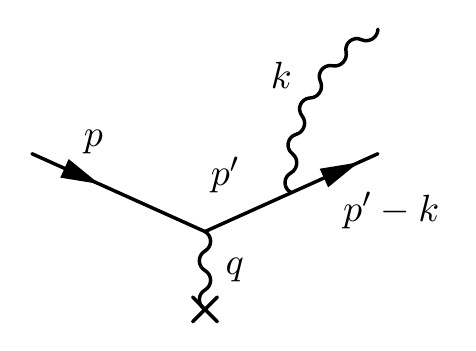}
\end{gathered}
\nonumber\\
& = \bar{u}^{s'}(p'-k) \left[-ie\gamma^{\mu}\right] \frac{i(\slashed{p}-\slashed{k}+m)}{(p-k)^2-m^2}\left[-ie\gamma^{\beta}\right]\varepsilon_{\beta}^{r*}(k)\ u^s(p)\left[-ieV^{\nu}\right]\frac{-ig_{\mu\nu}}{q^2}\nonumber\\
&\quad + \bar{u}^{s'}(p'-k) \left[-ie\gamma^{\beta}\right]\varepsilon_{\beta}^{r*}(k)\  \frac{i(\slashed{p}'+m)}{p'^2-m^2}\left[-ie\gamma^{\mu}\right] u^s(p)\left[-ieV^{\nu}\right]\frac{-ig_{\mu\nu}}{q^2} \nonumber\\
& = \frac{ie^3}{q^2}\bar{u}^{s'}(p'-k)\left[\frac{\gamma^0(\slashed{p}-\slashed{k}+m)\gamma^{\beta}}{(p-k)^2-m^2}+\frac{\gamma^{\beta}(\slashed{p}'+m)\gamma^0}{p'^{2}-m^2}\right]u^s(p)\ \varepsilon_{\beta}^{r*}.
\label{bremamp}
\end{align}

 We note that $p$ and $(p'-k)$ are on shell which means that $p^2=m^2$ and $(p'-k)^2= m^2$ which implies $p'^2 = m^2-m_{\gamma}^2 + 2p'\cdot k $. This allows us to rewrite the denominators of \Cref{bremamp} as $(p-k)^2-m^2\approx -2p\cdot k$ and $p'^2-m^2 \approx 2p'\cdot k$. The bremsstrahlung amplitude becomes
\begin{align}
i\mathcal{M}_{\texttt{B}}^f & = \frac{ie^3}{q^2}\bar{u}^{s'}(p'-k)\ S^{0\beta}\ u^s(p)\  \varepsilon_{\beta}^{r*},
\label{bremampm}
\end{align}
where
\begin{align}
S^{\alpha\beta} = \frac{\gamma^{\alpha}(\slashed{p}-\slashed{k}+m)\gamma^{\beta}}{(-2p\cdot k)}+\frac{\gamma^{\beta}(\slashed{p}'+m)\gamma^{\alpha}}{(2p'\cdot k)}.
\label{salphabeta}
\end{align}
The fact that the complete $p'-k$ is on shell means $|\vec{p}\,'-\vec{k}|= \sqrt{(E-\omega_k)^2-m^2}$, solving this equation for $|\vec{p}\,'|$ implies
\begin{equation}
|\vec{p}\,'| = k\cos \theta_{\gamma} + \sqrt{k^2(\cos \theta_{\gamma}-1) + (E-\omega_k)^2-m^2},
\label{modmom}
\end{equation}
where $\theta_{\gamma}$ is the angle between the vectors $\vec{p}\,'$ and $\vec{k}$.

The emission of a soft photon ($k\ll\Delta$) from either the incoming or the outgoing electrons can not be distinguished from these electrons, which causes an IR divergent part from these processes. The eikonal approximation is applicable in this case by which the photon momentum in the numerator of \Cref{bremampm} can be ignored, so we can write $(\slashed{p}-\slashed{k} + m) \approx (\slashed{p}+ m)$ and $u(p-k)\approx u(p)$. Using the identities $\slashed{p}\ u(p) = m\ u(p)$  and $\{\slashed{p},\gamma^{\mu}\} = 2 p^{\mu}$ we note
\begin{equation}
\begin{aligned}
(\slashed{p}-\slashed{k} + m)\gamma^{\alpha} \varepsilon_{\alpha}^{r*} u^s(p) &\approx 2p\cdot \varepsilon^{r*} u^s(p),\\
\bar{u}^{s'}(p')\gamma^{\alpha} \varepsilon_{\alpha}^{r*}(\slashed{p}'-\slashed{k} + m)& \approx \bar{u}^{s'}(p')\,2p\cdot \varepsilon^{r*}.
\label{bremampnum}
\end{aligned}
\end{equation}
Substituting \Cref{bremampnum} into \Cref{bremampm} gives us the amplitude of the emission of a soft photon from the incoming and the outgoing electrons as following 
\begin{align}
i\mathcal{M}_{\texttt{S}}^f & = \frac{i e^3}{q^2}\,\bar{u}^{s'}(p')\gamma^0\,u^s(p) \left(\frac{p'\cdot \varepsilon^{r*}}{p'\cdot k}-\frac{p\cdot \varepsilon^{r*}}{p\cdot k}\right)\nonumber\\
& = ie\, \mathcal{M}_{\texttt{0}}\,\left(\frac{p'\cdot \varepsilon^{r*}}{p'\cdot k}-\frac{p\cdot \varepsilon^{r*}}{p\cdot k}\right).
\label{softbremamp}
\end{align}
When we try to calculate the cross section, we need to integrate over the photon momentum $k$, and sum over the polarization $r$. The cross section is now 
\begin{equation}
\left(\frac{d\sigma}{d\Omega}\right)_{\texttt{S}}^f = \left(\frac{d\sigma}{d\Omega}\right)_{\texttt{0}} \cdot \int \frac{d^3 k }{(2\pi)^3}\, \frac{1}{2k} \sum_r e^2 \left|\frac{p'\cdot \varepsilon^{r*}}{p'\cdot k}-\frac{p\cdot \varepsilon^{r*}}{p\cdot k}\right|^2.
\label{sbremxsec}
\end{equation} 

The summation over all polarizations gives $\sum_r\varepsilon_{\alpha}^{r*} \varepsilon_{\beta}^r = - g_{\alpha\beta}+ \frac{k_{\alpha}k_{\beta}}{\omega_k^2}$, where $\omega_k = \sqrt{\vec{k}^2+m_{\gamma}^2}$ is the energy of the emitted photon. We note that the second term of the sum over polarizations gives no contribution to the cross section as $m_{\gamma}\rightarrow 0$ because of the Ward identity \cite{Mandl:1985bg}. We also note that the integral over $k$ will be only up to the energy resolution $\Delta$ as we interested in the emission of a soft photon in this section. Then \Cref{sbremxsec} becomes
\begin{equation}
\left(\frac{d\sigma}{d\Omega}\right)_{\texttt{S}}^f =\left(\frac{d\sigma}{d\Omega}\right)_{\texttt{0}} \frac{\alpha}{\pi}\int_0^{\Delta} dk\,\frac{k^2}{\omega_k}\int\frac{d\Omega_k}{4\pi}\left(\frac{2 p\cdot p'}{(p\cdot k)(p'\cdot k)}-\frac{m^2}{(p\cdot k)^2}-\frac{m^2}{(p'\cdot k)^2}\right).
\label{sbremxsecm}
\end{equation}

We should emphasize here that the soft photon approximation allows to simplify \Cref{modmom} to be $|\vec{p}\,'| = \sqrt{E^2-m^2}$. Now Let us divide the integral in \Cref{sbremxsecm} into the following three integrals
\begin{align}
\mathcal{A}_1 & = -\int_0^{\Delta} dk\,\frac{k^2}{\omega_k}\int\frac{d\Omega_k}{4\pi}\,\frac{m^2}{(p\cdot k)^2},\label{4.2.10a}\\
\mathcal{A}_2 & =-\int_0^{\Delta} dk\,\frac{k^2}{\omega_k}\int\frac{d\Omega_k}{4\pi}\,\frac{m^2}{(p'\cdot k)^2}, \label{4.2.10b}\\
\mathcal{A}_3 & = \int_0^{\Delta} dk\,\frac{k^2}{\omega_k}\int\frac{d\Omega_k}{4\pi}\,\frac{2 p\cdot p'}{(p\cdot k)(p'\cdot k)}.\label{sbremints}
\end{align}
The first integral $\mathcal{A}_1$ can be evaluated as follows
\begin{align}
\mathcal{A}_1 &= - \int_0^{\Delta} dk\,\frac{k^2}{\omega_k}\int_0^{2\pi}\frac{d\phi}{2\pi}\int_{-1}^1\frac{d\cos\theta}{2}\,\frac{m^2}{(E\,\omega_k-|\vec{p}|\,|\vec{k}|\, \cos\theta)^2}\nonumber\\
& = - \int_0^{\Delta}dk\frac{k^2}{\omega_k}\,\frac{m^2}{(E\,\omega_k)^2-(p\ k)^2}\nonumber\\
& = - \int_0^{\Delta}dk\,\frac{k^2}{\sqrt{k^2+m_{\gamma}^2}}\,\frac{m^2}{E^2\,m_{\gamma}^2+m^2\,k^2}\nonumber\\
& = \frac{1}{2}\log\left(\frac{E^2}{m^2}\right)-\frac{1}{2}\log\left(\frac{\Delta^2}{m_{\gamma}^2}\right)+ \mathcal{O}(m^2),
\label{sbreminta1}
\end{align}
where we used the relation $m^2 = E^2-|\vec{p}|^2$ to simplify the denominator. We find that $\mathcal{A}_2$ is exactly the same as $\mathcal{A}_1$, then we have
\begin{equation}
\mathcal{A}_1+\mathcal{A}_2 = \log\left(\frac{E^2}{m^2}\right)-\log\left(\frac{\Delta^2}{m_{\gamma}^2}\right)+\mathcal{O}(m^2).
\label{sbreminta1a2}
\end{equation}
To evaluate $\mathcal{A}_3$ we first need to evaluate the following integral
\begin{align}
\int \frac{d\Omega_k}{4\pi}\,\frac{2 p\cdot p'}{(p\cdot k)(p'\cdot k)}
& = \int_0^1 dx \int \frac{d\Omega_k}{4\pi}\,\frac{2 p\cdot p'}{\left[x(p\cdot k)+(1-x)(p'\cdot k)\right]^2}\nonumber\\
& = \int_0^1 dx \int \frac{d\Omega_k}{4\pi}\,\frac{2 p\cdot p'}{\left[E\,\omega_k-\vec{k}\cdot(x\vec{p}+(1-x)\vec{p}\,')\right]^2}.
\label{chiint}
\end{align} 
Evaluating the phase space integral and using the relations $q^2\approx 2m^2 - 2p\cdot p'$  and $\vec{p}^{\,2}= \vec{p}\,'^{\,2} \approx E^2-m^2$ to simplify the numerator and the denominator of \Cref{chiint}, the integral in \Cref{chiint} becomes
\begin{align}
\int_0^1 dx\,\frac{2p\cdot p'}{E^2\,\omega_k^2- k^2\,\left[x\vec{p}+(1-x)\vec{p}\,'\right]^2} & = \int_0^1 dx\,\frac{2m^2-q^2}{E^2\,m_{\gamma}^2+ k^2\left[m^2-x(1-x)q^2\right]}.
\label{chiintf}
\end{align}
The integral $\mathcal{A}_3$ is now
\begin{align}
\mathcal{A}_3 & = \frac{\alpha}{\pi}\int_0^1 dx\int_0^{\Delta} dk \frac{k^2}{\sqrt{k^2+m_{\gamma}^2}}\,\frac{2m^2-q^2}{E^2\, m_{\gamma}^2+ k^2 \left[m^2-x(1-x)q^2\right]}.
\label{sbreminta3}
\end{align} 

One can say that it is safe to set $m_{\gamma}\rightarrow 0$ in the denominator of \Cref{sbreminta3} as it does not diverge in this limit, but we will lose an important sub-leading collinear divergent term in the form of $\log m$. So it is very convenient to keep everything in that integral which makes it harder to calculate. We use the following trick to calculate this integral by first rewriting \Cref{sbreminta3} as
\begin{align}
\mathcal{A}_3 & = \frac{\alpha}{\pi}\, B\int_0^1 dx\int_0^{\Delta} dk \frac{k^2}{\sqrt{k^2+m_{\gamma}^2} \big( A k^2 +m_{\gamma}^2\big)},
\label{sbreminta1m}
\end{align}
where
\begin{equation}
\begin{aligned}
A &= \frac{m^2-x(1-x)q^2}{E^2},\\
B &= \frac{2m^2-q^2}{E^2}.
\end{aligned}
\end{equation}
Then we evaluate the integral over $k$, where we have
\begin{align}
\int_0^{\Delta} dk \frac{k^2}{\sqrt{k^2+m_{\gamma}^2} \big( A k^2 +m_{\gamma}^2\big)}& = \frac{1}{A}\int_0^{\Delta} dk \, \frac{1}{\sqrt{k^2+m_{\gamma}^2}}-\frac{m_{\gamma}^2}{A}\int_0^{\Delta}\frac{dk}{\big(A\,k^2+m_{\gamma}^2\big)\sqrt{k^2+m_{\gamma}^2}}\nonumber\\
& = \frac{1}{A}\int_0^{\Delta} dk \, \frac{1}{\sqrt{k^2+m_{\gamma}^2}}-\frac{1}{A}\int_0^{\infty}\frac{dz}{\big(A\,z^2+m_{\gamma}^2\big)\sqrt{z^2+m_{\gamma}^2}}\nonumber\\
 &=\frac{1}{A} \log\left(\frac{2 \Delta}{m_{\gamma}}\right)-\frac{1}{A\sqrt{1-A}}\log\left(\frac{1+\sqrt{1-A}}{\sqrt{A}}\right),
\label{etaint}
\end{align}
where we changed the variable $z = \frac{k}{m_{\gamma}}$ then the upper limit of the integral over $z$ will be $\frac{\Delta}{m_{\gamma}} \rightarrow \infty$ as $m_{\gamma}\rightarrow 0$. Then $\mathcal{A}_3$ becomes
\begin{align}
\mathcal{A}_3 &= \int_0^1 dx \left[\frac{B}{A} \log\left(\frac{2 \Delta}{m_{\gamma}}\right)-\frac{B}{A\sqrt{1-A}}\log\left(\frac{1+\sqrt{1-A}}{\sqrt{A}}\right)\right].
\label{sbreminta3m}
\end{align}
In the high energy limit $-q^2 \gg m^2$ the variable $A$ becomes much less than 1 and $\mathcal{A}_3$ becomes
\begin{align}
\mathcal{A}_3
& = \frac{1}{2}\int_0^1 dx \,\frac{B}{A} \log\left(\frac{A\, \Delta^2}{m_{\gamma}^2}\right)
\end{align}
Substituting the values of $A$ and $B$ we get
\begin{align}
\mathcal{A}_3 &= \frac{1}{2}\int_0^1 dx \,\frac{2m^2-q^2}{m^2-x(1-x)q^2} \log\left(\frac{\big(m^2-x(1-x)q^2\big)\, \Delta^2}{E^2\,m_{\gamma}^2}\right)\nonumber\\
& = \log\left(\frac{-q^2}{m^2}\right)\log\left(\frac{\Delta^2}{m_{\gamma}^2}\right)-\log\left(\frac{-q^2}{m^2}\right)\log\left(\frac{E^2}{m^2}\right)+\frac{1}{2}\log^2\left(\frac{-q^2}{m^2}\right)-\frac{\pi^2}{6}+\mathcal{O}(m^2,m_{\gamma}^2).
\label{sbreminta3f}
\end{align}
From \Cref{sbreminta1a2,sbreminta3f}, the differential cross section for the emission of soft photons will be
\begin{multline}
\left(\frac{d\sigma}{d\Omega}\right)_{\texttt{S}}^f = \left(\frac{d\sigma}{d\Omega}\right)_{\texttt{0}} \,\frac{\alpha}{\pi}\left[\log\left(\frac{-q^2}{m^2}\right)\log\left(\frac{\Delta^2}{m_{\gamma}^2}\right)-\log\left(\frac{-q^2}{m^2}\right)\log\left(\frac{E^2}{m^2}\right)+\frac{1}{2}\log^2\left(\frac{-q^2}{m^2}\right)\right.\\
\left.+\log\left(\frac{E^2}{m^2}\right)-\log\left(\frac{\Delta^2}{m_{\gamma}^2}\right) -\frac{\pi^2}{6}+\mathcal{O}(m^2,m_{\gamma}^2)\right].
\label{sbremxsecf}
\end{multline} 

\Cref{sbremxsecf} contains both soft and collinear IR divergences just like what we found from the contribution of the vertex correction. As we discussed above, according to the BN theorem, the contribution from the vertex and self energy corrections from \Cref{vlxsection} plus the emission of a soft photon from \Cref{sbremxsecf} should be free of IR singularities. The BN differential cross section summing the vertex correction and the soft photon emission, excluding the IR finite vacuum polarization \Cref{vpxsection} and the box corrections \Cref{boxxsection} for now, is given by
\begin{multline}
\left(\frac{d\sigma}{d\Omega}\right)_{\texttt{BN}} = \left(\frac{d\sigma}{d\Omega}\right)_{\texttt{0}}\left\{1+\frac{\alpha}{\pi}\left[\log\left(\frac{E^2}{\Delta^2}\right)\left(1-\log\left(\frac{-q^2}{m^2}\right)\right)+\frac{3}{2}\log\left(\frac{-q^2}{m^2}\right)-2 \right]\right\} 
\label{BN}.
\end{multline}
We see that all the soft IR divergences have indeed been canceled. However, collinear terms in the form of $\log m^2$ remain. These collinear logs diverge in the limits of either $m \rightarrow 0$ or $-q^2 \rightarrow\infty$. Kinoshita, Lee, and Nauenberg generalizes the BN theorem in order to cancel these collinear singularities.
\section{The Kinoshita-Lee-Nauenberg Theorem}
It was pointed out by T.~Kinoshita \cite{Kinoshita:1958ru} that taking the limit of the electron mass to zero produces additional divergences now known as collinear divergences. Kinoshita realized that these additional divergences are similar to the soft IR divergences since both types of divergences are associated with the vanishing of the mass of a particle. Kinoshita later investigated the cancellation of these mass singularities using the detailed properties of Feynman diagrams \cite{Kinoshita:1962ur}. 

Lee and Nauenberg \cite{Lee:1964is} proved a quantum mechanical theorem whereby all IR divergences, including collinear singularities, cancel, which became known as the Kinoshita-Lee-Nauenberg (KLN) theorem. The KLN theorem states that any physical observable for which all indistinguishable initial and final degenerate states are summed over is free of any IR divergences. According to the KLN theorem, collinear singularities are associated with additional degeneracies. These degeneracies are due to the emission of a photon collinearly with the electron as a final state or absorption of a photon collinearly in the initial state. As a starting point, one should then sum over the full final state degeneracies.
\subsection{Hard Collinear Final State Degeneracies}
The emission of a hard photon collinearly (i.e.\ the angle $\theta_{\gamma}$ is less than some angular resolution $\delta$) with the incoming or the outgoing electrons can also be indistinguishable, the word hard here used for the photons with energies $\Delta<k<E$. Such contribution from these diagrams is very important for the cancellation of the remaining collinear divergences. From \Cref{bremampm} we can write the amplitude squared of the final state bremsstrahlung as follows
\begin{align}
\left|\mathcal{M}_{\texttt{B}}^f\right|^2 & = \frac{e^6}{q^4}\,\bar{u}^{s'}(p')\, S^{0\beta}\, u^s(p)\, \bar{u}^s(p) \, S^{\alpha 0}u^{s'}(p')\,\varepsilon_{\alpha}^{r*}\,\varepsilon_{\beta}^r.
\label{bremampsquared}
\end{align}

Now we sum over all spins and polarizations using the identities $\sum_r\varepsilon_{\alpha}^{r*}\varepsilon_{\beta}^r = -g_{\alpha \beta}+\frac{k_{\alpha}k_{\beta}}{k^2}$ and $\sum_s u^s(p)\bar{u}^s(p) = \slashed{p}+m$ keeping in mind that the emission of a hard photon does not contain any soft IR divergences and it is finite as we send $m_{\gamma} \rightarrow 0$. This ensures that the second term of the first identity gives no contribution to the cross section so we can neglect it. Then we have
\begin{align}
\sum_{s',s}\sum_r\left|\mathcal{M}_{\texttt{B}}^f\right|^2  &= \frac{-e^6}{q^4}\, tr\left[(\slashed{p}'+m)\, S^{0\beta}\, (\slashed{p}+m) S_{\beta}^{\ 0}\right].
\label{hardbrem2}
\end{align}
Substituting \Cref{salphabeta} into \Cref{hardbrem2}, we get
\begin{align}\sum_{s',s}\sum_r\left|\mathcal{M}_{\texttt{B}}^f\right|^2
&= \frac{-e^6}{4q^4}\, \left(\frac{1}{(p\cdot k)^2}\, \text{tr}\left[(\slashed{p}'-\slashed{k}+m)\gamma^{0}(\slashed{p}-\slashed{k}+m)\gamma^{\beta}(\slashed{p}+m)\gamma_{\beta}(\slashed{p}-\slashed{k}+m)\gamma^{0}\right]\right.\nonumber\\
&\quad+\frac{1}{(p'\cdot k)^2}\,\text{tr}\left[(\slashed{p}'-\slashed{k}+m) \gamma^{\beta}(\slashed{p}'+m)\gamma^{0}(\slashed{p}+m)\gamma^{0}(\slashed{p}'+m)\gamma_{\beta}\right]\nonumber\\
& \quad-\frac{1}{(p\cdot k)(p'\cdot k)}\,\text{tr}\left[(\slashed{p}'-\slashed{k}+m)\gamma^0
(\slashed{p}-\slashed{k}+m)\gamma^{\beta}(\slashed{p}+m)\gamma^{0}(\slashed{p}'+m)\gamma_{\beta}\right]\nonumber\\
&\left. \quad- \frac{1}{(p\cdot k )(p'\cdot k)}\,\text{tr}\left[(\slashed{p}'-\slashed{k}+m)\gamma^{\beta}(\slashed{p}'+m)\gamma^{0}(\slashed{p}+m)\gamma_{\beta}(\slashed{p}-\slashed{k}+m)\gamma^{0}\right]\right)\nonumber\\
&= \frac{-e^6}{4q^4}\left(\frac{t_1}{(p\cdot k)^2}+\frac{t_2}{(p'\cdot k)^2} -\frac{t_3+t_4}{(p\cdot k)(p'\cdot k)}\right).
\label{bremampsquaredf}
\end{align}

Let us calculate the traces of \Cref{bremampsquaredf} and simplify them by neglecting the terms that are expected to give no contribution in the limits of $m_{\gamma}\rightarrow 0$ and $m\rightarrow 0$. We can also ignore the terms that become in $\mathcal{O}(\Delta,\delta)$ and do not contribute with any large logarithms. The traces $t_1$ and $t_2$ become
\begin{align}
t_1 &\approx 16 \left[m^2(2E^2-p\cdot p')-p\cdot k(2E\omega_k-p' \cdot k) + 2\omega_k^2(p\cdot k)\right],\label{t1}\\
t_2 &\approx 16 \left[m^2(2E^2-p\cdot p')-p'\cdot k(2E\omega_k-p \cdot k)\right],
\label{t2}
\end{align}
while $t_3$ and $t_4$ are similar and given by
\begin{align}
t_3=t_4 &\approx 16 \left[p\cdot p' (2E^2-p\cdot p')-p'\cdot k(2E^2- p\cdot p')- p\cdot p'(2E\omega_k-p\cdot k)\right.\nonumber\\
&\left.\hspace{6cm}+\omega_k^2 (p\cdot p') + E\omega_k (p'\cdot k-p\cdot k)\right].
\label{t3t4}
\end{align}

In order to avoid the double counting from the soft emission contribution, the integral limits over $k$ becomes $\Delta<k<E$ and the integral over $\theta_{\gamma}$ should include only small angles (i.e.\ $0<\theta_{\gamma}<\delta$), which implies $1-\frac{\delta^2}{2}<\cos \theta_{\gamma}<1$. The reason that we are including only hard photons with small angles is that we only include indistinguishable processes for the IR cancellation to happen. We note that the emission of a hard photon collinearly from the incoming electron can be easily distinguished. Thus we only consider the amplitude squared from the emission of a hard photon collinearly with the outgoing electron and the interference between the two diagrams. One can also see that the emission of a hard photon from the incoming electron collinearly with the outgoing electron contributes with terms in $\mathcal{O}(\delta^2)$ which makes them negligible. For the same reason, we can neglect the second term and the term $E \omega_k(p'\cdot k)$ of \Cref{t3t4}. Furthermore, we replace $\omega_k$ with $k$ since there are no soft IR divergences from the emission of a hard photon. Finally, we write the amplitude squared of the final state hard photon emission to be 
\begin{multline}
\sum_{s',s}\sum_r\left|\mathcal{M}_{\texttt{H}}^f\right|^2  =
\frac{-4e^6}{q^4}\left(\frac{m^2(2E^2-p\cdot p')-p'\cdot k(2E\ k-p\cdot k)}{(p'\cdot k)^2}\right. \\
\left.-\frac{2p\cdot p' (2E^2-p\cdot p') - 2p \cdot p'(2E\ k -p\cdot k)+2k^2 (p\cdot p')- 2 E\ k(p\cdot k)}{(p\cdot k)(p'\cdot k)}\right).
\label{hbrem}
\end{multline}
Then the contribution to the differential cross section from the final state for emission of hard and collinear is given by
\begin{align}
\left(\frac{d\sigma}{d\Omega}\right)_{\texttt{H}}^f &= \frac{2\alpha^2}{q^4}\left(\frac{\alpha}{2\pi}\right)\int k \ dk\ d\cos \theta_{\gamma} \ \frac{d\phi}{2\pi}\left[\frac{-m^2(2E^2-p\cdot p')+p'\cdot k(2E\ k-p\cdot k)}{(p'\cdot k)^2}\right. \nonumber\\
&\left.\quad+\frac{2p\cdot p' (2E^2-p\cdot p') - 2p \cdot p'(2E\ k -p\cdot k)+2k^2 (p\cdot p')- 2 E\ k(p\cdot k)}{(p\cdot k)(p'\cdot k)}\right].
\label{hbremxsec}  
\end{align}

Before we start evaluating the integrals in \Cref{hbremxsec}, we should remember that $(p'-k)$ is on shell which implies $(2E^2-p\cdot p') = 2E^2+\frac{q^2}{2}-(m^2+p'\cdot k)$. We also note that \Cref{modmom} can be expanded around $\theta_{\gamma}$ in the collinear limit ($0<\theta_{\gamma}<\delta$, with $\delta \ll 1$) to become $p'\approx E-\frac{m^2}{2(E-k)}$. With these informations in hand we split the integral in \Cref{hbremxsec} into three integrals, where the first integral is given by
\begin{align}
\mathcal{B}_1 &= \frac{-2\alpha^2}{q^4}\left(\frac{\alpha}{2\pi}\right)\int k \ dk\ d\cos \theta_{\gamma}\frac{d\phi}{2\pi}\ \frac{m^2(2E^2-p\cdot p')}{(p'\cdot k)^2}\nonumber\\
& = \frac{-2\alpha^2}{q^4}\left(\frac{\alpha}{2\pi}\right)\int_{\Delta}^E k \ dk \int_{1-\frac{\delta^2}{2}}^1 d\cos \theta_{\gamma}  \ \frac{m^2\left[2E^2+\frac{q^2}{2}-(m^2+p'\cdot k)\right]}{(p'\cdot k)^2}\nonumber\\
&= \left(\frac{d\sigma}{d\Omega}\right)_{\texttt{0}}\cdot \frac{\alpha}{2\pi} \int_{\Delta}^E \frac{-1}{E^2\ k}dk\int_{1-\frac{\delta^2}{2}}^1 d\cos\theta_{\gamma} \frac{m^2}{\left[1-\frac{p'}{E}\cos\theta_{\gamma}\right]^2}+\mathcal{O}(m^2) .
\label{f1int}
\end{align}
We now use the expansion of $|p'|$ around small angles to evaluate the integral over $\theta_{\gamma}$, which becomes
\begin{align}
\int_{1-\frac{\delta^2}{2}}^1 d\cos\theta_{\gamma} \frac{m^2}{\left[1-\left(1-\frac{m^2}{2E(E-k)}\right)\cos\theta_{\gamma}\right]^2}
& = \frac{4\delta^2E^2(E-k)^2}{2\delta^2E(E-k)-(\delta^2-2)m^2}\nonumber\\&=2E(E-k)+\mathcal{O}(m^2,\delta^2).
\label{f1partint}
\end{align}
The $\mathcal{B}_1$ integral evaluation is now straightforward to finally give
\begin{equation}
\mathcal{B}_1 = \left(\frac{d\sigma}{d\Omega}\right)_{\texttt{0}}\cdot \frac{\alpha}{2\pi}\left[\log\left(\frac{\Delta^2}{E^2}\right)-\frac{2\Delta}{E}+2+\mathcal{O}(m^2,\delta^2)\right].
\label{f1intf}
\end{equation}
The second integral of \Cref{hbremxsec} is given by
\begin{multline}
\mathcal{B}_2 = \frac{2\alpha^2}{q^4}\left(\frac{\alpha}{2\pi}\right)\int_{\Delta}^E k \ dk \int_{1-\frac{\delta^2}{2}}^1 d\cos \theta_{\gamma} \left[\frac{(2E\ k-p\cdot k)}{(p'\cdot k)}\right. \\
\left.+\frac{2p\cdot p' (2E^2-p\cdot p') - 2p \cdot p'(2E\ k -p\cdot k)}{(p\cdot k)(p'\cdot k)}\right].
\label{f2int}
\end{multline}

In the limit of $m,\theta_{\gamma}\rightarrow 0$, we can choose $k$ to be a portion of $p'$ such that $k\rightarrow \frac{k}{E}p'$. It is trivial to check that the remaining terms if we use this approximation are in order of $m^2$. This approximation allows us to write $(2E\ k - p\cdot k) = \frac{k}{E}(2E^2-p\cdot p')$ while $p\cdot k$ becomes $\frac{k}{E}(p\cdot p')$. Then $\mathcal{B}_2$ can be written as
\begin{align}
\mathcal{B}_2 = \left(\frac{d\sigma}{d\Omega}\right)_{\texttt{0}}\frac{\alpha}{2\pi}\int_{\Delta}^E k \ dk \int_{1-\frac{\delta^2}{2}}^1 d\cos \theta_{\gamma} \frac{1}{(p'\cdot k)}\left[\frac{k}{E}+2\left(\frac{E}{k}-1\right)\right]+\mathcal{O}(m^2).
\label{f2intm}
\end{align}
We recall the expansion of $|p'|$ around $\theta_{\gamma}$ to evaluate the integral over the angle, where we have
\begin{align}
\int_{1-\frac{\delta^2}{2}}^1 d\cos \theta_{\gamma} \frac{1}{(p'\cdot k)}& = \int_{1-\frac{\delta^2}{2}}^1 d\cos \theta_{\gamma} \frac{1}{\left[1-\left(1-\frac{m^2}{2E(E-k)}\right)\cos\theta_{\gamma}\right]}\nonumber\\
& = \log \left(\frac{\delta^2E(E-k)}{m^2}\right) +\mathcal{O}(m^2,\delta^2).
\label{f2partint}
\end{align}
The $\mathcal{B}_2$ integral becomes
\begin{align}
\mathcal{B}_2 & = \left(\frac{d\sigma}{d\Omega}\right)_{\texttt{0}}\frac{\alpha}{2\pi}\int_{\Delta}^E dk \left[\frac{k}{E^2}+\frac{2}{k}-\frac{2}{E}\right] \log \left(\frac{\delta^2E(E-k)}{m^2}\right)\nonumber\\
& = \left(\frac{d\sigma}{d\Omega}\right)_{\texttt{0}}\frac{\alpha}{2\pi}\left\{\log\left(\frac{\delta^2E^2}{m^2}\right)\left[\log\left(\frac{E^2}{\Delta^2}\right)-\frac{\Delta^2}{2E^2}+\frac{2\Delta}{E}-\frac{3}{2}\right]\right.\nonumber\\
&\left.\hspace{5cm}-\frac{\Delta^2}{2E^2}+\frac{2\Delta}{E}-\frac{\pi^2}{3}+\frac{5}{4}+\mathcal{O}(m^2,\delta^2)\right\}.\label{f3intf}
\end{align}

One can easily see that, using the collinear approximation used in $\mathcal{B}_2$, the remaining terms of \Cref{hbremxsec} vanishes where $2E\ k (p\cdot k) \rightarrow 2k^2 (p\cdot p')$. Then the contribution to the differential cross section from the final state emission of a collinear hard photon is given by
\begin{multline}
\left(\frac{d\sigma}{d\Omega}\right)_{\texttt{H}}^f = 
\left(\frac{d\sigma}{d\Omega}\right)_{\texttt{0}} \frac{\alpha}{2\pi}\left\{\log\left(\frac{\delta^2E^2}{m^2}\right)\left[\log\left(\frac{E^2}{\Delta^2}\right)-\frac{\Delta^2}{2E^2}+\frac{2\Delta}{E}-\frac{3}{2}\right]+\log\left(\frac{\Delta^2}{E^2}\right)\right.\\
\left.-\frac{\Delta^2}{2E^2}-\frac{\pi^2}{3}+\frac{13}{4}+\mathcal{O}(m^2,\delta^2)\right\}.
\label{hbremxsecf}
\end{multline} 

We note that \Cref{hbremxsecf} is not sufficient for the remaining collinear divergent terms in \Cref{BN} to cancel. LN included the initial state hard collinear degeneracies. In the next section, we give an overview of the remainder of the LN treatment.
\subsection{Hard Collinear Initial State Degeneracies}
\begin{figure}[h]
 \centering
\begin{subfigure}[t]{0.4\textwidth}
\begin{center}
\includegraphics[scale=1]{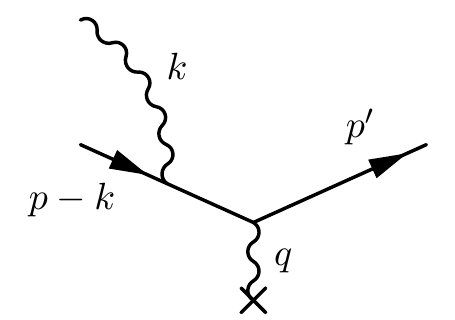}
\caption{}
\label{fig:ibrem1}
\end{center}
\end{subfigure}
\begin{subfigure}[t]{0.4\textwidth}
\begin{center}
\includegraphics[scale=1]{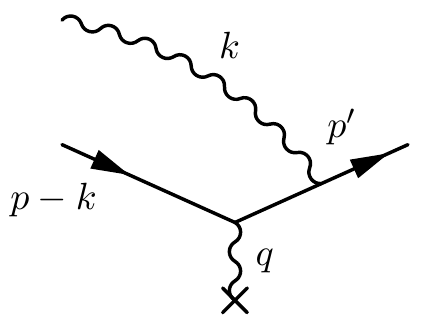}
\caption{}
\label{fig:ibrem2}
\end{center}
\end{subfigure}
\caption{Absorption of a single photon.}
\label{fig:ibrem}
\end{figure}

The application of the KLN theorem requires now including the absorption of a hard photon collinearly by the incoming and the outgoing electrons. We find that there is no difference between these calculations and the calculations made for the final states where we have $\left|\mathcal{M}_{\texttt{H}}^i\right|^2 = \left|\mathcal{M}_{\texttt{H}}^f\right|^2$. The only difference would be from the fact that the absorption of a hard photon collinearly by the outgoing electron can be easily distinguished from the incoming electron. Therefore we only include the amplitude squared of the diagram in \Cref{fig:ibrem1} and the interference between the two diagrams. Then the contribution from both the initial and final state of a collinear hard photon will be
\begin{multline}
\left(\frac{d\sigma}{d\Omega}\right)_{\texttt{H}} = 
\left(\frac{d\sigma}{d\Omega}\right)_{\texttt{0}} \frac{\alpha}{\pi}\left\{\log\left(\frac{\delta^2E^2}{m^2}\right)\left[\log\left(\frac{E^2}{\Delta^2}\right)-\frac{\Delta^2}{2E^2}+\frac{2\Delta}{E}-\frac{3}{2}\right]+\log\left(\frac{\Delta^2}{E^2}\right)\right.\\
\left.-\frac{\Delta^2}{2E^2}-\frac{\pi^2}{3}+\frac{13}{4}+\mathcal{O}(m^2,\delta^2)\right\}.
\label{totalhbrem}
\end{multline} 

LN \cite{Lee:1964is} did not write down the terms $\mathcal{O}\left(\left(\frac{\Delta}{E}\right)^{\alpha}\log\left(\frac{\delta^2E^2}{m^2}\right)\right)$, $\alpha = 1,2$, relying on the fact that $\Delta$ is very small. We note however that $\log\left(\frac{\delta^2E^2}{m^2}\right)$ can be a large logarithm. The authors of \cite{Lavelle:2005bt} showed that if one goes beyond the eikonal approximation in the soft emission calculations, then the new terms from non-eikonality cancel exactly the $\mathcal{O}\left(\left(\frac{\Delta}{E}\right)^{\alpha}\log\left(\frac{\delta^2E^2}{m^2}\right)\right)$ terms. We give explicitly the calculations for the differential cross section of the emission of a soft photon beyond eikonal approximation in \Cref{AppendixD}. \Cref{softbrem-be} shows that the $\mathcal{O}\left(\left(\frac{\Delta}{E}\right)^{\alpha}\log\left(\frac{\delta^2E^2}{m^2}\right)\right)$ terms from the hard and collinear emission and absorption cancel with the terms gained by removing the eikonal approximation leaving no finite pieces to affect the formula of the differential cross section. Adding the collinear absorption and emission contributions from \Cref{totalhbrem} to the BN treatment, given by \Cref{BN}, yields 
\begin{multline}
\left(\frac{d\sigma}{d\Omega}\right)_{\texttt{KLN}} = \left(\frac{d\sigma}{d\Omega}\right)_{\texttt{0}}\left\{1+\frac{\alpha}{\pi}\left[\log\left(\frac{-q^2}{\delta^2 E^2}\right)\left(\log\left(\frac{\Delta^2}{E^2}\right)+\frac{3}{2}\right)-\frac{\pi^2}{3}+\frac{5}{4}+\mathcal{O}(m^2,\delta^2) \right]\right\} 
\label{KLN},
\end{multline} 
which is indeed free of collinear singularities. Note that \Cref{KLN} represents the final result of the original LN paper including for the first time in the extant literature the $\mathcal{O}(1)$ terms; LN neglected these terms in their original treatment \cite{Lee:1964is}.

It is important to emphasize that the KLN theorem is a basic theorem of quantum mechanics. LN did not describe in general how to implement it order by order in QFT. However, LN gave an explicit example for Rutherford scattering, in which they included only hard degenerate initial and final states as well as soft degenerate final states. We can not emphasize enough that \textbf{LN did not include the contribution from soft degenerate initial states}. LN mistakenly relied on the fact that the IR soft divergences were canceled with the application of the BN theorem; i.e.\ LN only summed over degenerate soft final state photon emission. Here we quote from their original paper \emph{``In (20) the infrared divergence has already been eliminated by including the contributions due to emissions of soft photons.''} (20) in \cite{Lee:1964is} is equivalent to \Cref{BN}. However, LN discussed the degenerate soft-photon initial states, anticipating that these contributions are IR safe once the whole power series is taken into account.    

However, despite the KLN theorem requiring a sum over degenerate initial and final states, \Cref{KLN} neglects to sum over initial states with degenerate soft photons. \Cref{KLN} is thus not a result of a correct application of the KLN theorem; additionally, by treating the initial and final states differently, we have broken time reversal symmetry.

A naive attempt to treat the initial and final states symmetrically would be to add a single soft photon absorption, the time reversed process, of the single final state soft emission we have included. One can easily show that the contribution to the cross section from a single soft photon absorption is identical to the contribution from a single soft emission $\left(\frac{d\sigma}{d\Omega}\right)_{\texttt{S}}^{i}=\left(\frac{d\sigma}{d\Omega}\right)_{\texttt{S}}^{f}$ within the same energy resolution $\Delta$. Recall that $\left(\frac{d\sigma}{d\Omega}\right)_{\texttt{S}}^{f}$ contains both soft and collinear divergences. Therefore simply adding this naive contribution to the KLN cross section reintroduces uncancelled soft and collinear divergences. However, we have neglected additional sources of a soft degenerate initial state. 
 
\chapter{A Self-Consistent Implementation of the KLN Theorem} 

\label{Chapter5} 

The KLN theorem connects IR divergences to degeneracies. KLN ensures that summing all initial and final state degeneracies produces observables that are IR finite. We quote from Sterman \cite{Sterman:1994ce} on the KLN theorem: \emph{``For applications to high-energy scattering, its importance has thus far been more conceptual than practical, but it is a fundamental theorem of quantum mechanics and puts many specific results in perspective.''} Even though KLN is a basic theorem of quantum mechanics, it is not obvious how to implement it in QFT order by order in perturbation. Our goal is to find a proper understanding of how to implement the KLN theorem in QFT.

\begin{figure}[t]
 \centering
\includegraphics[scale=1]{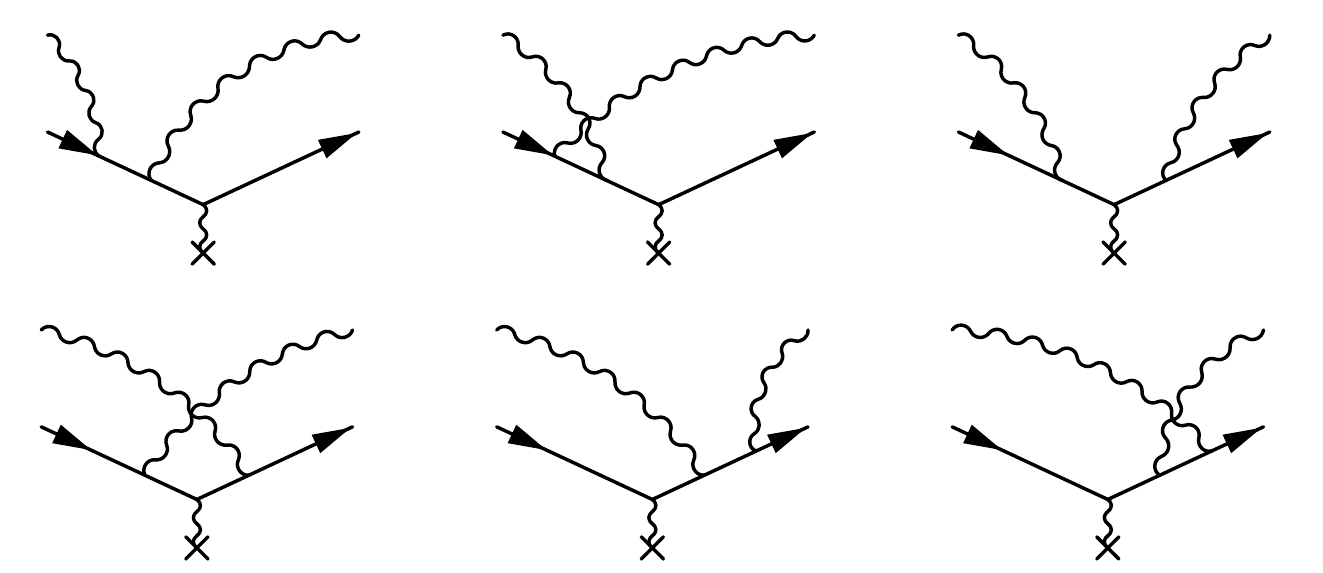}
\caption{Diagrams with both absorption and emission of a soft photon}
\label{fig:abs-em}
 \end{figure}

The non-cancelling IR divergences we found from the naive addition of $\left(\frac{d\sigma}{d\Omega}\right)_{\texttt{S}}^i$ implies that there are degenerate states we still need to sum over. The first type of diagrams that come in mind are the processes where we have both emission and absorption shown in \Cref{fig:abs-em}. Although the amplitude squared of these diagrams are higher order in perturbation expansion, the amplitudes from these diagrams can interfere with the amplitudes from the disconnected diagram shown in \Cref{fig:tdiscon} to produce a contribution in the same order.

M. Lavelle and D. McMullan \cite{Lavelle:2005bt} showed that including contributions from the disconnected amplitudes raises another issue, where one can add an arbitrary number of disconnected photons and the amplitude remains degenerate with the original state. Adding many disconnected photons forms a series that does not converge. In the next section, we give an overview of how previous work used the disconnected diagrams in the cancellation of IR divergences. We find that all of these previous attempts were incomplete.
\section{The Role of Disconnected Diagrams}
\begin{wrapfigure}{r}{0.35\textwidth}
  \begin{center}
\includegraphics[scale=1]{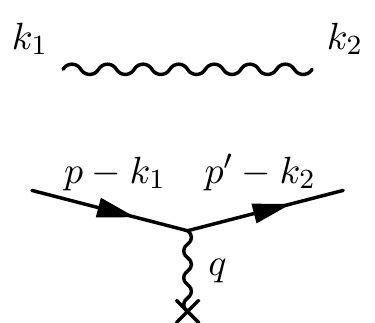}
  \caption{The disconnected diagram at tree level.}
\label{fig:tdiscon}  
  \end{center}
\end{wrapfigure}
Peskin and Schroeder \cite{Peskin:1995ev} claim that only fully connected amplitudes contribute to the cross section; Peskin and Schroeder claim that partially and fully disconnected amplitudes from part of the trivial $1$ in the $S$-matrix. However, one finds that when the electron line is connected to the source $p'\neq p$. Thus these diagrams do not contribute to the trivial $1$ in the $S$-matrix, in which $p=p'$. One also finds that these disconnected diagrams may interfere with other diagrams and form either a fully connected cut diagram in such a way that no disconnected part will be left out or partially disconnected cut diagrams. The contribution from these cut diagrams might provide the cancellation of the soft IR divergences from the initial states we seek. 
 
 We first include the contributions from the interference between the disconnected diagram shown in \Cref{fig:tdiscon} and absorption-emission diagrams in \Cref{fig:abs-em}. The Feynman rule for the disconnected photon is $(2\pi)^3\, 2\omega_{k_1}\delta^{(3)}(\vec{k}_1-\vec{k}_2) \,\delta_{\varepsilon_1 \varepsilon_2}$ where $\omega_{k_1}$ is the energy of the photon of momentum $k_1$,  $\varepsilon_1$ and $\varepsilon_2$ are the polarization vectors for the incoming photon of momentum $k_1$ and the outgoing photon of momentum $k_2$ respectively. 

\begin{figure}[h]
\centering
\begin{subfigure}[t]{0.23\textwidth}
\includegraphics[scale=1]{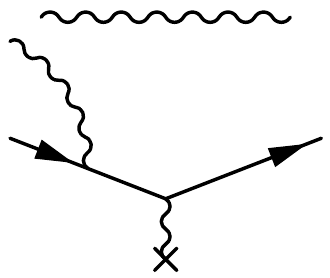}
\caption{}
\label{fig:abs-dis1}
\end{subfigure}
\begin{subfigure}[t]{0.23\textwidth}
\includegraphics[scale=1]{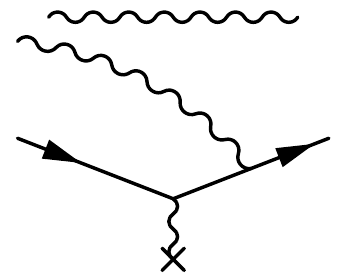}
\caption{}
\label{fig:abs-dis2}
\end{subfigure}
\begin{subfigure}[t]{0.23\textwidth}
\includegraphics[scale=1]{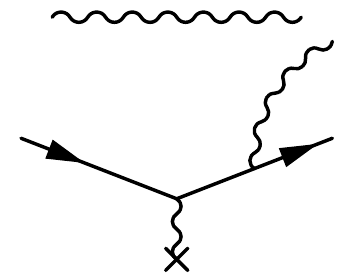}
\caption{}
\label{fig:em-dis1}
\end{subfigure}
\begin{subfigure}[t]{0.23\textwidth}
\includegraphics[scale = 1]{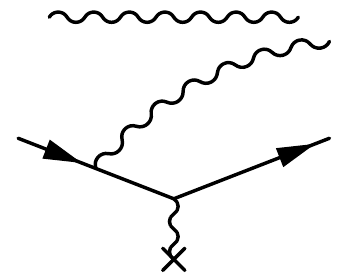}
\caption{}
\label{fig:em-dis2}
\end{subfigure}
\caption{The absorption and emission with a disconnected photon.}
\label{fig:if-disc}
\end{figure}

The amplitude due to the sum of the diagrams in \Cref{fig:abs-em}, taking into account that $k_1=k_2$ due to the delta function from the disconnected amplitude, is given by (See \autoref{AppendixC}) 
\begin{align}
i\mathcal{M}_{\texttt{AE}} = -i\mathcal{M}_{\texttt{0}} \,e^2 \left[\frac{p\cdot \varepsilon_1}{p\cdot k_1}-\frac{p^{\prime}\cdot \varepsilon_2}{p^{\prime}\cdot k_2}\right]^2,
\label{abs-em-amp}
\end{align}
then the contribution from the absorption-emission diagrams as derived in \Cref{abs-em-xsec-sep} is given by 
\begin{align}\left(\frac{d\sigma}{d\Omega}\right)_{\texttt{AE}} = - \left(\frac{d\sigma}{d\Omega}\right)_{\texttt{S}}^f- \left(\frac{d\sigma}{d\Omega}\right)_{\texttt{S}}^i.
\label{abs-em-xsec}
\end{align}
Adding \Cref{abs-em-xsec} to \Cref{KLN} plus $\left(\frac{d\sigma}{d\Omega}\right)_{\texttt{S}}^i$ cancels the BN  $\left(\frac{d\sigma}{d\Omega}\right)_{\texttt{S}}^f$, and we still have an IR unsafe result. Other degenerate states that may contribute to the cross section are the processes where we have an emission or absorption of a soft photon with a disconnected soft photon flying around as shown in \Cref{fig:if-disc}.

\begin{figure}[h]
\centering
\begin{subfigure}[t]{0.4\textwidth}
\begin{center}
\includegraphics{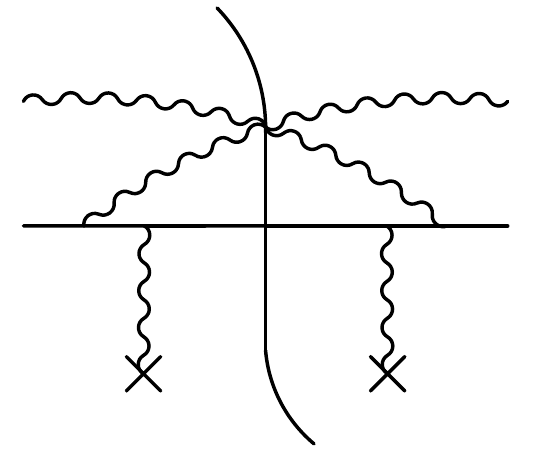}
\caption{}
\label{fig:fullcon}
\end{center}
\end{subfigure}
\begin{subfigure}[t]{0.4\textwidth}
\begin{center}
\includegraphics{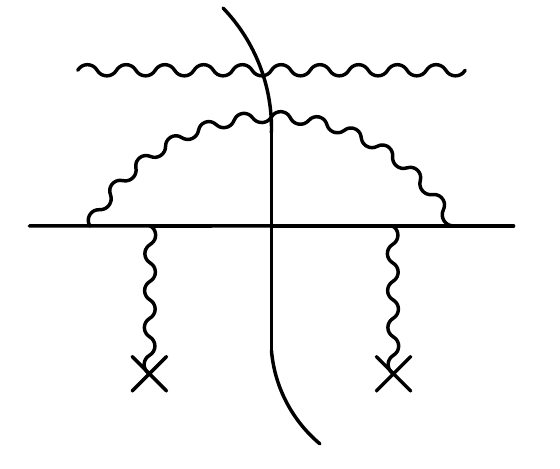}
\caption{}
\label{fig:partialcon}
\end{center}
\end{subfigure}
\caption{\protect\subref{fig:fullcon} represents a fully connected cut diagram while \protect\subref{fig:partialcon} is the partially connected cut diagram, both are produced from the amplitude squared of the last diagram in \Cref{fig:em-dis2}.}
\label{fig:cutdiagrams}
\end{figure}

Nonetheless, there are two possible contributions from the diagrams in \Cref{fig:if-disc}. The first can be obtained when $k_1 = k_2=k$ (assuming that $k$ is the momentum of the photon that is attached to the electron line and $k_1$ and $k_2$ are the incoming and the outgoing momenta, respectively, of the disconnected photon) and the cut diagram becomes a fully connected diagram as shown in \Cref{fig:fullcon}. The second can be constructed when $k_1 = k_2 \neq k$ and the cut diagram is called a partially connected cut diagram as shown in \Cref{fig:partialcon}. 

It can be easily verified that the contribution from the fully connected diagrams is exactly the same as the emission or absorption of a soft photon without any disconnected photons. We note that the delta function from the disconnected photon in the partially connected cut diagram contribute with $\delta^3(0)$. One can check that by calculating the amplitude squared from the diagram in \Cref{fig:tdiscon} where the contribution is proportional to $\mathcal{M}_{\texttt{0}}$ multiplied by an infinite factor in the form of $\delta^3(0)$, this infinite factor should be eliminated by the $S$-matrix normalization. However, including the fully connected contributions from the diagrams in \Cref{fig:if-disc} is not sufficient for the IR cancellation, it is also not clear how one could include the partially connected contributions. One can even add more than one disconnected photon and the amplitude remains degenerate. With the above as context, let us discuss some of the previous attempts to resolve the KLN crisis.
\subsection*{50 Years of Confusion, Inconsistency, and Incompleteness}
As was pointed out in the previous chapter, LN \cite{Lee:1964is} did not include the soft degenerate initial states. By treating the initial and final states differently, the LN treatment breaks time reversal symmetry. C.~De~Calan and G. Valent \cite{DeCalan:1972ya} discussed the cancellation of the IR divergences from the diagrams with incident photons. They included the disconnected diagrams from \Cref{fig:abs-em} as well as the contribution from \Cref{fig:em-dis1,fig:em-dis2}. However, they ignored the initial state absorption with a disconnected soft photon shown in \Cref{fig:abs-dis1,fig:abs-dis2} leading to the same time reversal symmetry breaking. C.~De~Calan and G. Valent also did not discuss the possibility of adding more than one disconnected soft photon.    

Doria, Frenkel, and Taylor \cite{Doria:1980ak} were the first to discover a non-cancelled soft IR divergences at two loops in non-Abelian gauge theory. The same problem was reported by Di'Lieto, Gendron, Halliday, and Sachrajda \cite{DiLieto:1980nkq}. T.~Muta and C.~A.~Nelson (MN) \cite{Muta:1981pe} highlighted the role of disconnected diagrams in the cancellation of the IR divergences by including the absorption-emission diagrams in \Cref{fig:abs-em} and the contribution from diagrams in \Cref{fig:abs-dis1,fig:abs-dis2}. MN relied on the fact that one can find the degenerate states by cutting the Kinoshita graphs in all possible ways: \emph{``Here we use the single-cut version of the Kinoshita diagram where the cut line refers only to the initial state.''} \cite{Muta:1981pe} Clearly, they did not consider the case when the cut line refers to the final state, which means that they did not consider, for example, contributions from amplitudes in \Cref{fig:em-dis1,fig:em-dis2}. MN might object to the need to include these final state emission amplitudes (with an additional disconnected photon) as MN claimed they were only concerned with cancelling the IR divergences from initial state photons.  The problem with this argument is that MN separates the cancellation of IR divergences when one considers initial or final states radiation.  In particular, we have seen previously, and will see again below, the importance of cut diagrams that include radiation in both the initial and final states.  Thus one can not consider the initial and final states radiation separately.  MN see this themselves, in fact, as some of the contributions to their initial states radiation include cut Kinoshita diagrams with radiation in the initial and final states.  Therefore by not considering the amplitudes in \Cref{fig:em-dis1,fig:em-dis2}, MN do not treat the initial and final states radiation symmetrically or fully; thus they do not fully or correctly implement the KLN theorem. 

A.~Axelrod and C.~A.~Nelson \cite{Axelrod:1985yi} followed the same treatment for a QCD parton like model. Sterman \cite{Sterman:1994ce} treated the initial and final states in a self-consistent way. It is not clear that he considered fully connected cut diagrams where the photon passed through either the initial state or final state cuts more than once. However, it is clear that Sterman's treatment does not include the partially connected cut diagrams like in \Cref{fig:partialcon}.
 
H.~F.~Contopanagos \cite{Contopanagos:1989pb} showed a cancellation of both soft and collinear divergences for the same process (one-loop corrections to electron scattering off an external potential) using different regularization schemes (massive and dimensional regularization). He followed exactly the LN treatment and thus did not include any soft initial state photon contributions. B. Mirza and M. Zarei \cite{Mirza:2006tk} also followed the LN treatment to show the cancellation of the soft and collinear divergences in noncommutative QED; they therefore also did not include soft initial state photon contributions

 M. Lavelle and D. McMullan \cite{Lavelle:2005bt} gave a very good review of the problem of inconsistent treatment of the soft and collinear divergences. In their own work, they included the contribution from diagrams in \Cref{fig:abs-em} and the fully connected contribution from diagrams in \Cref{fig:abs-dis1,fig:abs-dis2}. However, they did not consider the partially connected contributions, where they said \emph{``The disconnected contraction is ignored by Lee-Nauenberg and we will follow their
lead.''} However, they raised the issue of inconsistency by not including the contribution from diagrams in \Cref{fig:em-dis1,fig:em-dis2} as well as ignoring the possibility of having many disconnected photons. They concluded that urgent work is required to find a full systematic and consistent way to implement the KLN theorem. The authors found the same problem in asymptotically free theories such as the massless $\phi^3$ theory in six dimensions \cite{Lavelle:2010hq}.

The previous discussion leads us to the fact that the application of the KLN theorem requires including \emph{all} the initial and final degenerate states to avoid the confusion, inconsistency, and the incompleteness. Including all of these degeneracies forms an infinite series of diagrams with an arbitrary number of soft photons.
\section{The KLN Factorization Theorem (I-ASZ Treatment)}
Ito \cite{Ito:1981nq} and Akhoury,
Sotiropoulos, and Zakharov \cite{Akhoury:1997pb} (I-ASZ) constructed a series in an elegant way which includes all possible degenerate states with an arbitrary number of soft photons. Their series included the contributions from both partially and fully connected cut diagrams on the level of the transition probability. I-ASZ were able to rearrange the series in such a way that the disconnected contributions factorize and the total probability becomes free of any IR divergences. However, as it was pointed out by M.~Lavelle and D.~McMullan \cite{Lavelle:2005bt}, the I-ASZ final result leads to an identically zero contribution from both the tree level and NLO. In this section, we will briefly follow I-ASZ for the series construction. We will then show how to correct their treatment, leading to the fully correct implementation of the KLN theorem. 

We consider a general form of our process with $m$ incoming soft photons and $n$ outgoing soft photons:
\begin{align}
e^-+m \, \gamma \text{ (soft) } \rightarrow e^-+n \, \gamma \text{ (soft)},
\label{em-en} 
\end{align}
with an amplitude $\mathcal{M}_{mn}$, then the transition probability for the process becomes
\begin{align}
\mathcal{P}_{mn} = \frac{1}{m!}\frac{1}{n!} \sum_{i,f} \left|\mathcal{M}_{mn}\right|^2,
\label{pmn}
\end{align}
where $\mathcal{P}_{mn}$ contains contributions from both fully and partially connected cut diagrams and the sum over initial ($i$) and final $(f)$ states exist. The total Lee-Nauenberg probability will be
\begin{align}
\mathcal{P} & = \sum_{m,n=0}^{\infty} \mathcal{P}_{mn},
\label{p}
\end{align}
where the KLN theorem ensures that the quantity $\mathcal{P}$ is free of the IR (soft or collinear) divergences \cite{Lee:1964is}.

\begin{figure}[t]
\centering
\includegraphics[scale=0.75]{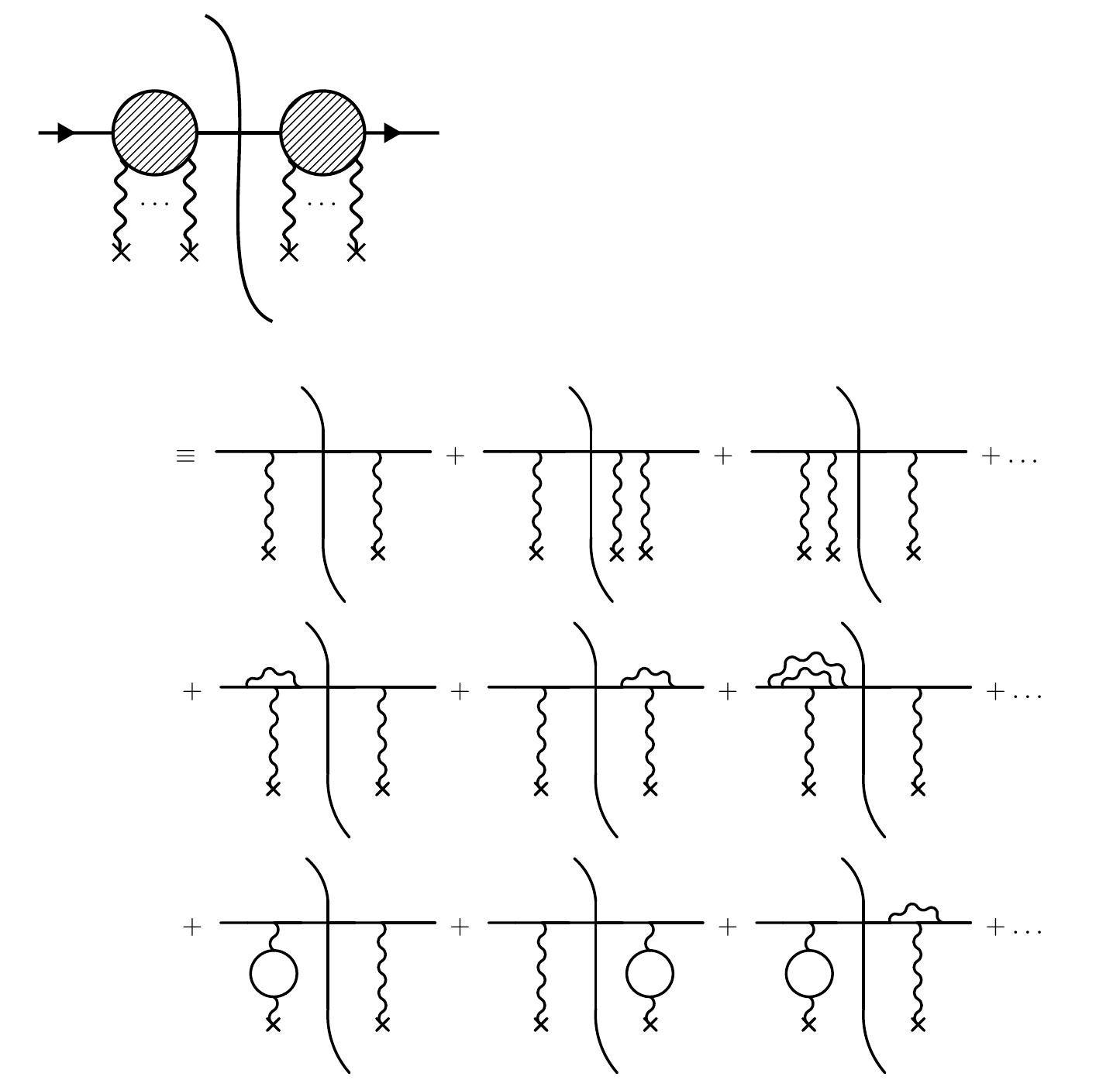}
\caption{A generic cut diagram of $\mathcal{P}_{00}$ on the L.H.S and its full expansion on the R.H.S}.
\label{fig:p00cut}
\end{figure}

 It is shown in \cite{Ito:1981nq,Akhoury:1997pb} that any cut diagram from $\mathcal{P}_{mn}$ at NLO can be constructed from four essential probabilities: $\mathcal{P}_{00}$, which is the cut diagram with no photons in the initial and final states (this may include the leading term, the vertex correction, the vacuum polarization, etc.); $\mathcal{P}_{10}$ and $\mathcal{P}_{01}$ which includes all cut diagrams with one soft photon in the initial and final states respectively; and $\widetilde{\mathcal{P}}_{11}$ which includes all cut diagrams from the absorption-emission diagrams. \Cref{fig:p00cut} shows a generic cut diagram for $\mathcal{P}_{00}$ and its expansion as an example of these essential probabilities. The fully connected cut diagrams are given by any of the previous basic probabilities while the partially connected ones are these probabilities multiplied by a number of $\delta$ functions according to the number of disconnected photons in the cut diagram level, so we can construct $\mathcal{P}_{mn}$ by splitting each cut diagram up into connected and disconnected parts. Then the quantity $\mathcal{P}_{mn}$ will be
\begin{align}
\mathcal{P}_{mn} = \sum_{i,j}\frac{1}{(m-i)!\,(n-j)!} \ \mathcal{D}(m-i,n-j)\, \mathcal{C}(i,j,\alpha,\beta),
\label{pmngeneral}
\end{align}
where the factor $(m-i)!(n-j)!$ is the number of ways to draw the equivalent cut diagram and the upper limit of the sum over $i$ and $j$ is the minimum of $m$ and $n$.
\begin{figure}
\centering
\includegraphics[scale=0.8]{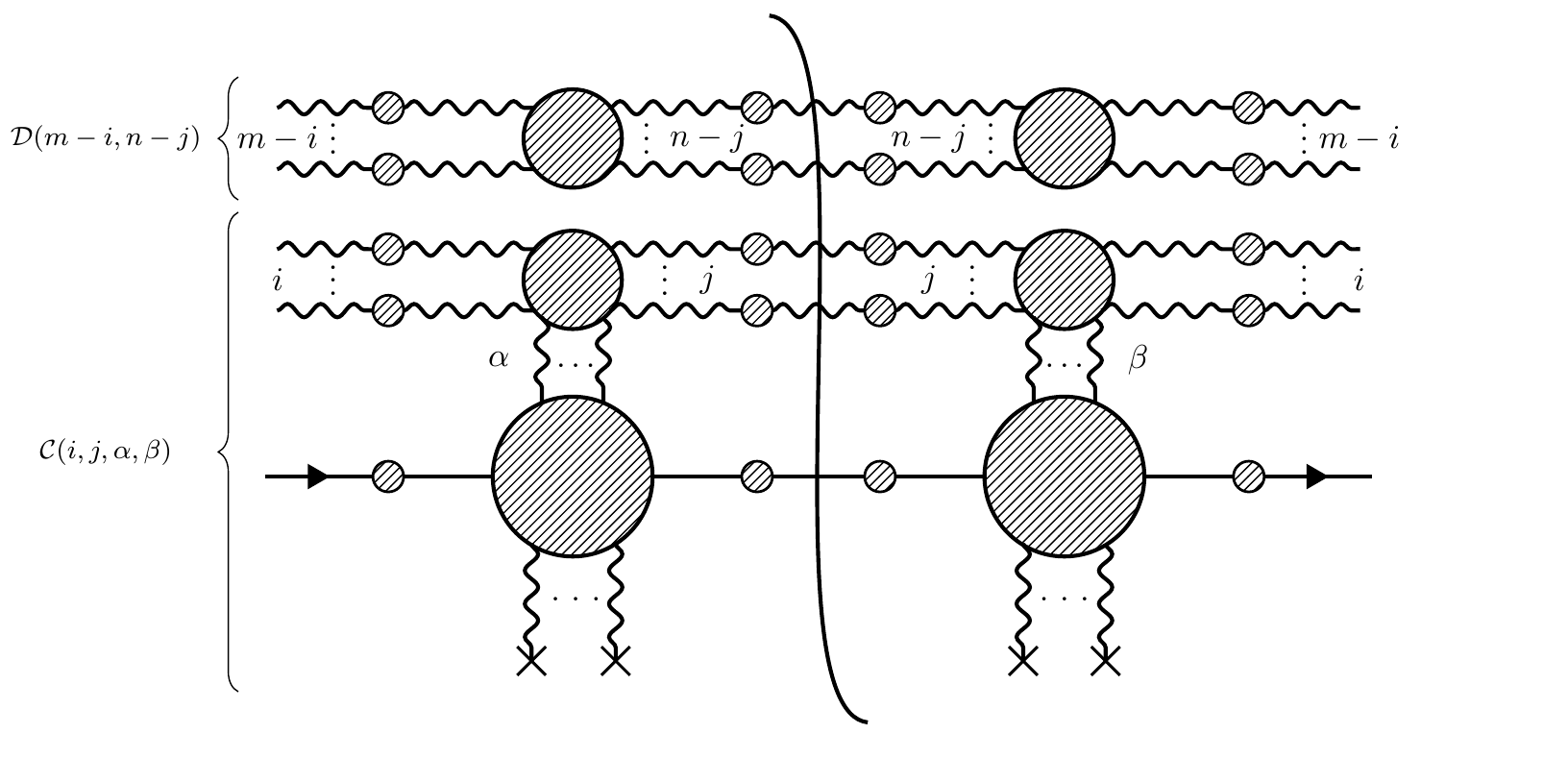}
\caption{A generic cut diagram with $m$ incoming and $n$ outgoing photons where the upper part describe the disconnected function $\mathcal{D}(m-i,n-j)$ while the lower part describe the connected function $\mathcal{C}(i,j,\alpha,\beta)$.}
\label{fig:genericcut}
\end{figure}

For a general cut diagram at any order in perturbation shown in \Cref{fig:genericcut}, we define the function $\mathcal{C}(i,j,\alpha,\beta)$ to describe the whole connected part, where $\alpha$ and $\beta$ are the numbers of photons that are connected directly to the hard part in the initial and the final states respectively, while $i$ and $j$ are the number photons in the initial and the final states that can be joined to $\alpha$ and $\beta$ to form a fully connected cut diagram. Notice that the little blobs within the photon and electron lines in \Cref{fig:genericcut} describe the correction from the LSZ formula. The medium sized blobs show the possibility of two gauge bosons to be joined together on the same side of the unitarity cut, e.g.\ for gluons in QCD. The possibility that we can get a fully connected cut diagram by adding more $i$ and $j$ photons can be understood as the photon passing through the unitarity cut more than once.

\begin{figure}[t]
\centering
\begin{adjustbox}{minipage=\linewidth,scale=1}
\hspace{4.5cm}
\begin{subfigure}[b]{0.3\textwidth}
\begin{center}
\includegraphics[scale=1]{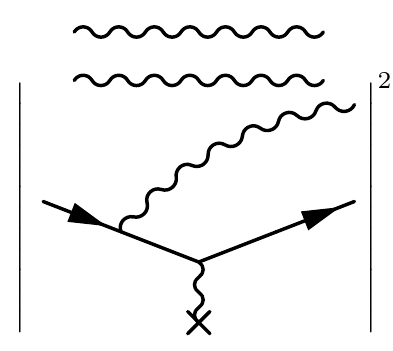}
\caption{}
\label{double-disc}
\end{center}
\end{subfigure}
\vspace{0.1cm}\\
\begin{subfigure}[b]{0.3\textwidth}
\begin{center}
\includegraphics[scale=0.8]{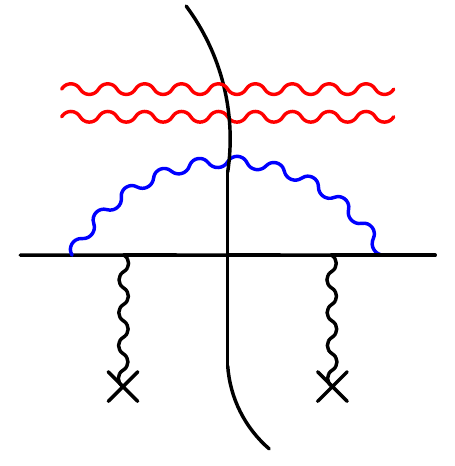}
\caption{}
\label{full-dis-cut}
\end{center}
\end{subfigure}
\begin{subfigure}[b]{0.3\textwidth}
\begin{center}
\includegraphics[scale=0.8]{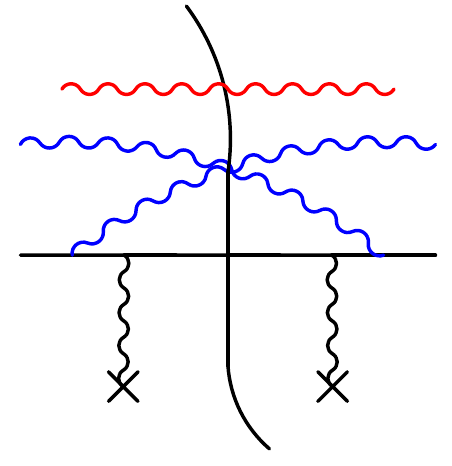}
\caption{}
\label{par-dis-cut}
\end{center}
\end{subfigure}
\begin{subfigure}[b]{0.3\textwidth}
\begin{center}
\includegraphics[scale=0.8]{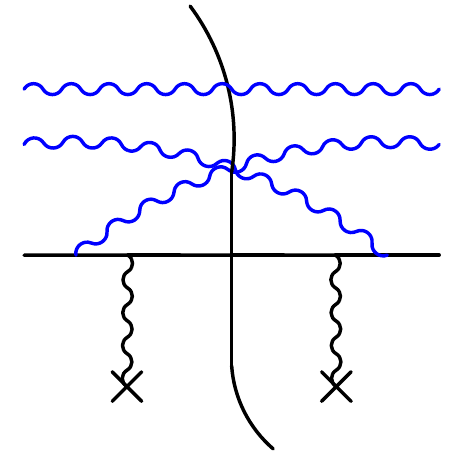}
\caption{}
\label{full-con-cut}
\end{center}
\end{subfigure}\\
\begin{subfigure}[b]{0.3\textwidth}
\begin{center}
\includegraphics[scale=0.65]{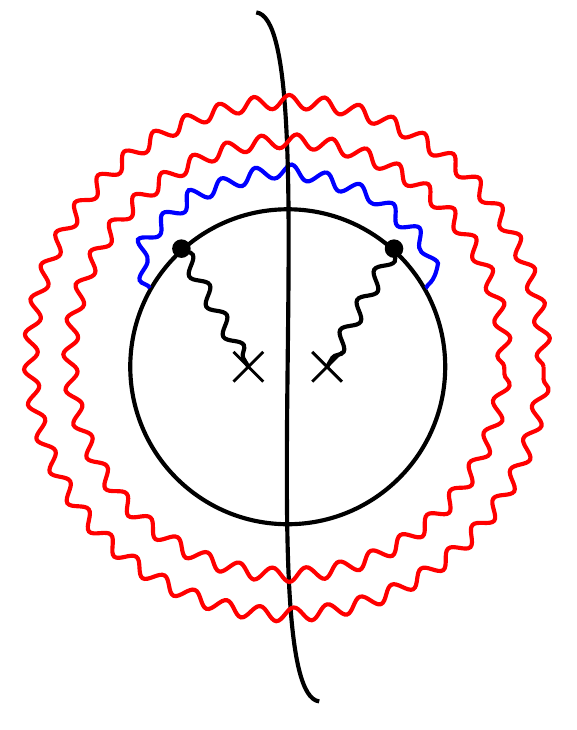}
\caption{}
\label{circ-cut1}
\end{center}
\end{subfigure}
\begin{subfigure}[b]{0.3\textwidth}
\begin{center}
\includegraphics[scale=0.65]{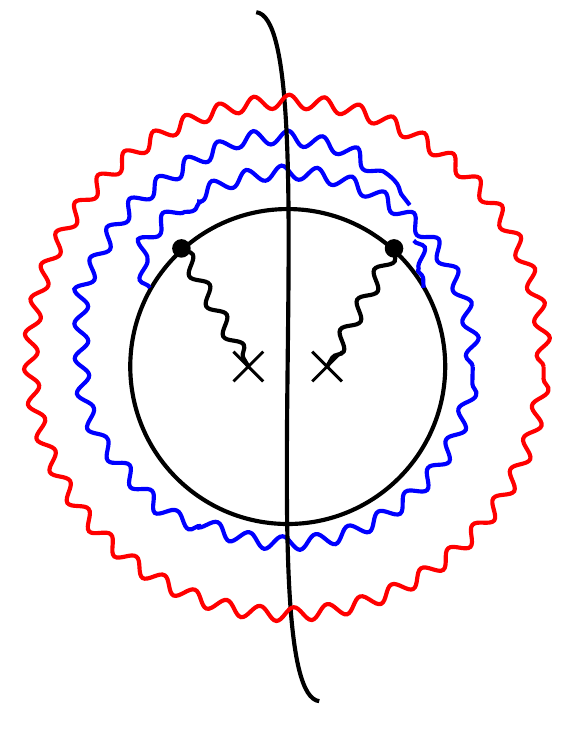}
\caption{}
\label{circ-cut2}
\end{center}
\end{subfigure}
\begin{subfigure}[b]{0.3\textwidth}
\begin{center}
\includegraphics[scale=0.65]{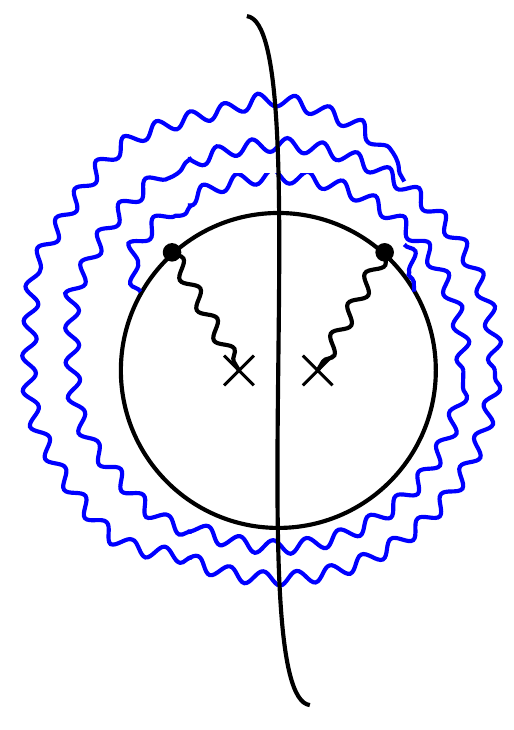}
\caption{}
\label{circ-cut3}
\end{center}
\end{subfigure}
\end{adjustbox}
\caption{\protect\subref{double-disc} is the amplitude squared of an emission of a soft photon from the incoming electron with two disconnected soft photons. \protect\subref{full-dis-cut}, \protect\subref{par-dis-cut}, and \protect\subref{full-con-cut} are the usual final state cut diagrams, while \protect\subref{circ-cut1}, \protect\subref{circ-cut2}, and \protect\subref{circ-cut3} are their corresponding circular cut diagrams.}
\label{p23general-cut}
\end{figure}

It is useful to draw circular cut diagrams to avoid the diagrammatic ambiguity; in particular to see how a disconnected amplitude with more than one disconnected photon can form a fully connected cut diagram. \Cref{p23general-cut} shows the cut diagrams that describe the probability $\mathcal{P}_{23}$ in which we specify the disconnected photons in red and the connected ones in blue. \Cref{double-disc} describes the amplitude squared of an emission of a soft photon from the incoming electron with two disconnected soft photons and its corresponding possible cut diagrams. \Cref{full-dis-cut} is the final state cut diagram when all the disconnected photons remain disconnected from the hard part, while its corresponding circular cut diagram is shown in \Cref{circ-cut1}. \Cref{par-dis-cut} is the final state cut diagram when one of the disconnected photons is joined to the photon emitted from the incoming electron to form a partially connected cut with only one disconnected photon, which corresponds to the circular diagram in \Cref{circ-cut2} in which one can see that the photon line passes the unitarity cut one more time. \Cref{full-con-cut} is the fully connected final state cut diagram, which happens when the photon passes the unitarity cut two more times leaving no disconnected photons as shown in the corresponding circular cut in \Cref{circ-cut3}.

We also define the function $\mathcal{D}(m-i,n-j)$ which describes the disconnected part in terms of $m-i$ incoming and $n-j$ outgoing soft photons joined together on the level of the amplitude squared. It is straightforward to see that $\mathcal{D}(0,0) = 1$ always and $\mathcal{D}(m,n) = 0$ for $m\neq n$ by definition. 

For the contributions at NLO we need to consider only the following four cases:
\begin{enumerate}
\item[(1)] $\alpha=\beta=0$, which implies $i=j=0$ and $\mathcal{C} = \mathcal{P}_{00}$ and then contributes with
\begin{align}
\mathcal{P}_{mn}^{(0)}= \frac{\mathcal{D}(m,n)}{m!\,n!}\  \mathcal{P}_{00},
\label{s0}
\end{align}
\item[(2)] $\alpha=\beta=1$, involving $j=i+1$ and $\mathcal{C} = \mathcal{P}_{01}$, this set of cut diagrams contributes with
\begin{align}
\mathcal{P}_{mn}^{(1)}= \sum_{i=0}^{\min(m,n-1)}\frac{\mathcal{D}(m-i,n-i-1)}{(m-i)!\ (n-i-1)!}\  \mathcal{P}_{01},
\label{s1}
\end{align}
\item[(3)] $\alpha=\beta=1$, this suggests that $i=j+1$ and $\mathcal{C} = \mathcal{P}_{10}$ while their contribution becomes
\begin{align}
\mathcal{P}_{mn}^{(2)}= \sum_{i=0}^{\min(m-1,n)}\frac{\mathcal{D}(m-i-1,n-i)}{(m-i-1)!\ (n-i)!}\ \mathcal{P}_{10},
\label{s2}
\end{align}
\item[(4)] $\alpha= 2,\ \beta=0$ or $\alpha = 0,\ \beta = 0$, which ensure that $i=j$ and $\mathcal{C} = \widetilde{\mathcal{P}}_{11}$ and the contributing term will be
\begin{align}
\mathcal{P}_{mn}^{(3)}= \sum_{i=0}^{\min(m-1,n-1)}\frac{\mathcal{D}(m-i-1,n-i-1)}{(m-i-1)!\ (n-i-1)!}\  \widetilde{\mathcal{P}}_{11}.
\label{s3}
\end{align}
\end{enumerate}
Now we put everything together and the transition probability at NLO of $m$ incoming and $n$ outgoing soft photons becomes
\begin{align}
\mathcal{P}_{mn} &= \frac{\mathcal{D}(m,n)}{m!\,n!}\, \mathcal{P}_{00}+\sum_{i=0}\frac{\mathcal{D}(m-i,n-i-1)}{(m-i)!\,(n-i-1)!}\, \mathcal{P}_{01}\nonumber\\&\quad+\sum_{i=0}\frac{\mathcal{D}(m-i-1,n-i)}{(m-i-1)!\,(n-i)!}\, \mathcal{P}_{10}+\sum_{i=0}\frac{\mathcal{D}(m-i-1,n-i-1)}{(m-i-1)!\,(n-i-1)!}\, \widetilde{\mathcal{P}}_{11}.
\label{pmn-nlo}
\end{align}

I-ASZ rearranged the series in \Cref{p} in such a way that the disconnected piece factors out and the sum becomes IR finite. We will show that the I-ASZ result is not physically acceptable. In order to understand their result and to motivate our correction, we will show in detail their rearrangement. I-ASZ made a use of the following identity
\begin{align}
\frac{\mathcal{D}(m,n)}{m!\,n!} = \sum_{i=0} \frac{\mathcal{D}(m-i,n-i)}{(m-i)!\,(n-i)!} -\sum_{i=1}\frac{\mathcal{D}(m-i-1,n-i-1)}{(m-i-1)!\,(n-i-1)!},
\label{I-ASZ-trick} 
\end{align}
which allows them to rewrite \Cref{pmn-nlo} to obtain
\begin{multline}
\mathcal{P} =\sum_{\substack{m=0 \\  n=0}} \sum_{i=0} \frac{\mathcal{D}(m-i,n-i)}{(m-i)!\,(n-i)!}\, \mathcal{P}_{00}+\sum_{\substack{m=0 \\  n=1}}\sum_{i=0}\frac{\mathcal{D}(m-i,n-i-1)}{(m-i)!\,(n-i-1)!}\, \mathcal{P}_{01}\\+\sum_{\substack{m=1 \\  n=0}}\sum_{i=0}\frac{\mathcal{D}(m-i-1,n-i)}{(m-i-1)!\,(n-i)!}\, \mathcal{P}_{10}+\sum_{\substack{m=1 \\  n=1}}\sum_{i=0}\frac{\mathcal{D}(m-i-1,n-i-1)}{(m-i-1)!\,(n-i-1)!}\, (\widetilde{\mathcal{P}}_{11}-\mathcal{P}_{00}) .
\label{I-ASZ-p}
\end{multline}
Shifting the indices over $n$ in the second term, $m$ in the third term and both in the last term of \Cref{I-ASZ-p} gives
\begin{align}
\mathcal{P} &= \sum_{m,n}\sum_{i}\frac{\mathcal{D}(m-i,n-i)}{(m-i)!\,(n-i)!}\left[\mathcal{P}_{00}+\mathcal{P}_{01}
+\mathcal{P}_{10}+(\widetilde{\mathcal{P}}_{11}-\mathcal{P}_{00})\right] \nonumber\\
& =\left[\mathcal{P}_{00}+\mathcal{P}_{01}
+\mathcal{P}_{10}+(\widetilde{\mathcal{P}}_{11}-\mathcal{P}_{00})\right] \sum_{m,n}\sum_{i}\frac{\mathcal{D}(m-i,n-i)}{(m-i)!\,(n-i)!}.
\label{I-ASZ-rearrang}
\end{align}
 
The result in \Cref{I-ASZ-rearrang} is very interesting since the quantity in the square bracket is IR finite and the disconnected piece has factored out and can be eliminated by the normalization of the $S$-matrix as the authors claimed. However, if one looks closely at \Cref{I-ASZ-rearrang}, we see that the contribution from $\mathcal{P}_{00}$ disappears! $\mathcal{P}_{00}$ includes the tree level contribution, in addition to other radiative corrections, and we are only left with the NLO corrections of $\mathcal{P}_{01}+\mathcal{P}_{10}+
\widetilde{\mathcal{P}}_{11}$. It is also worth emphasizing that our calculations for the probabilities in \autoref{AppendixC} show that
\begin{align} \mathcal{P}_{01}+ \mathcal{P}_{10}= - \widetilde{\mathcal{P}}_{11},
\label{p11}
\end{align}
which makes \Cref{I-ASZ-rearrang} precisely 0 at LO and NLO! The result in \Cref{p11} was also obtained in \cite{DeCalan:1972ya,Muta:1981pe}, and \Cref{p11} holds even if we have different energy resolutions for the initial and final states (i.e.\ $\Delta_i$ and $\Delta_f$) as we show in \autoref{AppendixC}.  Clearly \Cref{I-ASZ-rearrang} is a complete disaster.
\section{An Alternative Rearrangement}
The problem that the I-ASZ rearrangement leads to an identically 0 probability for the process $e^-+m \, \gamma  \rightarrow e^-+n \, \gamma$  shown in the previous section leads us to look for another rearrangement for the series in \Cref{p}. We are looking for a rearrangement with special features: physically sensible, IR safe, and retains the tree level contribution. We first perform an index shift on $i$ for the last term in \Cref{pmn-nlo} such that
\begin{align}
\sum_{\substack{m=1 \\  n=1}}\mathcal{P}_{mn}^{(3)} &= \sum_{\substack{m=1 \\  n=1}}\sum_{i=1}\frac{\mathcal{D}(m-i,n-i)}{(m-i)!\,(n-i)!}\, \widetilde{\mathcal{P}}_{11}.
\label{s4-ed}
\end{align}
Note that we can only apply an index shift on $i$ for the previous term to ensure that we get something in common from all the terms in such a way that the disconnected piece factors out. For the second term, we are only able to make an index shift on $n$ and then pull the $i=0$ term out of the series so that we can write
\begin{align}
 \sum_{\substack{m=0 \\  n=1}}\mathcal{P}_{mn}^{(1)}&= \sum_{\substack{m=0 \\  n=0}}\sum_{i=0}\frac{\mathcal{D}(m-i,n-i)}{(m-i)!\,(n-i)!}\, \mathcal{P}_{01}\nonumber\\& = \sum_{\substack{m=0 \\  n=0}}\frac{\mathcal{D}(m,n)}{m!\,n!}\, \mathcal{P}_{01} +\sum_{\substack{m=1 \\  n=1}}\sum_{i=1}\frac{\mathcal{D}(m-i,n-i)}{(m-i)!\,(n-i)!}\, \mathcal{P}_{01} .
\label{s1-ed}
\end{align} 

Now, we need to think about a rearrangement for the third term with a similar disconnected part as the previous two terms. The only way to do that is to perform a reverse index shift on $n$ followed by an index shift on $i$ to obtain
\begin{align}
\sum_{\substack{m=1 \\  n=0}}\mathcal{P}_{mn}^{(2)} &= \sum_{\substack{m=1 \\  n=1}}\sum_{i=0}\frac{\mathcal{D}(m-i-1,n-i-1)}{(m-i-1)!\,(n-i-1)!}\, \mathcal{P}_{10}\nonumber\\& = \sum_{\substack{m=1 \\  n=1}}\sum_{i=1}\frac{\mathcal{D}(m-i,n-i)}{(m-i)!\,(n-i)!}\, \mathcal{P}_{10}.
\label{s2-ed}
\end{align}
Finally, we put everything together and the result becomes
\begin{align}
\mathcal{P} = \sum_{\substack{m=0 \\  n=0}}\frac{\mathcal{D}(m,n)}{m!\,n!}\, (\mathcal{P}_{00}+\mathcal{P}_{01}) + \sum_{\substack{m=1 \\  n=1}}\sum_{i=1}\frac{\mathcal{D}(m-i,n-i)}{(m-i)!\,(n-i)!}\, (\mathcal{P}_{10}+\mathcal{P}_{01}+\widetilde{\mathcal{P}}_{11}),
\label{WA-rearrang}
\end{align}
\Cref{p11} allows us to omit the second term in \Cref{WA-rearrang}, we end up with the following result
\begin{align}
\mathcal{P}&= \sum_{m,n}\frac{\mathcal{D}(m,n)}{m!\,n!}\, (\mathcal{P}_{00}+\mathcal{P}_{01})\nonumber\\
& =  (\mathcal{P}_{00}+\mathcal{P}_{01})\sum_{m,n}\frac{\mathcal{D}(m,n)}{m!\,n!}.
\label{WA-prob}
\end{align}

With \Cref{WA-prob} we have accomplished our goals of finding a physically sensible rearrangement. We have factorized the infinite soft IR contributions and retained the LO contribution plus the NLO corrections. In fact, \Cref{WA-prob} is precisely the original BN result which means that, with this rearrangement, the more general KLN theorem reduces to the BN theorem. 

How did two rearrangements of the same series yield two completely different results? is a crucial question. If we think carefully about \Cref{pmn-nlo}, we realize that the series formally diverges. In particular, for large $m$ and $n$ the series just keeps adding terms to the partial sum over $i$ producing an infinite number either partially or fully connected cut diagrams. 

Now what we need to do is to rigorously prove mathematically that \Cref{WA-prob} is the unique and correct rearrangement of \Cref{pmn-nlo}. In order to make such a proof, we force \Cref{pmn-nlo} to converge and perform the manipulations under control. We introduce a convergence factor that becomes small for large $i$: we take
\begin{equation}
     \mathcal{D}(m-i,n-j) \;\; \rightarrow \;\; \mathcal{D}(m-i,n-j)e^{-(i+j)/\Lambda}
\end{equation}
with $\Lambda\gg1$, this allows us to sum over $m$ and $n$ up to a finite value $N$. Note that we will ultimately take our convergence factor $\Lambda \rightarrow \infty$ instead of the usual $\epsilon\rightarrow 0$; the reason for dividing by  a large number instead of multiplying by a small convergence factor will be obvious in a moment. We can also simplify \Cref{I-ASZ-rearrang,WA-rearrang} by replacing the double sum over $m$ and $n$ by a single sum over $n$ since $\mathcal{D}(m,n)$ is zero for $m\neq n$. Then we rewrite the original series to be
\begin{multline}
\mathcal{P}_{N}^{original}(\Lambda) =\sum_{n=0}^N \frac{\mathcal{D}(m,n)}{m!\,n!}\, \mathcal{P}_{00}+\sum_{n=0}^N\sum_{i=0}\frac{\mathcal{D}(m-i,n-i-1)}{(m-i)!\,(n-i-1)!}e^{-\frac{2i+1}{\Lambda}} \mathcal{P}_{01}\\+\sum_{n=0}^N\sum_{i=0}\frac{\mathcal{D}(m-i-1,n-i)}{(m-i-1)!\,(n-i)!}e^{-\frac{2i+1}{\Lambda}} \mathcal{P}_{10}+\sum_{n=0}^N\sum_{i=0}\frac{\mathcal{D}(m-i-1,n-i-1)}{(m-i-1)!\,(n-i-1)!}e^{-\frac{2i+2}{\Lambda}} \widetilde{\mathcal{P}}_{11}.
\label{original}
\end{multline}
\Cref{WA-rearrang} also becomes 
\begin{align}
\mathcal{P}_{N}^{KH}(\Lambda) &= \sum_{n=0}^N \frac{\mathcal{D}(n,n)}{(n!)^2}\,\left[\mathcal{P}_{00}+e^{-\frac{1}{\Lambda}}\mathcal{P}_{01}\right]\nonumber\\&\quad+\sum_{n=1}^N
\sum_{i=1}^n \frac{\mathcal{D}(n-i,n-i)}{[(n-i)!]^2}\, \left[e^{-\frac{2i+1}{\Lambda}}\mathcal{P}_{01}+e^{-\frac{2i-1}{\Lambda}}\mathcal{P}_{10}+e^{-\frac{2i}{\Lambda}}\widetilde{\mathcal{P}}_{11}\right],
\label{WA-re-con}
\end{align}
while \Cref{I-ASZ-rearrang} becomes
\begin{align}
\mathcal{P}_N^{\mathrm{I-ASZ}}(\Lambda)= \sum_{n=0}^N
\sum_{i=0}^n \frac{\mathcal{D}(n-i,n-i)}{[(n-i)!]^2}\, \left[e^{-\frac{2i}{\Lambda}} \mathcal{P}_{00}+
e^{-\frac{2i+1}{\Lambda}}(\mathcal{P}_{01}+\mathcal{P}_{10})+e^{-\frac{2i+2}{\Lambda}}(\widetilde{\mathcal{P}}_{11}-\mathcal{P}_{00})\right].
\label{I-ASZ-re-co}
\end{align}
The total probability can now be written in terms of the general $\mathcal{P}_{N}(\Lambda)$ as  
\begin{align}
\mathcal{P}= \lim_{\Lambda\rightarrow \infty}\  \lim_{N\rightarrow \infty}\ \mathcal{P}_{N}(\Lambda).
\label{p-con}
\end{align}
If we can swap the limits in \Cref{p-con}, with $\mathcal{P}_{N}(\Lambda)=\mathcal{P}_{N}^{KH}(\Lambda)$ then we proved that \Cref{WA-prob} is the unique and correct rearrangement.

Before we proceed with the proof, let us develop an intuition regarding the different rearrangements. We choose the disconnected function to be $\mathcal{D}(m,n) = m n\, \delta_{mn}$ and the values for the basic probabilities to be $\mathcal{P}_{00} = 0.5$, $\mathcal{P}_{01} = 0.15$, $\mathcal{P}_{10} = 0.25$, and $\widetilde{\mathcal{P}}_{11} = -0.4$ (Note that the choice of $\mathcal{D}(m,n)$ has no physical significance while the choice of $\mathcal{P}_{01}$, $\mathcal{P}_{10}$, and $\widetilde{\mathcal{P}}_{11}$ must satisfy \Cref{p11}). Thus we are able to calculate the total probability $\mathcal{P}$ and check its behavior for the different rearrangements.

\begin{figure}[t]
\centering
\begin{subfigure}[t]{0.45\textwidth}
\begin{center}
\includegraphics[scale=0.55]{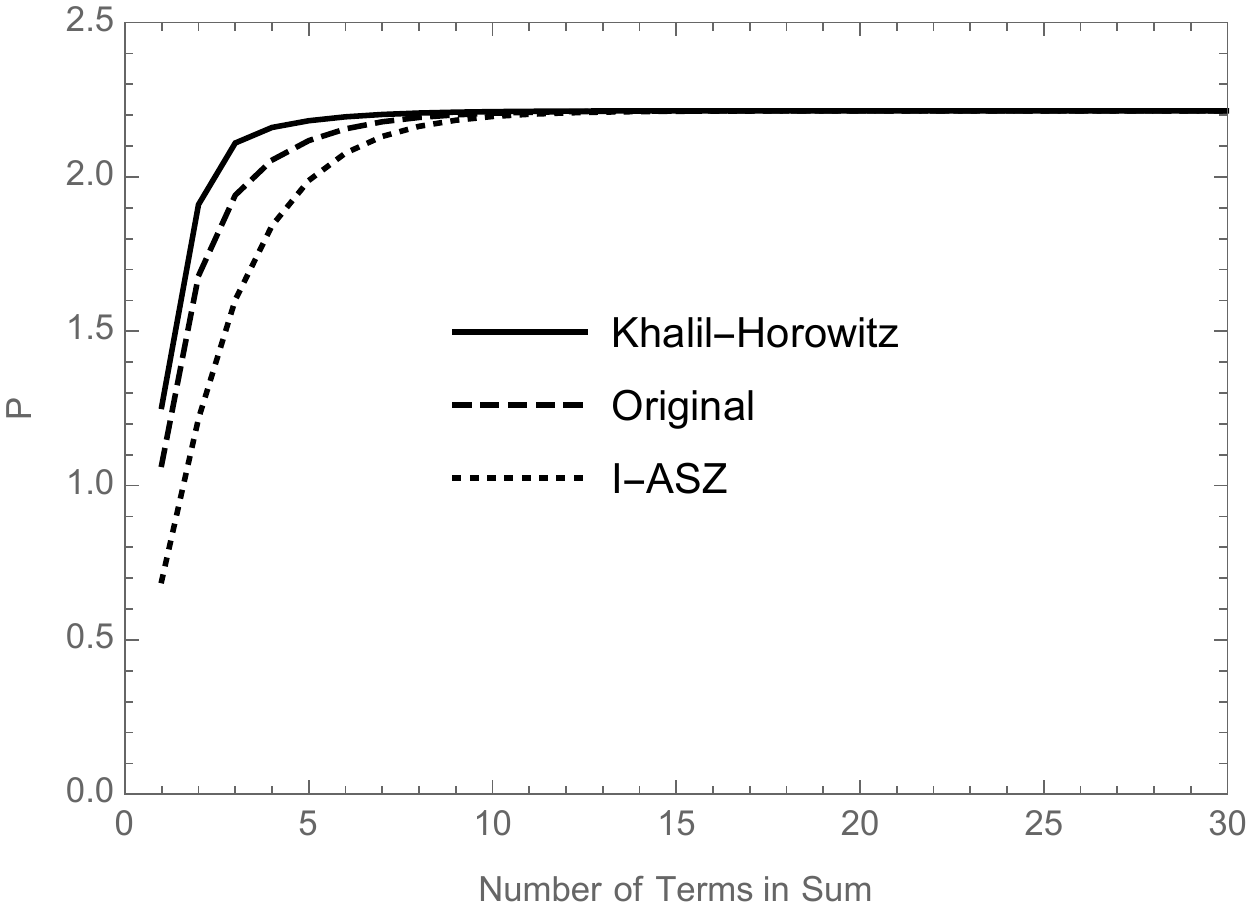}
\caption{$\Lambda$ = 4.}
\label{fig:lambda10}
\end{center}
\end{subfigure}
\begin{subfigure}[t]{0.45\textwidth}
\begin{center}
\includegraphics[scale=0.55]{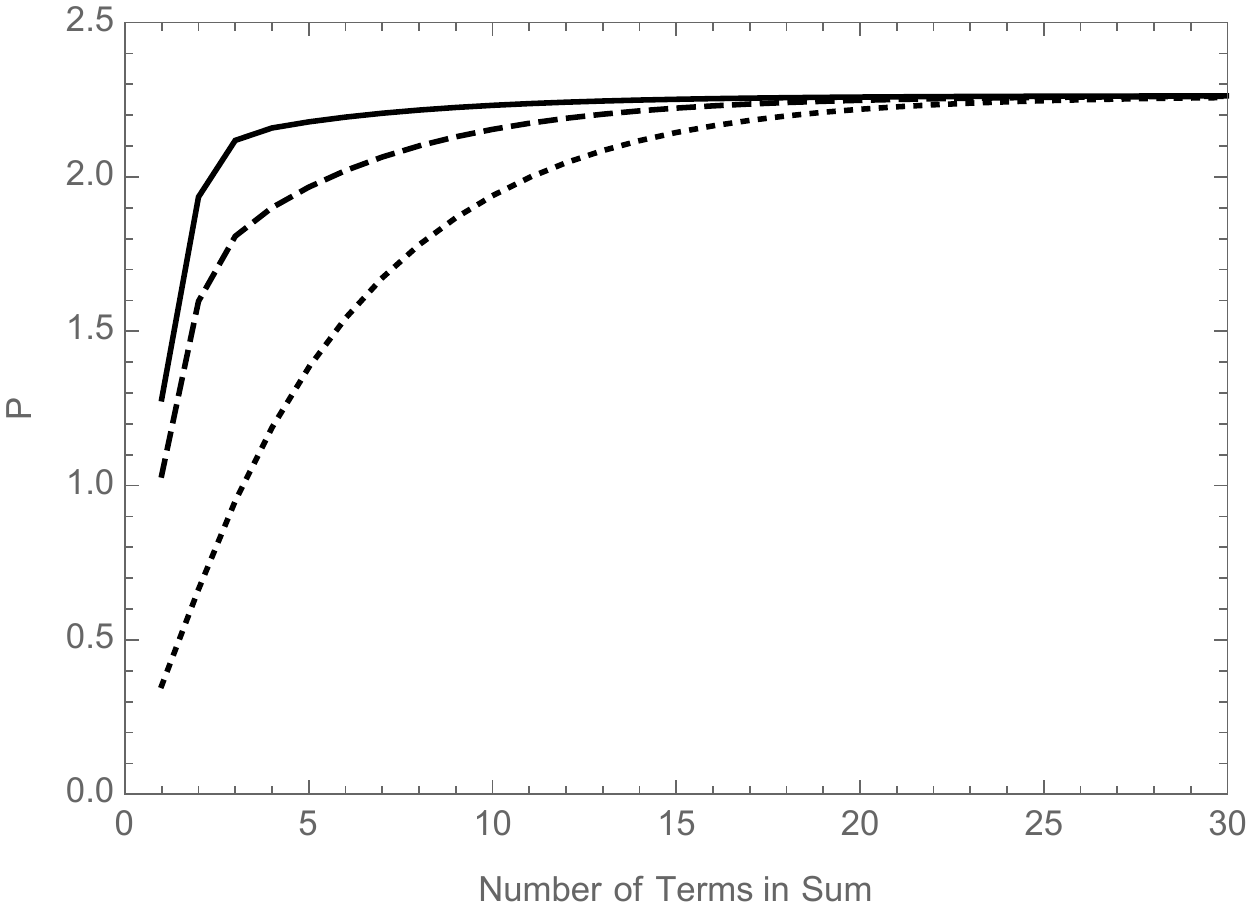}
\caption{$\Lambda$ = 10.}
\label{fig:lambda20}
\end{center}
\end{subfigure}
\begin{subfigure}[t]{0.45\textwidth}
\begin{center}
\includegraphics[scale=0.55]{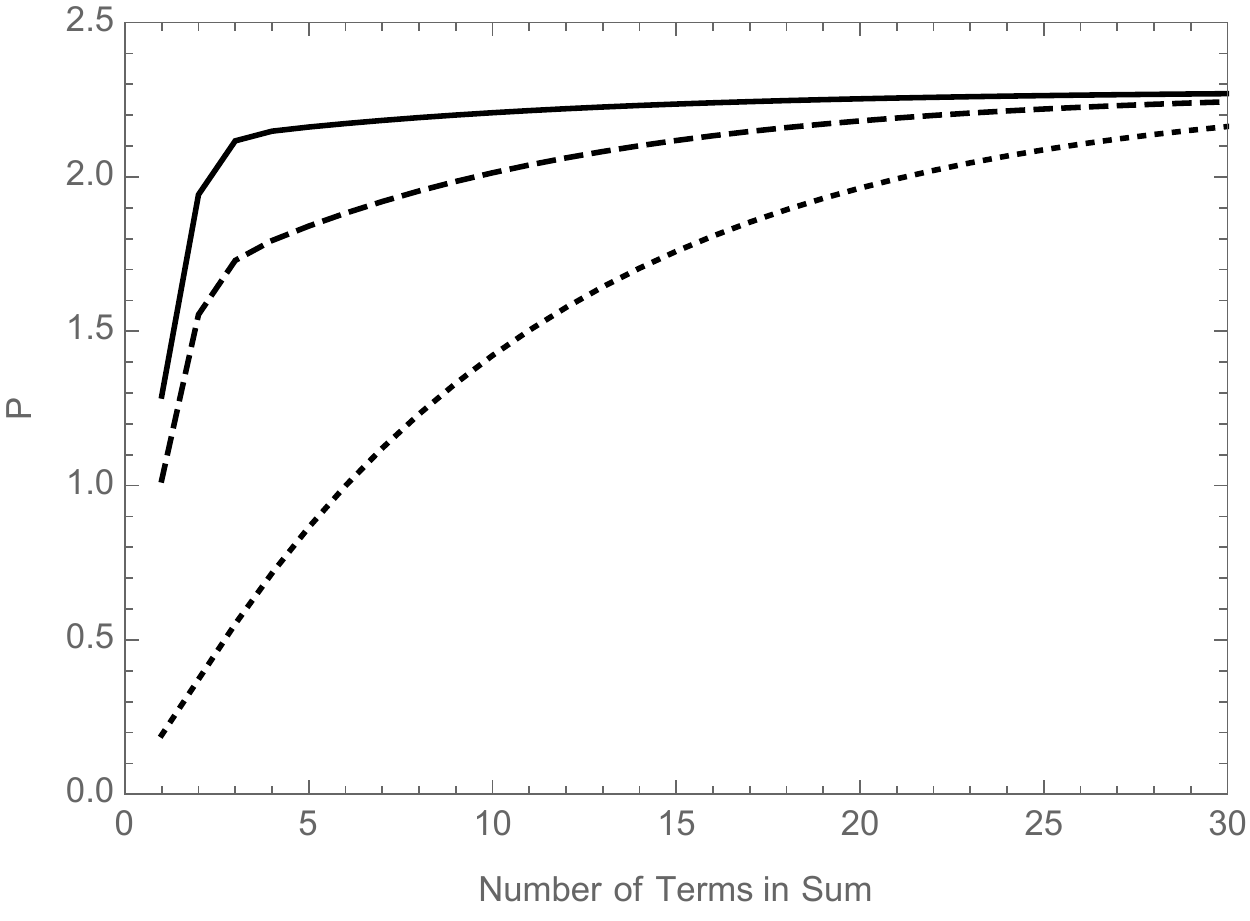}
\caption{$\Lambda$ = 20.}
\label{fig:lambda100}
\end{center}
\end{subfigure}
\begin{subfigure}[t]{0.45\textwidth}
\begin{center}
\includegraphics[scale=0.55]{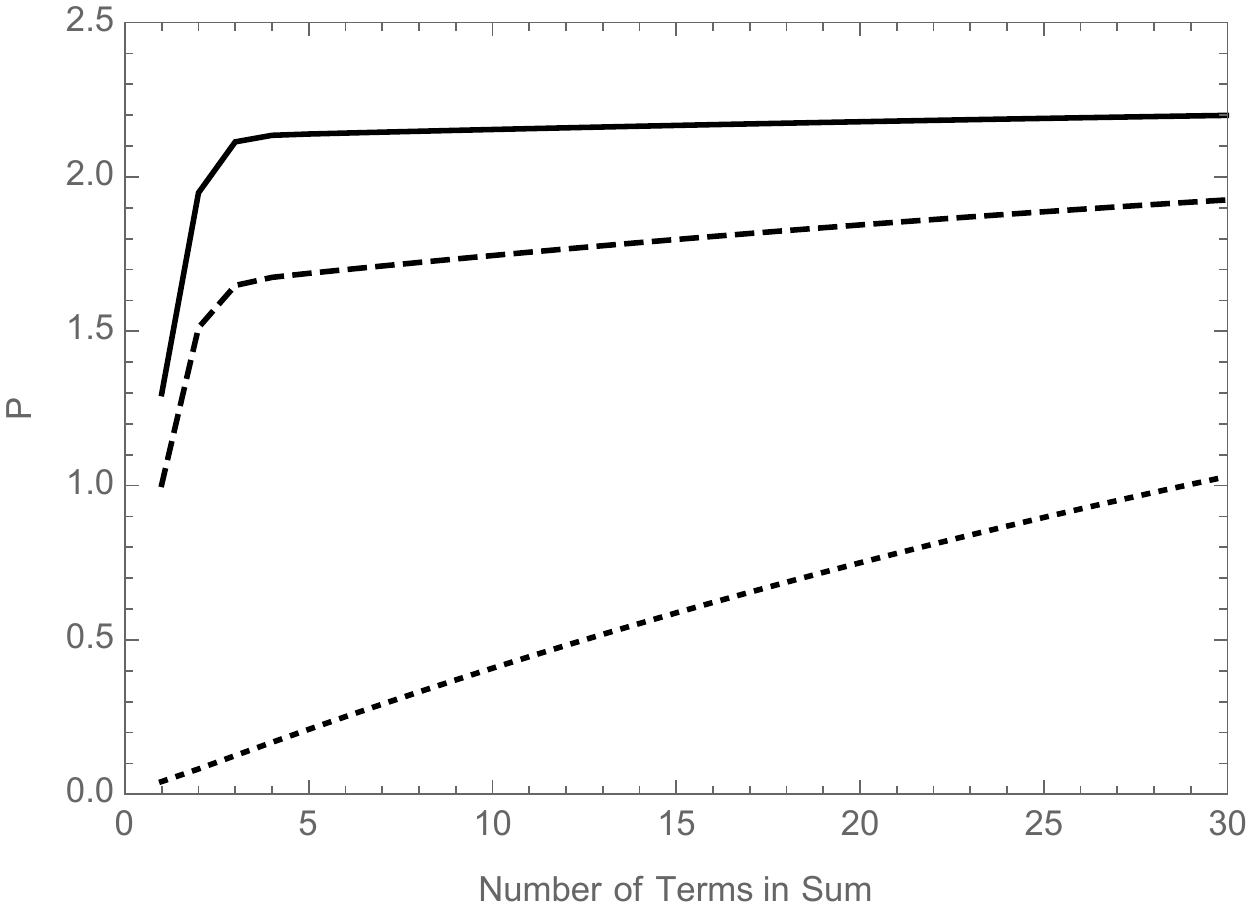}
\caption{$\Lambda$ = 100.}
\label{fig:lambda1000}
\end{center}
\end{subfigure}
\caption{The behavior of the total Lee-Nauenberg probability for the original series, I-ASZ rearrangement and our rearrangement at different values of the convergent factor $\Lambda$.}
\label{fig:totalp}
\end{figure}

\Cref{fig:totalp} shows the convergence properties of the original series, I-ASZ rearrangement, and our rearrangement. First, see that all the rearrangements converge to the same value, as they must. However, we see that the original series and the I-ASZ rearrangement converges very slowly as we increase the value of $\Lambda$, especially I-ASZ rearrangement is driven to 0; if we swap the order of limits, we find $\mathcal{P}^{I-ASZ} \equiv 0$. On the other hand, the number of terms needed for our rearrangement to converge is essentially independent of the convergent factor $\Lambda$. 
 \subsection{Proof of Uniqueness}
 
Swapping limits in an infinite series is a delicate procedure.  We are guaranteed from the Monotone Convergence Theorem that $P_N(\Lambda)$ converges to the same result independent of the order of limits taken should our partial sum 1) monotonically increase in $N$ for each $\Lambda$ and 2) monotonically increase in $\Lambda$ for each $N$ \cite{Knapp}. 

Let us show 1) first. For simplicity we choose $\Delta_i = \Delta_f = \Delta$ which allows the identity in \Cref{p11} to be $\mathcal{P}_{01}=\mathcal{P}_{10}=-\frac{1}{2}\widetilde{\mathcal{P}}_{11}$. Then we simplify \Cref{WA-re-con} to become
\begin{multline}
\mathcal{P}_{N}^{KH}(\Lambda) = \sum_{n=0}^N \frac{\mathcal{D}(n,n)}{(n!)^2}\,\left[\mathcal{P}_{00}+e^{-\frac{1}{\Lambda}}\mathcal{P}_{01}\right]\\+\sum_{n=1}^N
\sum_{i=1}^n \frac{\mathcal{D}(n-i,n-i)}{[(n-i)!]^2}\,2\mathcal{P}_{01}\, e^{-\frac{2i}{\Lambda}} \,\left[\cosh\left(\frac{1}{\Lambda}\right)-1\right].
\label{WA-re-cons}
\end{multline}
Expanding $\cosh\frac{1}{\Lambda} = 1+ \frac{1}{2\Lambda^2}+\mathcal{O}\left(\frac{1}{\Lambda^4}\right)$, equation \Cref{WA-re-cons} becomes
\begin{align}
\mathcal{P}_{N}^{KH}(\Lambda) &= \sum_{n=0}^N \frac{\mathcal{D}(n,n)}{(n!)^2}\,\left[\mathcal{P}_{00}+e^{-\frac{1}{\Lambda}}\mathcal{P}_{01}\right]+\sum_{n=1}^N
\sum_{i=1}^n \frac{\mathcal{D}(n-i,n-i)}{[(n-i)!]^2}\, \frac{1}{\Lambda^2}\,e^{-\frac{2i}{\Lambda}}\,\mathcal{P}_{01}.
\label{WA-re-conf}
\end{align}
Since $\mathcal{P}_{00}$, $\mathcal{P}_{01}$, $\Lambda$, and $\mathcal{D}(n,n)$ are all strictly positive, \Cref{WA-re-conf} clearly increases monotonically in $N$ for fixed $\Lambda$.  

To show 2), we take the derivative of \Cref{I-ASZ-re-co} with respect to $\Lambda$:
\begin{align}
    \frac{d\mathcal{P}_{N}^{KH}(\Lambda)}{d\Lambda} 
    = \sum_{n=0}^N\frac{\mathcal{D}(n,n)}{(n!)^2}\left[\frac{1}{\Lambda^2} \, e^{-\frac{1}{\Lambda}}\,\mathcal{P}_{01}\right] + \mathcal{O}\big( \frac{1}{\Lambda^4} \big).
    \label{WA-derivative}
\end{align}
Although one finds that the higher order in $1/\Lambda$ correction term is negative, for any $N$ we can find a $\Lambda$ large enough such that the first term, which is strictly positive, dominates.

We have thus proved that we may exchange limits for our rearranged formula \Cref{I-ASZ-re-co}, and we may evaluate the $\Lambda\rightarrow \infty$ limit first, yielding our main result in \Cref{WA-prob}
\begin{align}
\mathcal{P}&= \lim_{\Lambda\rightarrow \infty}\  \lim_{N\rightarrow \infty}\ \mathcal{P}_{N}^{KH}(\Lambda) = \lim_{N\rightarrow \infty}\  \lim_{\Lambda\rightarrow \infty}\ \mathcal{P}_{N}^{KH}(\Lambda)\nonumber\\
&= (\mathcal{P}_{00} + \mathcal{P}_{01})\sum_{n=0}^\infty \frac{\mathcal{D}(n,n)}{(n!)^2}.
\label{WAPf}
\end{align}
\subsection{Hard Collinear Contributions}
We note that the contribution from the emission or absorption of a hard collinear photon discussed in \autoref{Chapter4} may also have an arbitrary number of disconnected soft photons and remains degenerate with our original state, in such a way that none of the disconnected photons can be attached to the hard photon. Let us call the probabilities from the contribution of the absorption of hard and collinear photon $\mathcal{P}_{10}^{h,c}$ and the contribution from the emission of hard and collinear photon $\mathcal{P}_{01}^{hc}$. Then the contribution from the diagrams where a hard photon is taking part with an infinite number of disconnected photons becomes 
\begin{equation}
\mathcal{P}^{hc} = (\mathcal{P}_{10}^{hc}+\mathcal{P}_{01}^{hc})\sum_{n=0}^\infty \frac{\mathcal{D}(n,n)}{(n!)^2}.
\end{equation} 
Finally, the total probability contributions from \emph{all} the initial and final degenerate states is
\begin{equation}
\mathcal{P}_{tot}=(\mathcal{P}_{00} + \mathcal{P}_{01}+\mathcal{P}_{10}^{hc}+\mathcal{P}_{01}^{hc})\sum_{n=0}^\infty \frac{\mathcal{D}(n,n)}{(n!)^2}.
\label{WAPtot}
\end{equation}

\Cref{WAPtot} is an infinite transition probability, one may think this is counter to KLN, where KLN ensures that the total transition probability is free of IR divergences. However, the infinity is not related to $m_{\gamma}$ or $m_e$: if we had $m_{\gamma}<\Delta$, then infinity from $\mathcal{D}(n,n)$, with $n>0$, would remain even though $(\mathcal{P}_{00} + \mathcal{P}_{01}+\mathcal{P}_{10}^{hc}+\mathcal{P}_{01}^{hc})$ is IR finite. The infinity from $\mathcal{D}$ is from on-shell photons unrelated to the on-shell electrons, so the infinity from $\mathcal{D}$ should not be absorbed through the application of LSZ to the electron or through re-interpreting the electron with a QED version of parton distribution functions (PDFs). Additionally, LSZ or PDFs are applied at the amplitude level, but $\mathcal{D}$ is an infinity at the level of the amplitude squared; as shown earlier, some amplitudes with disconnected photons can yield fully connected cut diagrams that are finite. The infinity from $\mathcal{D}$ must therefore be cancelled in the physical observable, in this case the cross section. 
\subsection{Physical Cross Section}
The factorization of the disconnected parts from the initial states is very important for the cancellation of these pieces through the normalization of the $S$-matrix. One can write the differential cross section in terms of the incoming state $\ket{p}$, outgoing state $\bra{p'}$, and the vacuum states $\ket{0}$ without degenerate soft photons as

\begin{equation}
d\sigma \sim \left| _{out}\!\left<p'|p\right>_{in}\right|^2=\frac{\left| _{out}\!\left<p'|p\right>_{in}\right|^2}{\left| _{out}\!\left<0|0\right>_{in}\right|^2},
\label{sigma-norm}
\end{equation}
where $\left| _{out}\!\left<p'|p\right>_{in}\right|^2\sim \mathcal{P}_{00}$ and $\left| _{out}\!\left<0|0\right>_{in}\right|^2=1$. Including contributions from degenerate initial and final soft photon states, we have 
\begin{equation}
\begin{aligned}
\left| _{out,deg}\!\left<p'|p\right>_{in,deg}\right|^2&= \left| _{out}\!\left<p'|p\right>_{in}\right|^2\sum_{n} \frac{D(n,n)}{(n!)^2},\\
\left| _{out,deg}\!\left<0|0\right>_{in,deg}\right|^2 &=\left| _{out}\!\left<0|0\right>_{in}\right|^2 \sum_{n}\frac{D(n,n)}{(n!)^2},
\end{aligned}
\end{equation}
where $\left| _{out}\!\left<p'|p\right>_{in}\right|^2\sim \mathcal{P}_{00} + \mathcal{P}_{01}+\mathcal{P}_{10}^{hc}
+\mathcal{P}_{01}^{hc}$. Therefore the cross section including sum over all initial and final degenerate states is given by
\begin{equation}
d\sigma \sim \frac{\left| _{out,deg}\!\left<p'|p\right>_{in,deg}\right|^2}{\left| _{out,deg}\!\left<0|0\right>_{in,deg}\right|^2}=\frac{\left| _{out}\!\left<p'|p\right>_{in}\right|^2}{\left| _{out}\!\left<0|0\right>_{in}\right|^2}.
\label{sigma-norm-dis}
\end{equation}
Thus the disconnected part $\mathcal{D}$ cancels through the normalization process of the $S$-matrix in order to produce a physical cross section. \Cref{WAPtot,sigma-norm-dis} are the main results of this thesis.
\section{The Complete NLO Rutherford Cross Section}
Using our above results we may report the first calculation of NLO Rutherford scattering including the $\mathcal{O}(1)$ contribution. We combine the KLN result \Cref{KLN} with the vacuum polarization and the box diagram contributions, \Cref{vpxsection,boxxsection}, respectively. Putting all the pieces together, we find that Rutherford scattering complete complete to NLO is given by
\begin{multline}
\left(\frac{d\sigma}{d\Omega}\right)_{\texttt{NLO}} = \left(\frac{d\sigma}{d\Omega}\right)_{\texttt{0}}\Bigg\{1+\frac{\alpha}{\pi}\Bigg[\log\left(\frac{Q^2}{\delta^2 E^2}\right)\left(\log\left(\frac{\Delta^2}{E^2}\right)+\frac{3}{2}\right)+\frac{2}{3}\log\left(\frac{Q^2}{\mu_{\overline{\mathrm{MS}}}^2}\right)\\
-\pi^2\left(\frac{1}{\left(\frac{2E}{Q}+1\right)}+\frac{1}{3}\right)+\frac{5}{36}+\mathcal{O}(m^2,\delta^2) \Bigg]\Bigg\} 
\label{NLO-Rutherford}.
\end{multline} 
\Cref{NLO-Rutherford} is the second main result of this thesis.
\subsection{Size of LO Vs. NLO}

We wish to examine the magnitude of the NLO correction compared to the LO contribution. Since $Q^2 = 2E^2 (1-\cos\theta)$, where $\delta < \theta< \pi$, we have 
\begin{equation}
\delta^2E^2<Q^2<4E^2.\label{qmin-max}
\end{equation} 
We will take as representative the angular and energy resolutions from the CMS detector \cite{barney1998pedagogical}. We take $\delta = 50$ mrad, while $\Delta$ can be determined by
\begin{equation}
 \left(\frac{\Delta}{E}\right)^2=
 \left(\frac{S}{\sqrt{E}}\right)^2+\left(\frac{N}{E}\right)^2+C^2,
 \label{Delta-E}
\end{equation}  
where $S$ is the stochastic term, $N$ the noise term and $C$ the constant term. The values of the three parameters were determined by a electron test beam measurement to be $S = 0.028$ GeV$^{\frac{1}{2}}$, $N = 0.12$ GeV, and $C = 0.003$ \cite{barney1998pedagogical}.
\begin{figure}[h]
\centering
\includegraphics[scale=0.8]{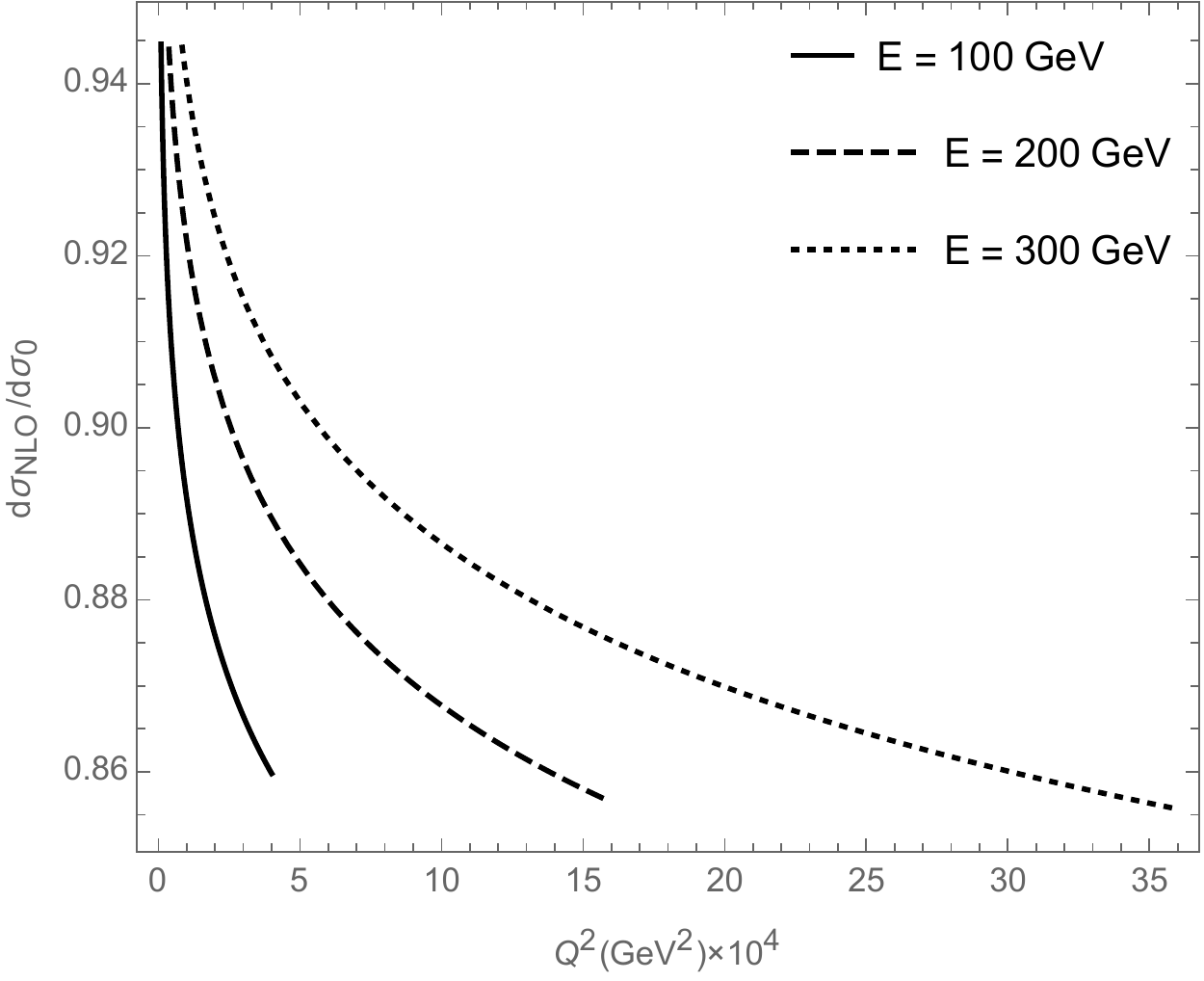}
\caption{The ratio of the NLO to the LO differential cross section of Rutherford scattering as function of the momentum transfer $Q^2$ at energies $E=100,\,200, \text{and } 300\, \text{GeV}$ plotted from $Q_{min}^2$ to $Q_{max}^2$ given by \Cref{qmin-max}.}
\label{fig:NLO-plot}
\end{figure}

\Cref{fig:NLO-plot} shows the effect of the NLO corrections to the differential cross section of Rutherford scattering at high energies. The plot shows that these corrections significantly affect the ratio of the NLO to the LO differential cross section at higher electron energies while it becomes less relevant at very high momentum transfer. One can also see that the $\mathcal{O}(1) \approx -3.15$ which is relatively small compared to the contributions from the logarithmic terms which become in order of $110-120$ at very high momentum transfer $Q^2$. 
\chapter{Renormalization Group}\label{Chapter6}
For the NLO corrections to Rutherford scattering cross section, we used dimensional regularization and the $\overline{\mathrm{MS}}$ renormalization scheme to eliminate the UV divergences. Consequently, we introduced an arbitrary mass scale $\mu$ in order to ensure that the electron charge $e$ remains dimensionless in $d$-dimensions. In fact, the physical observables must be independent of this unphysical parameter $\mu$.

\Cref{NLO-Rutherford} shows that the differential cross section at NLO is a function of $\mu$ which means that the renormalized quantities such as $e$ and $m$ are implicitly $\mu$ dependent. One can thus write for any physical observable $\mathcal{S}(p, p',m_0,e_0)$, in terms of the bare parameters and the physical particle momenta $p$ and $p'$ \cite{Sterman:1994ce}
\begin{equation}
\mu\frac{d}{d\mu}\mathcal{S}(p, p',m_0,e_0) = 0,
\label{ds0dmu}
\end{equation}
while in terms of the renormalized parameters \Cref{ds0dmu} becomes
\begin{equation}
\mu\frac{d}{d\mu}\mathcal{S}(p, p',m,e,\mu) = 0.
\label{dsrdmu}
\end{equation} 
We can now use the chain rule to rewrite \Cref{dsrdmu} as
\begin{align}
\left[\left.\mu\frac{\partial}{\partial\mu}\right|_{e,m}+\left.\beta(e)\frac{\partial}{\partial e}\right|_{\mu,m}-\left.\gamma_m(e,m)\ m \frac{\partial}{\partial m}\right|_{\mu,e}\right]\mathcal{S}(p, p',m,e,\mu)=0,
\label{rengroup}
\end{align}
where the dependence of the renormalized quantities on $\mu$ can be found from the derivatives given by
\begin{align}
\beta(e) \equiv \frac{\partial e(\mu)}{\partial\mu},\ \gamma_m(e,m) \equiv \frac{\mu}{m}\frac{\partial m(\mu)}{\partial \mu}.\label{mubeta-mugamma}
\end{align}

\Cref{rengroup} is known as the renormalization group equation where it gives the change in the renormalized quantities $e$ and $m$ as a function of $\mu$ \cite{Sterman:1994ce}. The dimensionless coefficients in the renormalization group equation depend on $e$ and $m$. However, using a mass independent scheme ($\overline{\mathrm{MS}}$) allows us to drop the mass dependence.

To find the dependence of the coupling $\alpha$ on $\mu$ from the $\beta$ function, we use first the fact that $(d/d\mu)[\text{bare quantities}]=0$. We then recall from \Cref{parrenorm} the relation between the bare charge $e_0$ and the renormalized charge $e$, where the Ward identity ensures that $Z_e = Z_{\psi}$, then we have $ e_0 = \mu^{\epsilon/2}\ Z_A^{-1/2}\ e$. We now differentiate $e_0$ with respect to $\mu$ to find
\begin{align}
\beta(e) = \mu\frac{de}{d\mu} = -\frac{\epsilon}{2}\ e+\frac{1}{2}\ \mu\ Z_A^{-1}\frac{dZ_A}{d\mu}\ e.
\label{beta1}
\end{align}
We can neglect the term $-\gamma_E+\log(4\pi)$ in $Z_A$ compared to the $\frac{1}{\epsilon}$ term to find 
\begin{align}
\frac{dZ_A}{d\mu} & = -\frac{1}{3\pi^2\epsilon}\,e\,\frac{de}{d\mu} = -\frac{e}{3\pi^2\epsilon\mu}\,\beta(e).
\label{dzadmu}
\end{align}
Substituting \Cref{dzadmu} into \Cref{beta1} and solving for $\beta$, we find
\begin{align}
\beta(e)  = -\frac{\epsilon}{2}\,e+\frac{e^3}{12\pi^2} +\mathcal{O}(e^5)\underset{\epsilon\rightarrow 0}{=}\frac{e^3}{12\pi^2}+\mathcal{O}(e^5).
\label{betae}
\end{align}
We can easily find the $\beta$-function in terms of the coupling constant $\alpha$ to be 
\begin{equation}
\beta(\alpha) = \frac{2\alpha^2}{3\pi}+\mathcal{O}(\alpha^3).
\label{betaalpha}
\end{equation}
It is straightforward now, using the $\mu$ dependence of the coupling constant given in \Cref{betaalpha}, to see that \Cref{NLO-Rutherford} satisfies the renormalization group equation given in \Cref{rengroup} at NLO where we have
\begin{align}
\frac{d}{d\mu_{\overline{\mathrm{MS}}}}\left(\frac{d\sigma}{d\Omega}\right)_{\texttt{NLO}}= \mathcal{O}(\alpha^4).
\label{dsigmadmu}
\end{align} 
\Cref{dsigmadmu} is a non-trivial check of our final complete NLO cross section formula \Cref{NLO-Rutherford}.

\chapter{Remarks and Conclusions}
\label{Chapter7} 

We calculated the first complete next-to-leading order, high-energy Rutherford elastic scattering cross section in the $\overline{\mathrm{MS}}$ renormalization scheme. We included all one loop contributions: the vertex, vacuum polarization, electron self-energy, and the box correction. We used dimensional regularization, the fictitious photon mass, and the electron mass to render the UV and IR (soft and collinear) divergences respectively finite. The regularized UV divergences were eliminated by the application of the $\overline{\mathrm{MS}}$ renormalization scheme.

We gave an overview of the BN and KLN theorems. The BN theorem states that summing over degenerate final states yields a result free of soft divergences. The KLN theorem states that summing over degenerate initial and final states yields a result free of all IR divergences. We described how all previous attempts to implement the KLN theorem were either incorrect or incomplete.

A self-consistent application of the KLN theorem requires a sum over all degenerate initial and final states to arrive at an IR safe cross section.  We included the full summation over all degenerate initial and final states including disconnected cut diagrams.  This summation is formally divergent; after introducing a convergence factor, we proved that our rearrangement of this summation allows one to safely exchange taking the limit of the convergence factor to infinity prior to the infinite sum limit. After taking the convergence factor to infinity, there was a complete cancellation of the IR divergences, and the infinite contribution from disconnected soft photons factored out. The infinite factor is not related to the absence of particle masses; hence, although infinite, we found the total transition probability is free of any IR divergence. Consistent with intuitive reasoning, the infinite factor from the soft initial state radiation will cancel in the physically observable cross section through the normalization of the $S$-matrix.

As was noted by Weinberg \cite{Weinberg:1995mt}, \emph{``no one has given a complete demonstration that the sums of transition rates that are free of infrared divergences are the only ones that are experimentally measurable.''} We believe that our work here can be expanded in the future to provide such a proof, especially for non-Abelian gauge theories. Additional possible future work includes investigating the application of the KLN theorem to the study of the electron mass singularity in the electron loop light-by-light contribution to the anomalous muon magnetic moment at order $\alpha^3$ \cite{Aldins:1969jz,Aldins:1970id}.

In the application of the KLN theorem to NLO Rutherford scattering, we arrived at the extremely nontrivial result that the summation over \emph{all} indistinguishable initial and final states is equivalent to the summation over \emph{only} the initial hard collinear and final soft, hard collinear, and soft and collinear degenerate states. 

Using the $\overline{\mathrm{MS}}$ renormalization scheme and the correct implementation of the KLN theorem gives us the complete NLO Rutherford scattering in \Cref{NLO-Rutherford}, which is finite as we send the mass of the electron to zero; equivalently, our result is valid up to arbitrarily large momentum exchange as long as $\alpha \log(1/\delta)\ll 1$ and $\alpha \log(1/\delta)\log(\Delta/E)\ll 1$. A non-trivial check of our result is that \Cref{NLO-Rutherford} satisfies the Callan-Symanzik equation. 

Future work will extend these calculations to derive a formula for the energy lost by high momentum particles propagating through weakly coupled plasmas in thermal field theory, including the next-to-leading-order corrections due to the emission of very high energy particles. This work will lead towards new insights in the dynamics of the QGP, in particular into the nature of the degrees of freedom at energy densities $(1 - 3)T_c$ from lattice QCD seen in \Cref{EDT}.


\appendix 


\chapter{Conventions and Integrals}\label{AppendixA} 

\section{Conventions}
Natural units:
\begin{equation}
\hbar = c = 1.\label{units}
\end{equation}
Electromagnetic coupling constant:
\begin{equation}
\alpha_e = \frac{e^2}{4\pi^2}.\label{alphae}
\end{equation}
Number of space-time dimensions:
\begin{equation}
d = 4-\epsilon.\label{dimensions}
\end{equation}
Bjorken-Drell metric:
\begin{equation}
g^{\mu\nu} = \text{diag}(1,-1,-1,-1),\label{metric}
\end{equation}
where in $d$-dimensions we have $g_{\mu\nu}g^{\mu\nu} = \delta^{\mu}_{\ \mu}= d$.\\
Dirac slash momentum:
\begin{equation}
\slashed{p} = \gamma^{\mu} p_{\mu}.\label{pslash}
\end{equation}
Clifford Algebra:
\begin{equation}
\{\gamma^{\mu},\gamma^{\nu}\}= 2 g^{\mu\nu}.\label{diracalgebra}
\end{equation}
Euler-Mascheroni constant: 
\begin{equation}
\gamma_E \approx 0.5772.\label{eulercons}
\end{equation}
\section{Properties of $\gamma$-matrices}
The traces of $\gamma$-matrices are given by \cite{Peskin:1995ev}
\begin{equation}
\begin{aligned}
\mathrm{tr}(1)& =4,\\
\mathrm{tr}(\text{any odd \# of $\gamma$'s})&= 0,\\
\mathrm{tr}(\gamma^{\mu}\gamma^{\nu}) &= 4 g^{\mu\nu},\\
\mathrm{tr}(\gamma^{\mu}\gamma^{\nu}\gamma^{\rho}\gamma^{\sigma} )&= 4\left[g^{\mu\nu}g^{\rho\sigma}-g^{\mu\rho}g^{\nu\sigma}+g^{\mu\sigma}g^{\nu\rho}\right].
\end{aligned}
\label{trgamma}
\end{equation} 
The contraction of $\gamma$-matrices in $d$-dimensions are given by \cite{Peskin:1995ev}
\begin{equation}
\begin{aligned}
\gamma^{\mu}\gamma_{\mu} &= d,\\
\gamma^{\mu}\gamma^{\nu}\gamma_{\mu} &= (d-2)\gamma^{\nu},\\
\gamma^{\mu}\gamma^{\nu}\gamma^{\rho}\gamma_{\mu} &= 4g^{\nu\rho}-(4-d) \gamma^{\nu}\gamma^{\rho}.
\end{aligned}
\label{contrgamma}
\end{equation} 
\section{Feynman Parameters}
The Feynman parameters trick used to combine the propagator denominators, the general identity is given by \cite{Peskin:1995ev}
\begin{equation}
\frac{1}{D_1D_2\dots D_n}=\int_0^1dx_1\dots dx_n\delta\left(\sum x_i-1\right)\frac{(n-1)!}{\left[x_1D_1+x_2D_2+\dots x_nD_n\right]^n}.
\label{feynparg}
\end{equation}
A special case when we have only two denominators will be
\begin{equation}
\frac{1}{D_1D_2}=\int_0^1dx\frac{1}{\left[x D_1+(1-x)D_2\right]^2}.
\label{feynpar2}
\end{equation}
\section{Integrals in $d$-dimensions}
\begin{itemize}
\item[\emph{a)}] \emph{Scalar Integrals:}
\par 
The standard scalar integrals are given by \cite{Peskin:1995ev}
\begin{equation}
\begin{aligned}
\int \frac{d^d\ell}{(2\pi)^d}\frac{1}{\left(\ell^2-M^2+i\epsilon\right)^n}& = \frac{(-1)^n i}{(4\pi)^{d/2}}\frac{\Gamma\left(n-\frac{d}{2}\right)}{\Gamma(n)}\left(M^2\right)^{\frac{d}{2}-n},\\
\int \frac{d^d\ell}{(2\pi)^d}\frac{\ell^2}{\left(\ell^2-M^2+i\epsilon\right)^n}& = \frac{(-1)^{n-1} i}{(4\pi)^{d/2}}\frac{d}{2}\frac{\Gamma\left(n-\frac{d}{2}-1\right)}{\Gamma(n)}\left(M^2\right)^{1+\frac{d}{2}-n}.
\end{aligned}
\label{scalarellint}
\end{equation}
\item[\emph{b)}] \emph{Tensor Integrals:}
\par
The standard tensor integrals are given by \cite{Peskin:1995ev}
\begin{equation}
\int \frac{d^d\ell}{(2\pi)^d}\frac{\ell^{\mu}}{\left(\ell^2-M^2+i\epsilon\right)^n}= 0,
\label{tensorellint1}
\end{equation}
which implies that the contribution from $\mathcal{\ell}$ terms in the numerator vanishes.
 \begin{equation}
\int \frac{d^d\ell}{(2\pi)^d}\frac{\ell^{\mu}\ell^{\nu}}{\left(\ell^2-M^2+i\epsilon\right)^n}= \frac{(-1)^{n-1} i}{(4\pi)^{d/2}}\frac{g^{\mu\nu}}{2}\frac{\Gamma\left(n-\frac{d}{2}-1\right)}{\Gamma(n)}\left(M^2\right)^{1+\frac{d}{2}-n}.
\label{tensorellint2}
\end{equation}
\end{itemize}
The following expansions may be done for $\Gamma$-function around $\epsilon = 0$ as $d\rightarrow 4$ which are very useful in special cases where $n = 2$ or $3$
\begin{equation}
\begin{aligned}
\Gamma(\epsilon) &= \frac{1}{\epsilon}-\gamma_E + \mathcal{O}(\epsilon),\\
(M^2)^{-\epsilon} & = 1-\epsilon\log M^2+\mathcal{O}(\epsilon^2).
\end{aligned}
\label{intapprox}
\end{equation}

\chapter{Feynman Rules}\label{AppendixB} 

\section{Feynman Rules for the Bare Lagrangian}
\begin{figure}[h]
\centering
\includegraphics[scale=1]{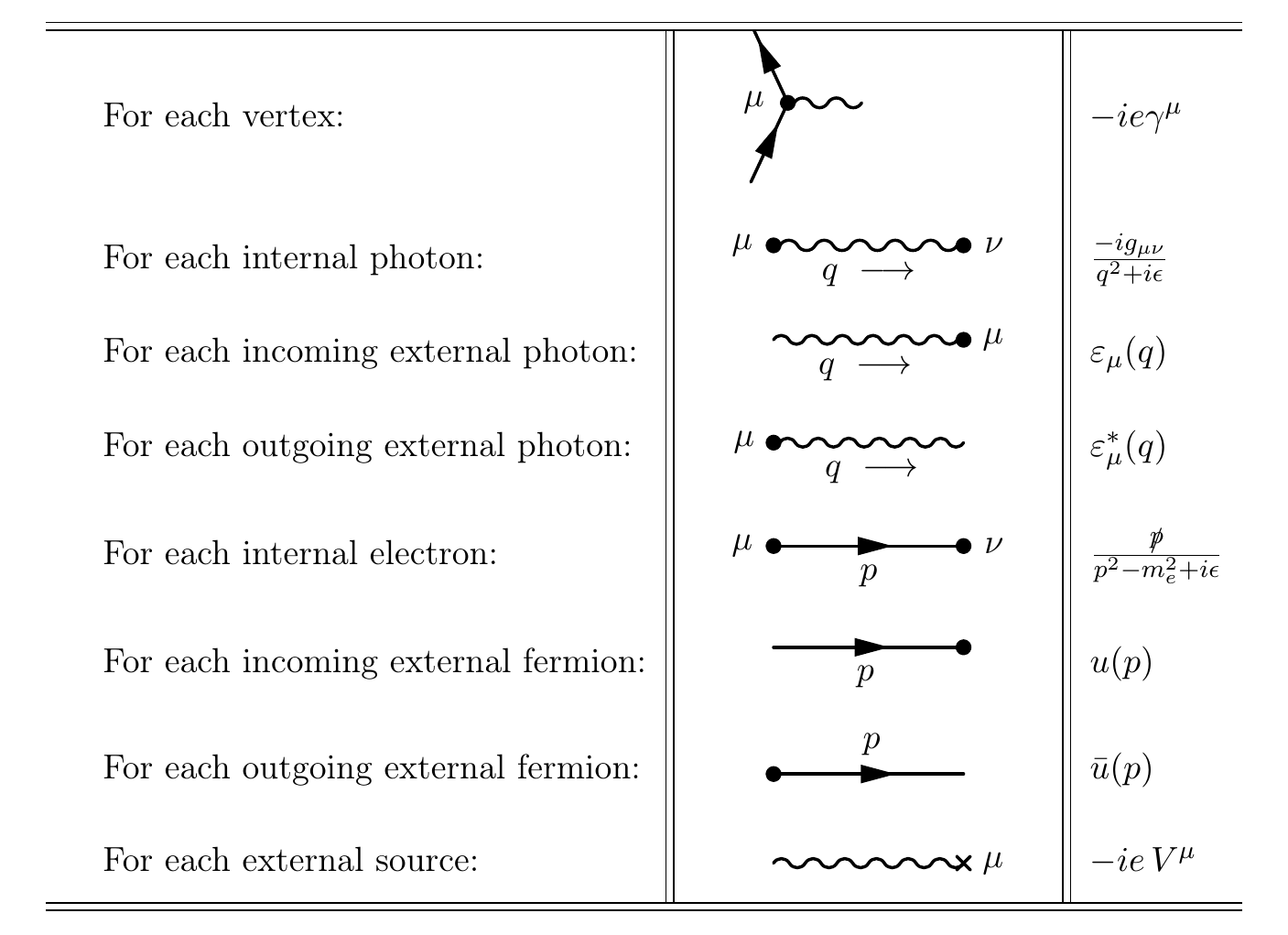}
\caption{Feynman rules of the bare Lagrangian.}
\label{table1}
\end{figure}

We provide in \Cref{table1} the Feynman rules for the system described by the Lagrangian density given by \Cref{qedlagrangian}. These are similar to the rules for the normal QED Lagrangian defined in \cite{Peskin:1995ev}
\section{Feynman Rules for the Renormalized Lagrangian}
\begin{figure}[h]
\centering
\includegraphics[scale=1]{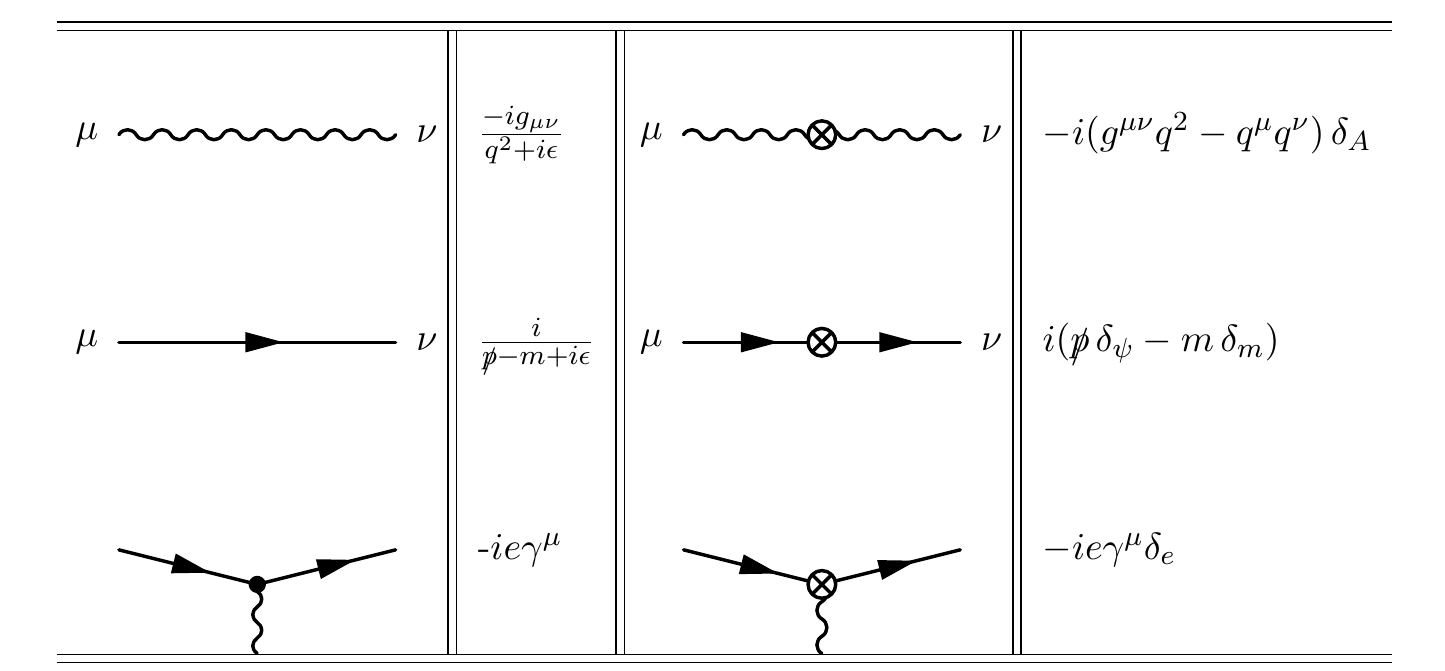}
\caption{Feynman rules for the renormalized QED Lagrangian.}
\label{renormrules}
\end{figure}

\Cref{renormrules} shows the complete Feynman rules as result of the renormalzation procedure \cite{Peskin:1995ev}. We see that the rule in the top gives an insight to the photon field renormalization, while the middle rule shows the renormalization of both the electron field and the mass. The last rule is related directly to the remaining counter term which describes the renormalization of the electric charge.
\chapter{Contributions from Disconnected Diagrams}\label{AppendixC}
\section{Absorption-emission contribution}
We consider here the contribution from the absorption-emission diagrams when they interfere with the tree level amplitude with a disconnected soft photon. We first write the amplitude of each pair of diagrams shown in \Cref{fig:abs-em}. The first pair of diagrams amplitude is given by
\begin{align}
i\mathcal{M}_{\texttt{ae}}^{(1)} & =\hspace{.5cm} \begin{gathered}
\includegraphics[scale=1]{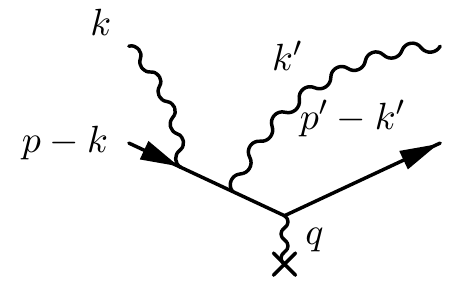}
\end{gathered}\hspace{0.5cm}+\hspace{.5cm}
\begin{gathered}
\includegraphics[scale=1]{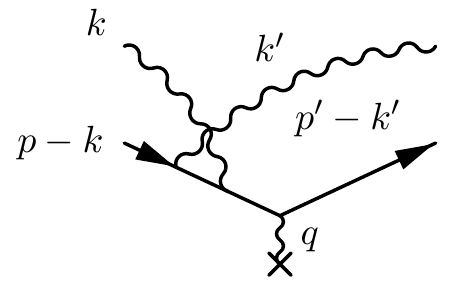}
\end{gathered}\nonumber\\
& = \bar{u}^{s'}(p'-k') \left[-ie\gamma^{\mu}\right] \frac{i(\slashed{p}-\slashed{k'}+m)}{(p-k')^2-m^2}\left[-ie\gamma^{\alpha}\right]\varepsilon_{\alpha}^{'*}(k') \frac{i(\slashed{p}+m)}{p^2-m^2}\left[-ie \gamma^{\beta}\right] \varepsilon_{\beta}(k) \nonumber\\ & \quad\times u^s(p-k)\left[-ieV^{\nu}\right]\frac{-ig_{\mu\nu}}{q^2} + \bar{u}^{s'}(p'-k') \left[-ie\gamma^{\mu}\right]\frac{i(\slashed{p}-\slashed{k}'+m)}{(p-k')^2-m^2}\left[-ie\gamma^{\alpha}\right]\varepsilon_{\alpha}(k)\nonumber\\&\quad\times\frac{i(\slashed{p}-\slashed{k}-\slashed{k}'+m)}{(p-k-k')^2-m^2}  \left[-ie\gamma^{\beta}\right]\varepsilon_{\beta}^{'*}(k') u^s(p-k)\left[-ieV^{\nu}\right]\frac{-ig_{\mu\nu}}{q^2}.
\label{abs-em1}
\end{align}

In the soft photon limit, using the Eikonal approximation, we can write $\slashed{p}-\slashed{k}'+m \approx \slashed{p}+m$,
$\slashed{p}-\slashed{k}-\slashed{k}'+m \approx \slashed{p}+m$, $\bar{u}^{s'}(p'-k')\approx\bar{u}^{s'}(p')$, and $u^s(p-k) \approx u^s(p)$ which allow us to rewrite the numerators of \Cref{abs-em1} in the form of either $(\slashed{p}+m)\slashed{\varepsilon}\ u^s(p) \approx 2(p\cdot \varepsilon)\ u^s(p)$ and $(\slashed{p}+m)\slashed{\varepsilon}^{'*}\ u^s(p) \approx 2(p\cdot \varepsilon^{'*})\ u^s(p)$. Then \Cref{abs-em1} becomes 
\begin{multline}
i\mathcal{M}_{\texttt{ae}}^{(1)}  = \frac{4ie^4}{q^2}\ \bar{u}^{s'}(p')\gamma^0 u^s(p)\ (p\cdot \varepsilon)(p\cdot \varepsilon^{'*})\left(\frac{1}{\left[(p-k')^2-m^2\right]\left[p^2-m^2\right]}\right.\\
\left.+\frac{1}{\left[(p-k')^2-m^2\right]\left[(p-k-k')^2-m^2\right]}\right).
\label{abs-em1s}
\end{multline}
Since $p-k$ is on-shell momenta, we can write $p^2 \approx 2p\cdot k + m^2$. Then we can simplify the denominators of \Cref{abs-em1s}, where we write the denominator of the first term as
\begin{align}
\left[(p-k')^2-m^2\right]\left[p^2-m^2\right] & = \left[p^2-2p \cdot k'-m^2\right]\left[p^2-m^2\right]\nonumber\\
& \approx 2p\cdot k \left[2p\cdot (k-k')\right],
\label{abs-em1-den1}
\end{align}
while the denominator of the second term will be
\begin{align}
\left[(p-k')^2-m^2\right]\left[(p-k-k')^2-m^2\right]& = \left[p^2-2p\cdot k' -m^2 \right]\left[-2k'\cdot (p-k)\right]\nonumber\\
&\approx -2p\cdot k' \left[2p\cdot (k-k')\right].
\label{abs-em1-den2}
\end{align}

We note that the denominators in \Cref{abs-em1-den1,abs-em1-den2} diverge separately as we send $k\rightarrow k'$ which come out from the delta function in the amplitude of the diagram in \Cref{fig:tdiscon}. However, these divergences disappear in each pair of diagrams where they describe a physical process together. We then write the amplitude given in \Cref{abs-em1s} to be
\begin{align}
i\mathcal{M}_{\texttt{ae}}^{(1)}  = \frac{-ie^4}{q^2} \ u^{s'}(p')\gamma^0u^s(p)\left[\frac{(p\cdot \varepsilon)(p\cdot \varepsilon^{'*})}{(p\cdot k)(p\cdot k')}\right].
\label{abs-em1f}
\end{align}

The calculations of the amplitude for each pair is very similar to what we have done here, where we can easily find in the soft limit and using the on-shell conditions that 
\begin{align}
i\mathcal{M}_{\texttt{ae}}^{(2)}& =\hspace{0.5cm} \begin{gathered}
\includegraphics[scale=1]{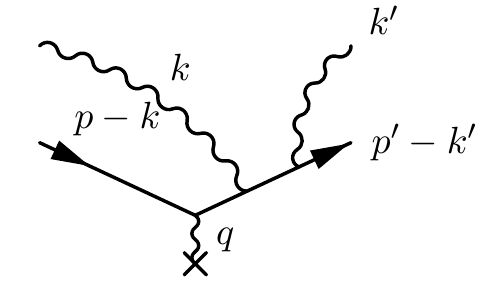}
\end{gathered}\hspace{.5cm}+\hspace{0.5cm}
\begin{gathered}  
\includegraphics[scale=1]{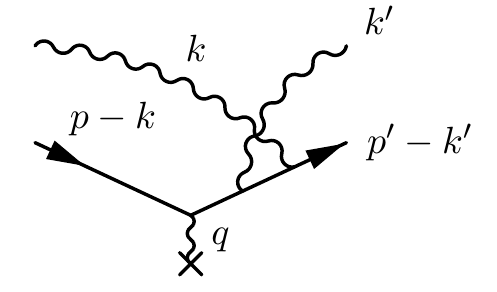}
\end{gathered}
\nonumber\\
&= \frac{-ie^4}{q^2} \ u^{s'}(p')\gamma^0 u^s(p)\left[\frac{(p'\cdot \varepsilon)(p'\cdot \varepsilon^{'*})}{(p'\cdot k)(p'\cdot k')}\right].
\label{abs-em2}
\end{align}
Similarly, we also find the amplitude of the last pair is given by
\begin{align}
i\mathcal{M}_{\texttt{ae}}^{(3)}& =\hspace{.5cm}
\begin{gathered}  
\includegraphics[scale=1]{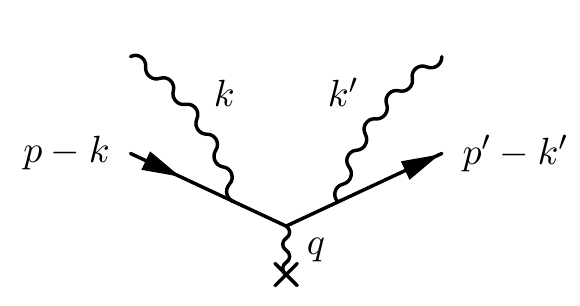}
\end{gathered} \hspace{.5cm}+\hspace{0.5cm}
\begin{gathered}
\includegraphics[scale=1]{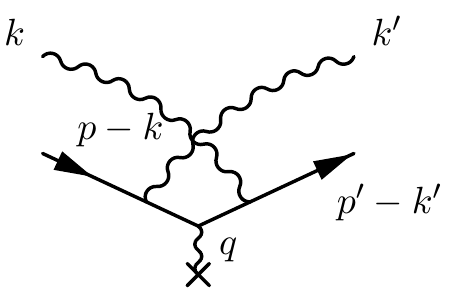}
\end{gathered}
\nonumber\\
&= \frac{-ie^4}{q^2} \ u^{s'}(p')\gamma^0 u^s(p)\left[\frac{(p\cdot \varepsilon)(p'\cdot \varepsilon^{'*})}{(p'\cdot k)(p\cdot k')}+\frac{(p'\cdot \varepsilon)(p\cdot \varepsilon^{'*})}{(p'\cdot k)(p\cdot k')}\right].
\label{abs-em3}
\end{align}
Let us now collect all the amplitudes that contain both an absorption and emission of a soft photon by adding \Cref{abs-em1f,abs-em2,abs-em3}, where we keep in mind that the delta functions from the disconnected photon ensure that $k'\rightarrow k$ and $\varepsilon' \rightarrow \varepsilon$, we find 
\begin{align}
i\mathcal{M}_{\texttt{ae}} &= \frac{-ie^4}{q^2}  u^{s'}(p')\gamma^0 u^s(p)\left[\frac{(p\cdot \varepsilon)^2}{(p\cdot k)^2}+\frac{(p'\cdot \varepsilon)^2}{(p'\cdot k)^2}-\frac{2(p\cdot \varepsilon)(p'\cdot \varepsilon)}{(p\cdot k)(p'\cdot k)}\right]\nonumber\\
&= -i\mathcal{M}_{\texttt{0}} e^2 \left[\frac{(p\cdot \varepsilon)}{(p\cdot k)}-\frac{(p'\cdot \varepsilon)}{(p'\cdot k)}\right]^2.
\label{abs-em-all}
\end{align}
We recall the amplitude of the diagram in \Cref{fig:tdiscon}
\begin{align}
i\mathcal{M}_{\texttt{d}}^{(1)} = i\mathcal{M}_{\texttt{0}} (2\pi)^3\ 2\omega_k \delta^{(3)}(k-k')\ \delta_{\varepsilon\varepsilon'},
\label{tdis-amp}
\end{align}
where the superscript $(1)$ refers that there is only one disconnected photon. The interference between the tree amplitude with a disconnected photon and the absorption-emission diagrams is given by
\begin{align}
\left|\mathcal{M}_{\texttt{AE}}\right|^2 = 2 \mathcal{M}_{\texttt{ae}}^* \mathcal{M}_{\texttt{d}}^{(1)} = -2e^2 \left|\mathcal{M}_{\texttt{0}}\right|^2\left[\frac{(p\cdot \varepsilon)}{(p\cdot k)}-\frac{(p'\cdot \varepsilon)}{(p'\cdot k)}\right]^2(2\pi)^3\ 2\omega_k \delta^{(3)}(k-k')\ \delta_{\varepsilon\varepsilon'}.
\label{abs-em-amp-sq}
\end{align}
Summing over all the polarizations and integrating over the photon momenta $k$ and $k'$ to find the contribution to the differential cross section where the delta function will absorb the integral over $k'$ leaving only 
\begin{align}
\left(\frac{d\sigma}{d\Omega}\right)_{\texttt{AE}} & = \left(\frac{d\sigma}{d\Omega}\right)_{\texttt{0}}(-2e^2) \int \frac{d^3k}{(2\pi)^3 2\omega_k} \sum_{\text{pol}}\left[\frac{(p\cdot \varepsilon)}{(p\cdot k)}-\frac{(p'\cdot \varepsilon)}{(p'\cdot k)}\right]^2\nonumber\\
&= -2\left(\frac{d\sigma}{d\Omega}\right)_{\texttt{S}}^i.
\label{abs-em-xsec-app}
\end{align}

We note that, for simplicity, we assumed that the energy resolution for the initial and final states are the same. In case of different energy resolutions, $\Delta_i$ and $\Delta_f$ for the initial and final states respectively, we can easily find that 
\begin{align}
\left(\frac{d\sigma}{d\Omega}\right)_{\texttt{AE}} & = \frac{1}{32\pi^2}\left[\frac{1}{2}\int_0^{\Delta_i}\frac{d^3k}{(2\pi)^32 \omega_k}\int_0^{\Delta_f}\frac{d^3k'}{(2\pi)^3 2\omega_{k'}}\sum_{\text{pol}}\left|\mathcal{M}_{\texttt{AE}}\right|^2\right.\nonumber\\&\left.\hspace{2cm}+\frac{1}{2}\int_0^{\Delta_f}\frac{d^3k'}{(2\pi)^3 2\omega_{k'}}\int_0^{\Delta_i}\frac{d^3k}{(2\pi)^32 \omega_k}\sum_{\text{pol}}\left|\mathcal{M}_{\texttt{AE}}\right|^2\right]\nonumber\\
& = -\left[\left(\frac{d\sigma}{d\Omega}\right)_{\texttt{S}}^i+\left(\frac{d\sigma}{d\Omega}\right)_{\texttt{S}}^f\right].
\label{abs-em-xsec-sep}
\end{align}
We can also write from \Cref{abs-em-amp-sq} the probability $\widetilde{\mathcal{P}}_{11}$ to be 
\begin{align}
\widetilde{\mathcal{P}}_{11} & = \left[\frac{1}{2}\int_0^{\Delta_i}\frac{d^3k}{(2\pi)^32 \omega_k}\sum_{\text{pol}}\left|\mathcal{M}_{\texttt{AE}}\right|^2+\frac{1}{2}\int_0^{\Delta_f}\frac{d^3k'}{(2\pi)^3 2\omega_{k'}}\sum_{\text{pol}}\left|\mathcal{M}_{\texttt{AE}}\right|^2\right]\nonumber\\& = -\left[\mathcal{P}_{10}+\mathcal{P}_{01}\right].
\label{p11-app}
\end{align}
One can easily show that the choice we made for the equivalent initial and final energy resolutions is reasonable and the result in \Cref{WA-prob} will not be affected by choosing different values.
\section{Emission with a disconnected photon}
Now we consider the contribution from the fully connected cut diagram for a process of an emission of a soft photon with one disconnected soft photon, where the amplitude of such a process in the soft limit photon is given by
\begin{align}
\mathcal{M}_{\texttt{01}}^{(1)}& =\quad\begin{gathered}
\includegraphics[scale=1]{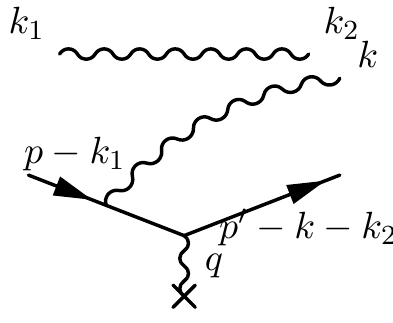}
\end{gathered} \hspace{0.5cm}+\hspace{0.5cm}\begin{gathered}
\includegraphics[scale=1]{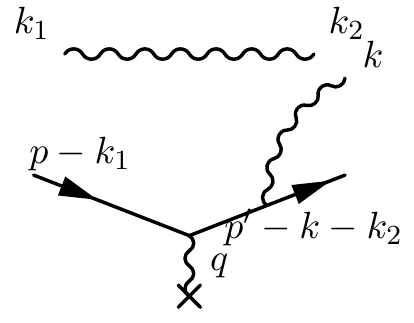}
\end{gathered}\nonumber\\
& = -ie \mathcal{M}_{\texttt{0}}\left[\frac{p\cdot \varepsilon}{p\cdot k}-\frac{p'\cdot \varepsilon}{p'\cdot k}\right]\ (2\pi)^3 2\omega_{k_1} \delta^{(3)}(k_1-k_2) \ \delta_{\varepsilon_1\varepsilon_2}.
\label{m01-d}
\end{align}

In order to get the fully connected contribution from these diagrams, one should write the complex conjugate of the amplitude in \Cref{m01-d} such that $k$ is replaced by $k_1$ and vice versa and so the polarizations of the two photons, then the complex conjugate of the amplitude becomes
\begin{align}
\left[\mathcal{M}_{\texttt{01}}^{(1)}\right]^*& = ie\mathcal{M}_{\texttt{0}}^* \left[\frac{p\cdot \varepsilon_1}{p\cdot k_1}-\frac{p'\cdot \varepsilon_1}{p'\cdot k_1}\right] \ (2\pi)^3 2\omega_{k} \delta^{(3)}(k-k_2) \ \delta_{\varepsilon\varepsilon_2}.
\label{m01-ds}
\end{align} 
The delta functions in \Cref{m01-d,m01-ds} ensure that $k = k_1 = k_2$ and $\varepsilon=\varepsilon_1=\varepsilon_2$ to get the fully connected contributions. Squaring the amplitude, we get
\begin{align}
\left[\left|\mathcal{M}_{\texttt{01}}\right|^{(1)}\right]^2&=  e^2 \left|\mathcal{M}_{\texttt{0}}\right|^2 \left|\frac{p\cdot \varepsilon}{p\cdot k}-\frac{p'\cdot \varepsilon}{p'\cdot k}\right|^2\ (2\pi)^6\  2\omega_{k_1}\ 2\omega_{k_2} \delta^{(3)}(k_1-k_2)\ \delta^{(3)}(k-k_2).
\label{m01-d-sq}
\end{align}
Then the contribution to the differential cross section is given by
\begin{align}
\left(\frac{d\sigma}{d\Omega}\right)_{\texttt{01}}^{(1)} & = \frac{1}{32\pi^2}\int\frac{d^3k}{(2\pi)^3 2\omega_k}\int\frac{d^3k_1}{(2\pi)^3 2\omega_{k_1}}\int\frac{d^3k_2}{(2\pi)^3 2\omega_{k_2}} \sum_{pol}\left[\left|\mathcal{M}_{\texttt{01}}\right|^{(1)}\right]^2\nonumber\\
& = \left(\frac{d\sigma}{d\Omega}\right)_{\texttt{0}} \int \frac{d^3k}{(2\pi)^3 2\omega_k} \sum_{\text{pol}}e^2\left|\frac{(p\cdot \varepsilon)}{(p\cdot k)}-\frac{(p'\cdot \varepsilon)}{(p'\cdot k)}\right|^2\nonumber\\
&= \left(\frac{d\sigma}{d\Omega}\right)_{\texttt{S}}^f.
\label{em-d-xsec}
\end{align}

We note that all the fully connected contributions from the diagrams that have an emission of a soft photon with a number of disconnected soft photons can be given by \Cref{em-d-xsec}. Similarly, the fully connected contributions from the absorption of a soft photon with a number of disconnected soft photons are given by the absorption of a soft photon contributions.
\chapter{Soft bremsstrahlung beyond eikonal approximation}\label{AppendixD}

In order to make sure that the $\left(\frac{\Delta}{E}\right)^{\alpha}\log\left(\frac{E^2}{m^2}\right)$, $\alpha = 1, 2$, from the emission and absorption of a hard and collinear photon contributions cancels, we redo the calculations of the soft bremsstrahlung contribution without the eikonal approximation. We can start from equation \Cref{bremampsquaredf}, where we have not used the eikonal approximation for simplifying the scattering amplitude. However, we perform the integrals over all scattering angles $\theta_{\gamma}$ and only soft photons (i.e$.\ 0<k<\Delta$). From \Cref{t1,t2,t3t4}, we can rewrite the differential cross section for the final state soft bremsstrahlung
\begin{multline}
\left(\frac{d\sigma}{d\Omega}\right)_{\texttt{s}}^f = \frac{-2\alpha^2}{q^4}\left(\frac{\alpha}{2\pi}\right)\int \frac{k^2}{\omega_k} \ dk\ d\cos \theta_{\gamma} \ \frac{d\phi}{2\pi}\left[\frac{m^2(2E^2-p\cdot p')}{(p\cdot k)^2}-\frac{(2E\ \omega_k-p'\cdot k)}{(p\cdot k)}+\frac{2\omega_k^2}{(p\cdot k)}\right.\\
+\left.\frac{m^2(2E^2-p\cdot p')}{(p'\cdot k)^2}-\frac{(2E\ \omega_k-p\cdot k)}{(p'\cdot k)}-\frac{2p\cdot p' (2E^2-p\cdot p')}{(p\cdot k)(p'\cdot k)}+\frac{2(E^2-p\cdot p')}{(p\cdot k)}\right. \\
\left.+\frac{2p \cdot p'(2E\ \omega_k -p\cdot k)}{(p\cdot k)(p'\cdot k)}-\frac{2\omega_k^2 (p\cdot p')}{(p\cdot k)(p'\cdot k)}+\frac{ 2 E\ \omega_k}{(p'\cdot k)}-\frac{ 2 E\ \omega_k}{(p\cdot k)}\right].
\label{sbrem-be}  
\end{multline}
Let us now rewrite \Cref{sbrem-be} by labelling  the integrals to be $I_1+ I_2+\ldots+I_{11}$. The integrals in $I_1$ and $I_4$ are similar to the integrals calculated in $\mathcal{A}_1$ and $\mathcal{A}_2$ in \Cref{sbreminta1a2}. Then we have 
\begin{align}
I_1+I_4  = \left(\frac{d\sigma}{d\Omega}\right)_{\texttt{0}}\cdot \frac{\alpha}{\pi}\left[\log\left(\frac{E^2}{m^2}\right)-\log\left(\frac{\Delta^2}{m_{\gamma}^2}\right)+\mathcal{O}(m^2,m_{\gamma}^2)\right].
\label{I2I4-be}
\end{align}

The second integral of \Cref{sbrem-be} is given by
\begin{align}
I_2= \frac{2\alpha^2}{q^4}\left(\frac{\alpha}{2\pi}\right)\int \frac{k^2}{\omega_k} \ dk\ d\cos \theta_{\gamma} \ \frac{d\phi}{2\pi} \frac{2E\,\omega_k-p'\cdot k }{p\cdot k}.
 \label{I2int-be}
\end{align}
We note that $I_2$ is finite as $m_{\gamma}\rightarrow 0$. So we can safely replace $\omega_k$ by $k$. In order to solve the $I_2$ integral, we choose a reference frame in which
$p^{\mu}\equiv(E,0,0,p)$, $p'^{\mu}\equiv (E,0,p_y,p_z)$, and $k^{\mu} \equiv (\omega_k,k\sin\theta_\gamma\cos\phi,
k\sin\theta_\gamma\sin\phi,k\cos\theta_\gamma)$. Then we have $|\vec{p}|^2 \approx|\vec{p}\,'|^2 \approx \sqrt{E^2-m^2}$ which implies that $p_y^2+p_z^2 \approx p^2 $. We also have $q^2 = (p'-p)^2 = -p_y^2-(p_z-p)^2 = -2p^2+2p\,p_z$ which implies that $p_z = \frac{q^2}{2p}+p$. With all of the previous tools in our hands we can rewrite
\begin{equation}
\begin{aligned}
p\cdot k &= Ek -pk\cos\theta_\gamma = Ek\left[1-\frac{p}{E}\cos\theta_\gamma\right],\\
p'\cdot k&= Ek-p_yk\sin\theta_\gamma\sin\phi-p_zk\cos\theta_\gamma = Ek\left[1-\left(\frac{p_y}{E}\sin\theta_\gamma\sin\phi+\frac{p_z}{E}\cos\theta_\gamma\right)\right].
\end{aligned}
\label{kin1-be}
\end{equation}
Substituting \Cref{kin1-be} into \Cref{I2int-be}, we also not that the term with $\sin\phi$ will disappear due to the $\phi$ integral. Then we find
\begin{align}
I_2&=\frac{2\alpha^2}{q^4}\left(\frac{\alpha}{2\pi}\right)\int k \ dk\ d\cos \theta_{\gamma} \ \frac{d\phi}{2\pi}\frac{\left[2E\,k -E\,k\left(1-\frac{(p+q^2/2p)}{E}\cos\theta_{\gamma}\right)\right]}{E\,k(1-\frac{p}{E}\cos\theta_{\gamma})}\nonumber\\
& = \frac{2\alpha^2}{q^4}\left(\frac{\alpha}{2\pi}\right)\int k \ dk\ d\cos \theta_{\gamma} \ \frac{d\phi}{2\pi}\left[\frac{2}{(1-\frac{p}{E}\cos\theta_{\gamma})}+\frac{q^2}{2E^2}\frac{\cos\theta_{\gamma}}{\frac{p}{E}(1-\frac{p}{E}\cos\theta_{\gamma})}-1\right].
\label{I2int1-be}
\end{align}
The last term in the previous equation will be in $\mathcal{O}(\Delta^2)$ with no expected large logs in front, so it can be neglected. The angular integrals in \Cref{I2int1-be} can be easily evaluated as following
\begin{align}
\eta_1 &= \int_{-1}^{1}d\cos\theta_{\gamma}\frac{1}{1-\sqrt{1-\frac{m^2}{E^2}}\cos\theta_{\gamma}} = \log\left(\frac{4E^2}{m^2}\right)+\mathcal{O}(m^2),\label{I2J1}\\
\eta_2 & = \frac{q^2}{2E^2}\int_{-1}^{1}d\cos\theta_{\gamma}\frac{\cos\theta_{\gamma}}{\sqrt{1-\frac{m^2}{E^2}}\left[1-\sqrt{1-\frac{m^2}{E^2}}\cos\theta_{\gamma}\right]} =\frac{q^2}{2E^2} \left[\log\left(\frac{4E^2}{m^2}\right)-2\right]+\mathcal{O}(m^2).\label{I2J2}
\end{align} 
Again, we can neglect the $-2$ term in $\eta_2$ as it will produce a term in $\mathcal{O}(\Delta^2)$. Substituting \Cref{I2J1,I2J2} into \Cref{I2int1-be} and perform the $k$ integral we get
\begin{align}
I_2 = \left(\frac{d\sigma}{d\Omega}\right)_{\texttt{0}}\cdot \frac{\alpha}{2\pi}\left[\frac{\Delta^2}{2E^2}\log\left(\frac{4E^2}{m^2}\right)+\mathcal{O}(m^2,m_{\gamma}^2)\right].
\label{I2f-be}
\end{align}
Similarly, for $I_5$ by choosing the appropriate reference frame, we will get exactly the same result as in $I_2$. Then we have
\begin{align}
I_2+I_5 = \left(\frac{d\sigma}{d\Omega}\right)_{\texttt{0}}\cdot \frac{\alpha}{\pi}\left[\frac{\Delta^2}{2E^2}\log\left(\frac{4E^2}{m^2}\right)+\mathcal{O}(m^2,m_{\gamma}^2)\right].
\label{I2I5-be}
\end{align}

We can combine $I_6$ and $I_7$ and recall that $p\cdot p' = m^2 -q^2/2+p'\cdot k$; and we find
\begin{align}
I_6+I_7 &=  \frac{2\alpha^2}{q^4}\left(\frac{\alpha}{2\pi}\right)\int \frac{k^2}{\omega_k} \ dk\ d\cos \theta_{\gamma} \frac{d\phi}{2\pi}\frac{2(p\cdot p'-p'\cdot k)(2E^2 - p\cdot p')}{(p\cdot k)(p'\cdot k)}\nonumber \\ 
& = \frac{2\alpha^2}{q^4}\left(\frac{\alpha}{2\pi}\right)\int \frac{k^2}{\omega_k} \ dk\ d\cos \theta_{\gamma}\frac{(2m^2-q^2)[2E^2+q^2/2-m^2-p'\cdot k]}{(p\cdot k)(p'\cdot k)}\nonumber\\
&= \left(\frac{d\sigma}{d\Omega}\right)_{\texttt{0}}\cdot \frac{\alpha}{2\pi}\int_0^{\Delta} \frac{k^2}{\omega_k}dk\int_{-1}^1d\cos\theta_{\gamma}\frac{2m^2-q^2}{(p\cdot k)(p'\cdot k)}\nonumber\\&\qquad\qquad-\frac{2\alpha^2}{q^4}\left(\frac{\alpha}{2\pi}\right)\int_0^{\Delta} k\,dk\int_{-1}^1d\cos\theta_{\gamma}\frac{2m^2-q^2}{(p\cdot k)}.
\label{I6I71-be}
\end{align}
The integral in the first term of \Cref{I6I71-be} is similar to the $\mathcal{A}_3$ integral given by \Cref{sbreminta3f}, while the second term can be found be evaluating the following integral
\begin{align}
\int_{-1}^1 d\cos\theta_{\gamma}\frac{2m^2-q^2}{\left[1-\frac{p}{E}\cos\theta_{\gamma}\right]} &= -q^2 \int_{-1}^1 d\cos\theta_{\gamma}\frac{1}{\left[1-\frac{p}{E}\cos\theta_{\gamma}\right]}+\mathcal{O}(m^2)\nonumber\\
& = -q^2 \log\left(\frac{4E^2}{m^2}\right)+\mathcal{O}(m^2)
\label{I6I7part2}
\end{align} 
Substituting \Cref{sbreminta3f,I6I7part2} into \Cref{I6I71-be} and performing the $k$ integral, we find
\begin{multline}
I_6+I_7 = \left(\frac{d\sigma}{d\Omega}\right)_{\texttt{0}} \,\frac{\alpha}{\pi}\left[\log\left(\frac{-q^2}{m^2}\right)\log\left(\frac{\Delta^2}{m_{\gamma}^2}\right)-\log\left(\frac{-q^2}{m^2}\right)\log\left(\frac{E^2}{m^2}\right)+\frac{1}{2}\log^2\left(\frac{-q^2}{m^2}\right)\right.\\
\left. -\frac{\pi^2}{6}\right]-\frac{2\alpha^2}{q^4}(q^2/2)\frac{\alpha}{\pi}\left[\frac{\Delta}{E}\log\left(\frac{4E^2}{m^2}\right)\right]+\mathcal{O}(m^2,m_{\gamma}^2).
\label{I6I7-be}
\end{multline}

We now split the $I_8$ integral into two parts $I_8^{(1)}$ and $I_8^{(2)}$ to be easy to calculate, where we have
\begin{align}
I_8& =\frac{-2\alpha^2}{q^4}\left(\frac{\alpha}{2\pi}\right)\int k \ dk\ d\cos \theta_{\gamma} \ \frac{d\phi}{2\pi} \frac{2(p\cdot p')(2E\, k - p\cdot k)}{(p\cdot k)(p'\cdot k)}\nonumber\\
&  =\frac{-2\alpha^2}{q^4}\left(\frac{\alpha}{2\pi}\right)\int k \ dk\ d\cos \theta_{\gamma} \frac{(2m^2-q^2 +2p'\cdot k)(2E\, k -p\cdot k)}{(p\cdot k)(p'\cdot k)}\nonumber\\
& = I_8^{(1)}+I_8^{(2)},
\end{align}
where
\begin{align}
I_8^{(1)}& = \frac{-2\alpha^2}{q^4}\left(\frac{\alpha}{2\pi}\right)\int k \ dk\ d\cos \theta_{\gamma}\left[\frac{2m^2-q^2}{(p\cdot k)(p'\cdot k)}+\frac{2}{p\cdot k}\right]\nonumber\\
&= \frac{-\alpha^2}{q^4}(2E^2)\frac{\alpha}{\pi}\int_0^{\Delta} \frac{dk}{E}\left[2\log\left(\frac{-q^2}{m^2}\right)+\frac{k}{E}\log\left(\frac{4E^2}{m^2}\right)\right]+\mathcal{O}(m^2)\nonumber\\
& = \frac{-\alpha^2}{q^4}(2E^2)\frac{\alpha}{\pi}\int_0^{\Delta} \frac{dk}{E}\left[2\log\left(\frac{4E^2}{m^2}\right)+2\log\left(\frac{-q^2}{4E^2}\right)+\frac{k}{E}\log\left(\frac{4E^2}{m^2}\right)\right]+\mathcal{O}(m^2)
\nonumber\\
& = \frac{-2\alpha^2}{q^4}(2E^2)\frac{\alpha}{\pi}\log\left(\frac{4E^2}{m^2}\right)\left[\frac{2\Delta}{E}+\frac{\Delta^2}{2E^2}\right]+\mathcal{O}(m^2,\Delta). 
\label{I81-be}
\end{align}
Up to a correction of $\mathcal{O}(\Delta)$, we may replace $\log(-q^2)$ by $\log(4E^2)$ as we did in the last line of \Cref{I81-be}. The second part of $I_8$ is given by
\begin{align}
I_8^{(2)}& = \frac{2\alpha^2}{q^4}\left(\frac{\alpha}{\pi}\right)\int_0^{\Delta}k\, dk\int_{-1}^1 d\cos\theta_{\gamma}\frac{m^2-q^2/2}{(p'\cdot k)}\nonumber\\
& = \frac{-2\alpha^2}{q^4}(q^2/2)\frac{\alpha}{\pi}\left[\frac{\Delta}{E}\log\left(\frac{4E^2}{m^2}\right)\right]+\mathcal{O}(m^2).
\label{I82-be}
\end{align}
We can now recombine $I_6$, $I_7$, and $I_8$, which gives
\begin{multline}
I_6+I_7+I_8 =\left(\frac{d\sigma}{d\Omega}\right)_{\texttt{0}} \,\frac{\alpha}{\pi}\left[\log\left(\frac{-q^2}{m^2}\right)\log\left(\frac{\Delta^2}{m_{\gamma}^2}\right)-\log\left(\frac{-q^2}{m^2}\right)\log\left(\frac{E^2}{m^2}\right)+\frac{1}{2}\log^2\left(\frac{-q^2}{m^2}\right)\right.\\
\left.-\frac{2\Delta}{E}\log\left(\frac{4E^2}{m^2}\right) -\frac{\pi^2}{6}\right]-\frac{2\alpha^2}{q^4}(2E^2)\frac{\alpha}{\pi}\left[\frac{\Delta^2}{2E^2}\log\left(\frac{4E^2}{m^2}\right)\right]+\mathcal{O}(m^2,m_{\gamma}^2). 
\label{I6I7I8-be}
\end{multline}

The integral integrals $I_3$ and $I_9$ are finite as $m_{\gamma}\rightarrow 0$. Then we can write $I_9$ as follows
\begin{align}
I_9 &= \frac{2\alpha^2}{q^4}\left(\frac{\alpha}{\pi}\right) \int_0^{\Delta}k^3 dk\int_{-1}^1d\cos\theta_{\gamma}\frac{p\cdot p'}{(p\cdot k)(p'\cdot k)}\nonumber\\
& =  \frac{2\alpha^2}{q^4}\left(\frac{\alpha}{\pi}\right) \int_0^{\Delta}k^3 dk\int_{-1}^1d\cos\theta_{\gamma}\frac{m^2-q^2/2+ p'\cdot k}{(p\cdot k)(p'\cdot k)}\nonumber\\
& =  \frac{2\alpha^2}{q^4}\left(\frac{\alpha}{2\pi}\right) \int_0^{\Delta}k^3 dk\int_{-1}^1d\cos\theta_{\gamma}\frac{2m^2-q^2}{(p\cdot k)(p'\cdot k)}-I_3. 
\label{I91-be}
\end{align}
We can simplify the integral in the first term of \Cref{I91-be} in the same way we did for simplifying $\mathcal{A}_3$ in \Cref{chiintf}. Then we have
\begin{align}
I_3+I_9 & = \frac{2\alpha^2}{q^4}\left(\frac{\alpha}{\pi}\right) \int_0^{\Delta}k dk \int_0^1 dx\frac{2m^2-q^2}{\left[m^2-x(1-x)q^2\right]}\nonumber\\
&= \frac{2\alpha^2}{q^4}(2E^2)\frac{\alpha}{\pi}\left[\log\left(\frac{-q^2}{m^2}\right)\int_0^{\Delta}\frac{k}{E^2}dk\right]+\mathcal{O}(m^2).
\end{align} 
Again, the numerator inside the logarithm is not important as we are only interested for terms that becomes logarithmically large as $m\rightarrow 0$. So we can replace $-q^2$ by $4E^2$, we finally find
\begin{align}
I_3+I_9 = \frac{2\alpha^2}{q^4}(2E^2)\frac{\alpha}{\pi}\left[\frac{\Delta^2}{2E^2}\log\left(\frac{4E^2}{m^2}\right)\right]+\mathcal{O}(m^2).
\label{I3I9-be}
\end{align}

We note that \Cref{I3I9-be} will cancel with the last term of \Cref{I6I7I8-be}. One can easily see that the $I_{10}$ and $I_{11}$ terms give the same contribution with a relative sign difference, which means that $I_{10}+I_{11}=0$. Finally, we substitute \Cref{I2I4-be,I2I5-be,I6I7I8-be,I3I9-be} into \Cref{sbrem-be}, and we find
\begin{multline}
\left(\frac{d\sigma}{d\Omega}\right)_{\texttt{s}}^f= \left(\frac{d\sigma}{d\Omega}\right)_{\texttt{0}} \,\frac{\alpha}{\pi}\left[\log\left(\frac{-q^2}{m^2}\right)\log\left(\frac{\Delta^2}{m_{\gamma}^2}\right)-\log\left(\frac{-q^2}{m^2}\right)\log\left(\frac{E^2}{m^2}\right)+\frac{1}{2}\log^2\left(\frac{-q^2}{m^2}\right)\right.\\
\left.-\frac{2\Delta}{E}\log\left(\frac{4E^2}{m^2}\right)+\frac{\Delta^2}{2E^2}\log\left(\frac{4E^2}{m^2}\right) -\frac{\pi^2}{6}\right]+\mathcal{O}(m^2,m_{\gamma}^2).
\label{softbrem-be}
\end{multline}

\printbibliography[heading=bibintoc]
---------------------------------------------------------------

\end{document}